\documentclass[10pt]{article}
\usepackage{graphicx}
\usepackage[T1]{fontenc}
\usepackage[utf8]{inputenc}
\usepackage{amssymb}
\usepackage{amsfonts}
\usepackage{dsfont}
\usepackage{mathtools}
\usepackage{amsthm}
\usepackage{amsmath}
\usepackage{relsize}
\usepackage{textcomp}
\usepackage{eurosym}
\usepackage{stmaryrd}
\usepackage{xcolor}
\usepackage[multiple]{footmisc}
\usepackage{pdflscape}
\usepackage{enumerate}
\usepackage{subcaption}

\usepackage{bigints}
\usepackage{geometry}
\allowdisplaybreaks[4]

\begin{document}

\title{Liquidity Dynamics in RFQ Markets and Impact on Pricing}
\author{Philippe \textsc{Bergault}\footnote{Université Paris Dauphine-PSL, Ceremade, 75116 Paris, France, bergault@ceremade.dauphine.fr.} \and Olivier \textsc{Guéant}\footnote{Université Paris 1 Panthéon-Sorbonne, UFR 27 Mathématiques et Informatique, CNRS, Centre d'Economie de la Sorbonne, Paris, France, olivier.gueant@univ-paris1.fr.}}
\date{}

\maketitle
\setlength\parindent{0pt}

\begin{abstract}
To assign a value to a portfolio, it is common to use Mark-to-Market prices. However, how should one proceed when the securities are illiquid? When transaction prices are scarce, how can one use all the available real-time information? In this article, we address these questions for over-the-counter (OTC) markets based on requests for quotes (RFQs). We extend the concept of micro-price, which was recently introduced for assets exchanged through limit order books in the market microstructure literature, and incorporate ideas from the recent literature on OTC market making. To account for liquidity imbalances in RFQ markets, we use an approach based on bidimensional Markov-modulated Poisson processes. Beyond extending the concept of micro-price to RFQ markets, we introduce the new concept of Fair Transfer Price. Our concepts of price can be used to value securities fairly, even when the market is relatively illiquid and/or tends to be one-sided.

\vspace{2mm}
\medskip
\noindent{\bf Key words:} fair price, imbalance, liquidity, market making, RFQ markets.
\vspace{2mm}
\end{abstract}

\section{Introduction}

We are all used to seeing real-time stock prices scrolling on TV or blinking on our computer and cellphone screens. However, we seldom ask ourselves what these prices actually represent or should represent. Do they correspond to the prices of the last trades? Are they some form of mid-prices? From which exchange(s) or venue(s) do they come? In fact, the very notion of real-time prices raises many questions.\\

For liquid securities traded through limit order books (LOBs), a wide variety of real-time price concepts have been proposed under different names such as mid-price, efficient price, fair price, micro-price, and so on. Each of these concepts comes with its own desired or undesired properties.\\

The first notion that naturally arises in the case of LOBs is that of the mid-price \(\frac{S^b + S^a}{2}\), where \(S^b\) is the best bid price and \(S^a\) is the best offer or ask price. This notion is simple but suffers from several limitations. If we consider that a good notion of price should result from a nowcasting procedure, the above notion of mid-price does not use all the available information in the LOB, particularly the available volumes. Additionally, it evolves discontinuously and may suffer from a form of bid-ask bounce when limits are depleted by trades (though a less severe form of bid-ask bounce than in the case of last trade prices). Moreover, if an asset can be traded on several venues, the mid-price ceases to be defined unambiguously: it could be defined, for instance, as the mid-price on the main venue or as the average between the best bid prices across venues and the best ask prices across venues. Questions also arise when prices are not reliable because orders are not firm due to last look practices (a typical feature in foreign exchange markets, see~\cite{oomen2017last}). Despite these problems, mid-prices are widely used and are adequate for many applications.\\

The most famous extension of mid-price is that of the weighted mid-price (also called imbalance-based mid-price) defined as 
\(
\frac{V^a}{V^b+V^a} S^b + \frac{V^b}{V^b+V^a} S^a
\)
where \(V^b\) and \(V^a\) are the volumes available in the LOB at the best bid and best ask prices respectively. This weighted mid-price is related to the saying ``the price is where the volume is not'' (see \cite{delattre_estimating_2013}) that has inspired a lot of the approaches discussed below. Although it suffers from numerous flaws (discontinuity, counterintuitive sensitivity to price improvement in some cases, excessive noise, etc.) this weighted mid-price is widely used. It is indeed attractive since the imbalance between the volumes posted at the best bid and at the best ask is known to be a good predictor of the price of the next trade or of the next (mid-)price move. One can cite \cite{gould_queue_2016} for an empirical study, \cite{cont_price_2013} for a simple expression of the probability of an upward move conditional on these volumes in a simple Markovian model for the dynamics of a limit order book, and \cite{cartea2018enhancing} for an example of use of volume imbalance in trading strategies.\\

Measures of imbalance based on \(V^b\) and \(V^a\) just have to be monotone in \(V^b/V^a\) and can therefore take a variety of forms. In an attempt to generalize the price formation model of \cite{madhavan_why_1997} to large-tick assets, Bonart and Lillo proposed in \cite{bonart_continuous_2018} an extension of the above weighted mid-price in which they replaced volumes at the best limits by their squares,\footnote{To account for make-take fees on some platforms, they also propose to replace bid and ask prices in the formulas by rebate-adjusted prices.} \textit{i.e.},
\(
\frac{{V^a}^2}{{V^b}^2+{V^a}^2} S^b + \frac{{V^b}^2}{{V^b}^2+{V^a}^2} S^a.
\)
They argue, based on theoretical and empirical grounds, that the quadratic version is preferable to the linear one, especially for assets with bid-ask spreads (almost always) equal to one tick -- so-called large-tick assets.\\

Many other notions of mid-price can be proposed along the above lines. One can indeed easily extend the above definitions beyond top-of-book prices and volumes, or consider several venues. Another commonly seen method consists in regressing signed cumulated volumes in the LOB on prices and defining an extended mid-price as the intersection between the regression line and the price axis. In all cases, these notions are only heuristics and deserve a micro-foundation.\\

In the specific case of large-tick assets, different notions have also emerged in the academic literature. Delattre \textit{et al.} introduced in \cite{delattre_estimating_2013} an interesting approach in which they assume that there exists an unobservable ``efficient'' price and deduce the location of that price through the order flow at the bid. More precisely, they consider limit orders sent with a probability that is a monotone function of the distance between that unobserved efficient price and the (observed) bid. Using historical data, they estimate that function in a nonparametric way and then deduce from the current (in fact recent) order flow an estimation of the efficient price. Two-sided extensions of this approach, where one uses both the bid and the ask sides, could be imagined and would share much with the modeling approach of the trading flow typically used in OTC market making models. Robert and Rosenbaum proposed in \cite{robert_new_2011} another route to estimate an efficient price for large-tick assets that does not rely on the volumes at the best limits in the LOB but rather on transaction prices only. The main idea underlying their approach is that if a transaction occurs and changes one of the best limits in the LOB, then the efficient price must be close enough to the transaction price. Their paper is one of the applications of the concept of uncertainty zones, which has also been used for the optimal choice of tick sizes (see \cite{dayri2015large} and \cite{baldacci2023bid} for a recent paper).\\

In an attempt to provide a general framework for defining notions of real-time price, Stoikov proposed in~\cite{stoikov_micro-price_2018} the concept of micro-price.\footnote{See \cite{stoikov_microstructure_2021} for a recent multi-asset extension.} This micro-price is defined as the long-term expectation of the (classical) mid-price conditional on all the information currently available. In other words, it relies on a long-term limit to eliminate microstructural noise.\footnote{A vast literature exists regarding the filtering of microstructural noise. However, the aim of that literature is more that of estimating volatility at the high-frequency level rather than effectively constructing a denoised price.} Similar ideas were present in the paper \cite{lehalle_limit_2017} by Lehalle and Mounjid, who, however, restricted the conditioning to the value of the current mid-price and imbalance.\footnote{In fact, the idea could be traced back to \cite{jaisson_liquidity_2015}.} The general framework proposed by Stoikov leads to various notions of price depending on the assumptions made regarding the random variables at stake. In particular, the notions of mid-price and weighted mid-price are outcomes of the approach for simple models of the LOB dynamics. An important advantage of this approach, beyond its versatility, is that the micro-price is always, by definition, a martingale.\\

Many concepts have been introduced in the case of markets organized around LOBs, and these concepts are commonly used by practitioners in the equity world. The case of RFQ markets, however, has always attracted less research. In fact, several questions arise naturally when it comes to RFQ markets, especially regarding the available information.\\

On some markets, post-trade transparency is enforced, and both dealers and clients\footnote{In this paper, we use the word client to designate a liquidity-taker, \textit{i.e.}, any market participant who is not a dealer (as in the expression ``dealer-to-client segment''). It is, of course, not the client of a specific dealer, although we are going to use a dataset of RFQs sent to a specific dealer.} can have access -- at least theoretically -- to a consolidated tape of transactions. This is the case in the US corporate bond markets with TRACE data (see \cite{dick-nielsen_liquidity_2009, dick-nielsen_how_2014} for relevant statistical methods to exploit TRACE data), but the situation is different in the European market despite recent efforts. The problem is, in fact, the fragmented nature of information and, as in all OTC markets, the lag in reporting.\\

Beyond transaction prices and volumes, clients usually have access to the prices streamed by dealers on electronic platforms.\footnote{In the case of the European market for corporate bonds, the main multi-dealer-to-client platforms are those of Bloomberg, MarketAxess, and Tradeweb.} However, streamed prices are only indicative and for a given size. As far as dealers are concerned, the information available to them depends on the market. In the case of corporate bonds, dealers do not have access to the prices streamed by competitors, but they have access to composite prices provided by multi-dealer-to-client electronic platforms (CBBT for Bloomberg, CP+ for MarketAxess, etc.) or can create their own composites from multiple sources. These prices have many drawbacks, but they often constitute a useful first estimate. Beyond indicative prices, dealers have access to a lot of information through their customer flows. In the case of corporate bond markets, requests for quotes (RFQs) constitute, for a market maker with a decent market share, the main source of information beyond composite prices. The information content of client flows is indeed very important: (i)~the side/sign of RFQs (\textit{i.e.}, the willingness to buy or to sell) indicates the sentiment of clients on each asset or, more generally, on assets with similar characteristics (sector of the issuer and maturity in the case of corporate bonds), and (ii) client decisions to trade at the price quoted by the dealer, at a better or identical price proposed by another dealer, or not to trade, inform about competition, but also about the demand curve of clients and, therefore, about the current (unobservable) price or its distribution.\footnote{One limitation is that some requests are sent without the intention to trade (for instance, to value a portfolio). However, on multi-dealer-to-client platforms, dealers know whether the requests they answered led to a transaction with a competitor.}\\

The use of RFQ data to estimate a real-time price in corporate bond markets is not new in the literature. A multivariate approach based on particle filtering has been proposed in \cite{gueant_mid-price_2018}, which exploited information from a proprietary database of RFQs sent to a dealer and trades in the dealer-to-dealer segment of the market. This particle filtering approach is interesting in that it is Bayesian and therefore provides a distribution for real-time prices.\\

In this paper, we propose two new ideas that both rely on a novel approach to model the flow of RFQs and its complex dynamics. In many OTC market making models, requests are modeled by Poisson processes: they arrive randomly, and the probability of occurrence of an RFQ is constant over time -- we call this probability the intensity, which is the infinitesimal probability of an RFQ occurrence per unit of time. To model varying liquidity, we assume in this paper that RFQs arrive randomly with an intensity that is itself a stochastic process: a simple continuous-time Markov chain with only a few states. In technical terms, we model the flow of RFQs at the bid and ask sides by a bidimensional Markov-modulated Poisson process (MMPP).\footnote{See \cite{fischer1993markov} for an overview of MMPPs and their historical applications in telecommunications.}\\

Our first idea consists in defining a micro-price \textit{à la} Stoikov using the information contained in the flow imbalance. More precisely, we assume that the price process drifts proportionally to the difference between the intensity at the ask and the intensity at the bid. When the intensities at the bid and the ask are the same, the micro-price is nothing but the current price. However, imbalance leads to a micro-price above or below the current price depending on the side of the imbalance. The exact value of the micro-price depends, of course, on the proportionality factor and on the joint dynamics of intensities.\\

Our second idea is inspired by the recent literature on OTC market making (see the reference books \cite{cartea2015algorithmic,gueant2016financial} for an overview of the recent market making literature). When two agents want to agree on a price, they can resort to a neutral third party. However, if the seller requests a price from a market maker, they will get the bid price quoted by that market maker. If, instead, the buyer requests a price from a market maker, they will get the ask price quoted by that market maker. If we assume that this third party is aware of the flow imbalances in the market, it is then natural to regard the average between these two prices as a fair price, especially when the market maker has zero inventory.\\

In market making models \textit{à la} Avellaneda-Stoikov \cite{avellaneda_high-frequency_2008} (see also \cite{bergault_closed-form_2021, bergault2021size, gueant2016financial, gueant2017optimal} for presentations more consistent with OTC markets), trading flows depend on the distance of the dealer's quotes to an exogenous reference price. If trading flows (or intensities in mathematical models) at the bid and ask are the same, then the optimal bid and ask prices of a market maker with no inventory should be symmetric around the reference price, which is therefore a fair transfer price. However, when a market maker is aware of asymmetries in the trading flows, they skew their quotes even in the absence of inventory. As a consequence, the average between the optimal bid and ask quotes ceases to coincide with the reference price. Nonetheless, it remains a fair transfer price given the current context in terms of liquidity. We therefore propose an extension of existing market making models to incorporate MMPPs and obtain a new model in which the average between the bid and ask quotes (in the absence of inventory) defines a fair transfer price that can be used to value or transfer securities even when the market is illiquid and/or tends to be one-sided.\\

In Section~2, we introduce the modeling framework for the flow of RFQs and present a statistical technique for the estimation of the model parameters. In Section~3, we present a notion of micro-price inspired by that of Stoikov, but rooted in our model for the flow of RFQs, and introduce our notion of Fair Transfer Price. Section~4 discusses numerical methods, presents numerous numerical examples, and analyzes them. Appendix~A present two important extensions of our model for the flow of RFQs that are used in the paper. The first extension is linked to an exchangeability assumption between the intensities at the bid and the ask. This assumption means that there is no structural asymmetry between the bid and the ask: liquidity can, of course, be asymmetric from time to time, with a higher intensity on one side, but this is only transitory and the same could happen on the other side with the same probability. The second extension allows us to go multi-asset.

\section{A modelling framework for the flow of RFQs}

\subsection{Introduction and notation}

In OTC markets based on RFQs, the number of requests received by a dealer can vary significantly. It can also be high on one side and low on the other, highlighting the crucial role of dealers who hold inventory and bridge the gap between different phases.\\

To model the dynamics of liquidity, the basic idea is to regard the number of RFQs received by a dealer on a given asset at the bid and at the ask as two point processes. Of course, Poisson processes are not sufficient: the intensities \((\lambda^b_t)_t\) (for the bid) and \((\lambda^a_t)_t\) (for the ask) must be stochastic processes. In quantitative finance, the most commonly used extensions of Poisson processes are Hawkes processes. Hawkes processes are indeed very good at modeling events that may happen in clusters. However, they are self-excited processes and, in a market with limited post-trade transparency, we argue that it is odd to assume that an RFQ sent by a client is the consequence of an RFQ sent by another client. Instead of using Hawkes processes, we assume that intensities are continuous-time Markov chains with values in a finite set and use the concept of Markov-modulated Poisson process. Because liquidity shocks can sometimes be symmetric and sometimes asymmetric, we consider more precisely a bidimensional MMPP: the intensity\footnote{Throughout the paper, we call this process an intensity process in spite of it being bidimensional.} process \((\lambda_t)_t = (\lambda^b_t, \lambda^a_t)_t\) is a continuous-time Markov chain taking values in \(\{\lambda^{1,b}, \ldots, \lambda^{m_b,b}\} \times \{\lambda^{1,a}, \ldots, \lambda^{m_a,a}\}\) with transition (or rate) matrix \(Q \in M_{m_b m_a}\).\footnote{In what follows, we order the states in lexicographic order: $$(\lambda^{1,b}, \lambda^{1,a}), \ldots, (\lambda^{1,b}, \lambda^{m_a,a}), \ldots, (\lambda^{m_b,b}, \lambda^{1,a}), \ldots, (\lambda^{m_b,b}, \lambda^{m_a,a}).$$ The case in which the two intensity processes are considered in an independent manner is a specific one and corresponds, for the chosen order, to \(Q = Q^b \otimes I_{m_a} + I_{m_b} \otimes Q^a\) where \(Q^b\) and \(Q^a\) are the transition matrices associated with \((\lambda^b_t)_t\) and \((\lambda^a_t)_t\) respectively and \(\otimes\) denotes the tensor (or Kronecker) product.}\\

In what follows, we focus on the estimation of the intensities $\lambda^{1,b}, \ldots, \lambda^{m_b,b}$ and $\lambda^{1,a}, \ldots, \lambda^{m_a,a}$ and the coefficients of the transition matrix $Q$. The method we propose is inspired by the EM algorithm proposed in~\cite{ryden_em_1996} but generalized to the more complex case of a bidimensional MMPP. We also present two important extensions in Appendix \ref{ext}.

\subsection{Estimation of the parameters}
\label{EMsec}
\subsubsection{Likelihood of a sample}
\label{LLsec}
Our goal in the next paragraphs is to compute the likelihood of a sequence of RFQ times $t_1< \ldots< t_N$ with sides $\frak s_1, \ldots, \frak s_N$, where the sides are encoded as elements of $\{b,a\}$ for bid and ask.\\

Let us denote by $(N^{\text{RFQ}, b}_t)_t$ and $(N^{\text{RFQ}, a}_t)_t$ the processes counting the number of RFQs at the bid and at the ask respectively, and let us consider the function
$$\mathcal{G} : t \mapsto (\mathcal G^{(j_b-1) m_a+j_a, (k_b-1) m_a+k_a}(t))_{1\le j_b,k_b\le m_b,1\le j_a,k_a\le m_a}$$
where $$\mathcal G^{(j_b-1) m_a+j_a, (k_b-1) m_a+k_a}(t) = \mathbb P(N^{RFQ,b}_t = 0, N^{RFQ,a}_t = 0,  \lambda_t = (\lambda^{k_b,b}, \lambda^{k_a,a})  | \lambda_0 = (\lambda^{j_b,b}, \lambda^{j_a,a})).$$

We have for $h>0$, $1\le j_b,k_b\le m_b$ and $1\le j_a,k_a\le m_a$:
\begin{eqnarray*}
&&\mathcal G^{(j_b-1) m_a+j_a, (k_b-1) m_a+k_a}(t+h)\\
&=& \mathbb P(N^{RFQ,b}_{t+h} = 0, N^{RFQ,a}_{t+h} = 0,  \lambda_{t+h} = (\lambda^{k_b,b}, \lambda^{k_a,a})  | \lambda_0 = (\lambda^{j_b,b}, \lambda^{j_a,a}))\\
&=& \sum_{l_b=1}^{m_b} \sum_{l_a=1}^{m_a}  \mathbb P(N^{RFQ,b}_{t+h} = 0, N^{RFQ,a}_{t+h} = 0,  \lambda_{t+h} = (\lambda^{k_b,b}, \lambda^{k_a,a}), \lambda_{t} = (\lambda^{l_b,b}, \lambda^{l_a,a})  | \lambda_0 = (\lambda^{j_b,b}, \lambda^{j_a,a}))\\
&=& \sum_{l_b=1}^{m_b} \sum_{l_a=1}^{m_a} \mathcal G^{(j_b-1) m_a+j_a, (l_b-1) m_a+l_a}(t) \mathbb P\left(N^{RFQ,b}_{t+h} = 0, N^{RFQ,a}_{t+h} = 0,  \lambda_{t+h} = (\lambda^{k_b,b}, \lambda^{k_a,a})\right.\\
&& \qquad\qquad\qquad\qquad\qquad\qquad\qquad\qquad \left| \left. N^{RFQ,b}_{t} = 0, N^{RFQ,a}_{t} = 0,  \lambda_{t} = (\lambda^{l_b,b}, \lambda^{l_a,a})\right)\right. \\
&=& \mathcal G^{(j_b-1) m_a+j_a, (k_b-1) m_a+k_a}(t) \left(1+Q_{(k_b-1) m_a+k_a, (k_b-1) m_a+k_a}h +o(h)\right)\\
&&\qquad\qquad\qquad\qquad\qquad\qquad\times\left(1-\lambda^{k_b,b} h + o(h)\right)\left(1-\lambda^{k_a,a} h + o(h)\right)\\
&&+ \sum_{1\le l_b \le m_b, 1\le l_a \le m_a, (l_b,l_a) \neq (k_b,k_a)}  \mathcal G^{(j_b-1) m_a+j_a, (l_b-1) m_a+l_a}(t) \left(Q_{(l_b-1) m_a+l_a, (k_b-1) m_a+k_a}h + o(h)\right).
\end{eqnarray*}

This leads to the following differential equation:
$$\frac{d\ }{dt}\mathcal G^{(j_b-1) m_a+j_a, (k_b-1) m_a+k_a}(t) = \mathcal G^{(j_b-1) m_a+j_a, (k_b-1) m_a+k_a}(t) \left(Q_{(k_b-1) m_a+k_a, (k_b-1) m_a+k_a} - \lambda^{k_b,b} - \lambda^{k_a,a}\right)$$$$ + \sum_{1\le l_b \le m_b, 1\le l_a \le m_a, (l_b,l_a) \neq (k_b,k_a)}  \mathcal G^{(j_b-1) m_a+j_a, (l_b-1) m_a+l_a}(t) Q_{(l_b-1) m_a+l_a, (k_b-1) m_a+k_a} $$
which, in matrix form, writes $$\mathcal G'(t) =  \mathcal G(t) \left(Q - \tilde{\Lambda}^b - \tilde{\Lambda}^a\right)$$
where $\Lambda^b = \text{diag}(\lambda^{1,b}, \ldots, \lambda^{m_b,b})$, $\Lambda^a = \text{diag}(\lambda^{1,a}, \ldots, \lambda^{m_a,a})$, $\tilde{\Lambda}^b = \Lambda^b \otimes I_{m_a}$ and $\tilde{\Lambda}^a = I_{m_b} \otimes \Lambda^a$.\\

As $\mathcal G(0)$ is the identity matrix $I_{m_bm_a}$, we conclude that $$\mathcal G(t) = \exp\left(\left(Q - \tilde{\Lambda}^b - \tilde{\Lambda}^a\right)t\right).$$

By Markov property, for $s \ge 0$, if we assume that $\lambda_s$ is distributed according to a distribution represented by a column vector $\pi_s$ (in $\mathbb{R}^{m_bm_a}$), then for $1\le j_b \le m_b, 1\le j_a \le m_a$, $\pi_s'\exp((Q - \tilde{\Lambda}^b - \tilde{\Lambda}^a)t)e^{(j_b-1)m_b+j_a}$ is the probability that there was no RFQ between time $s$ and time $s+t$ and the intensity process is equal to $(\lambda^{j_b,b}, \lambda^{j_a,a})$  at time $s+t$.\footnote{$(e^1, \ldots, e^{m_b m_a})$ is the canonical basis of $\mathbb R^{m_bm_a}$.}\\

If we assume that $\lambda_0$ is distributed according to a distribution represented by a column vector $\pi_0$, then the likelihood of the whole sample writes

\begin{eqnarray*}
&&\mathcal{L}(Q,\Lambda^b,\Lambda^a| t_1, \ldots, t_N, \frak s_1, \ldots \frak s_N)\\
&=& \pi_0'\left(\prod_{n=1}^N \exp\left(\left(Q - \tilde{\Lambda}^b - \tilde{\Lambda}^a\right)(t_n-t_{n-1})\right) \tilde{\Lambda}^{\frak s_n}\right)e
\end{eqnarray*}
where $t_0 = 0$ and $e= \sum_{j_b=1}^{m_b}\sum_{j_a=1}^{m_a} e^{(j_b-1)m_a+j_a} = (1, \ldots, 1)'$.\\

Maximizing the above likelihood expression is not straightforward. Instead, we propose in the next paragraph an EM algorithm in which the hidden variables correspond to the trajectory of the unobservable intensity process.

\subsubsection{An EM algorithm}
\label{EMsubsec}
Let us consider as hidden variables a sequence of times $0=\tau_0 < \ldots < \tau_P (\le t_N)$ corresponding to transitions of the process $(\lambda_t)_t$ and a sequence of couples $(s^b_0, s^a_0), \ldots, (s^b_P, s^a_P)$ in $\{1, \ldots, m_b\}\times\{1, \ldots, m_a\}$ such that $(\lambda^b_t, \lambda^a_t) = (\lambda^{s^b_p,b}, \lambda^{s^a_p,a})$ over $[\tau_p, \tau_{p+1})$ (where, by convention $\tau_{P+1} = t_N$).\\

The likelihood of $t_1< \ldots< t_N$, $\frak s_1, \ldots \frak s_N$, $\tau_1 < \ldots < \tau_{P}$, $s^b_0, \ldots, s^b_P$ and $s^a_0, \ldots, s^a_P$ is
\begin{eqnarray*}
&&\mathcal{L}(Q,\Lambda^b, \Lambda^a | t_1, \ldots, t_N, \frak s_1, \ldots \frak s_N, \tau_1, \ldots, \tau_{P+1}, s^b_0, \ldots, s^b_P, s^a_0, \ldots, s^a_P)\\
&=&(\pi_0)_{(s^b_0-1)  m_a +s^a_0} \left(\prod_{p=0}^{P-1} Q_{(s^b_p-1) m_a +s^a_p,(s^b_{p+1}-1)m_a + s^a_{p+1}} \exp\left({Q_{(s^b_p-1) m_a +s^a_p,(s^b_p-1) m_a +s^a_p}(\tau_{p+1} - \tau_p)}\right)\right)\\
&& \times \exp\left({Q_{(s^b_p-1) m_a +s^a_p,(s^b_p-1) m_a +s^a_p}(\tau_{P+1} - \tau_P)}\right)\left(\prod_{p=0}^{P} \left(\lambda^{s^b_p,b}\right)^{c^b_p} \exp\left({-\lambda^{s^b_p,b}(\tau_{p+1} - \tau_p)}\right)\right)\\
&&\times\left(\prod_{p=0}^{P} \left(\lambda^{s^a_p,a}\right)^{c^a_p} \exp\left({-\lambda^{s^a_p,a}(\tau_{p+1} - \tau_p)}\right)\right)
\end{eqnarray*}
where $c^b_p = \text{Card}(\{n | t_n \in [\tau_p, \tau_{p+1}), \frak{s}_n = b\})$ and $c^a_p = \text{Card}(\{n | t_n \in [\tau_p, \tau_{p+1}), \frak{s}_n = a\})$.\\

The associated log-likelihood writes
\begin{eqnarray}&&\log\left((\pi_0)_{(s^b_0-1)  m_a +s^a_0}\right) + \sum_{p=0}^{P-1} \log\left(Q_{(s^b_p-1) m_a +s^a_p,(s^b_{p+1}-1)m_a + s^a_{p+1}}\right)\nonumber\\ &&+ \sum_{p=0}^{P} \left(Q_{(s^b_p-1) m_a +s^a_p,(s^b_{p}-1)m_a + s^a_{p}} -\lambda^{s^b_p,b} -\lambda^{s^a_p,a}\right)(\tau_{p+1} - \tau_p)+ \sum_{p=0}^{P} c^b_p \log(\lambda^{s^b_p,b}) + \sum_{p=0}^{P} c^a_p \log(\lambda^{s^a_p,b})\nonumber\\
&=& \log\left((\pi_0)_{(s^b_0-1)  m_a +s^a_0}\right) + \sum_{\substack{1\le j_b \le m_b\\1\le j_a \le m_a}}\sum_{\substack{1\le k_b \le m_b\\1\le k_a \le m_a\\(j_b,j_a) \neq (k_b, k_a)}} \tilde{n}^{(j_b,j_a),(k_b,k_a)}\log(Q_{(j_b-1)m_a+j_a,(k_b-1)m_a+k_a})\nonumber\\
&& + \sum_{\substack{1\le j_b \le m_b\\1\le j_a \le m_a}} \tilde{T}^{(j_b, j_a)} \left(Q_{(j_b-1)m_a+j_a,(j_b-1)m_a+j_a} - \lambda^{j_b,b} -\lambda^{j_a,a}\right)\nonumber\\
&&+ \sum_{\substack{1\le j_b \le m_b\\1\le j_a \le m_a}} \tilde{n}^b_{(j_b,j_a)} \log(\lambda^{j_b,b}) + \sum_{\substack{1\le j_b \le m_b\\1\le j_a \le m_a}} \tilde{n}^a_{(j_b,j_a)} \log(\lambda^{j_a,a}) \nonumber\\
&=& \log\left((\pi_0)_{(s^b_0-1)  m_a +s^a_0}\right) + \sum_{\substack{1\le j_b \le m_b\\1\le j_a \le m_a}}\sum_{\substack{1\le k_b \le m_b\\1\le k_a \le m_a\\ (k_b, k_a) \neq (j_b,j_a)}} \tilde{n}^{(j_b,j_a),(k_b,k_a)}\log(Q_{(j_b-1)m_a+j_a,(k_b-1)m_a+k_a})\nonumber\\
&& - \sum_{\substack{1\le j_b \le m_b\\1\le j_a \le m_a}} \left(\left(\sum_{\substack{1\le k_b \le m_b\\1\le k_a \le m_a\\(k_b, k_a) \neq (j_b,j_a)}} Q_{(j_b-1)m_a+j_a,(k_b-1)m_a+k_a}\right) + \lambda^{j_b,b} +\lambda^{j_a,a}\right)\tilde{T}^{(j_b, j_a)} \nonumber\\
&&+ \sum_{\substack{1\le j_b \le m_b\\1\le j_a \le m_a}} \tilde{n}^b_{(j_b,j_a)} \log(\lambda^{j_b,b}) + \sum_{\substack{1\le j_b \le m_b\\1\le j_a \le m_a}} \tilde{n}^a_{(j_b,j_a)} \log(\lambda^{j_a,a})
\label{llc2}
\end{eqnarray}
where:
\begin{itemize}
\item for $1 \le j_b, k_b \le m_b$ and $1 \le j_a, k_a \le m_a$ with $(j_b,j_a) \neq (k_b,k_a)$, $\tilde{n}^{(j_b,j_a),(k_b,k_a)}$ is the number of transitions of the intensity process $(\lambda_t)_t$ from $(\lambda^{j_b,b}, \lambda^{j_a,a})$ to $(\lambda^{k_b,b}, \lambda^{k_a,a})$ over the time interval $[0,t_N]$,
\item for $1\le j_b \le m_b$ and $1\le j_a \le m_a$, $\tilde{T}^{(j_b,j_a)}$ is the total time spent by the intensity process $(\lambda_t)_t$ in $(\lambda^{j_b,b}, \lambda^{j_a,a})$ over the time interval~$[0,t_N]$,
\item for $1\le j_b \le m_b$ and $1\le j_a \le m_a$, $\tilde{n}_{(j_b,j_a)}^b$ and $\tilde{n}_{(j_b,j_a)}^a$ are the number of RFQs at the bid and at the ask respectively over the time interval $[0,t_N]$ while the intensity process $(\lambda_t)_t$ is in $(\lambda^{j_b,b}, \lambda^{j_a,a})$.
\end{itemize}

The EM algorithm consists in iteratively computing the expectation of the log-likelihood expression \eqref{llc2} conditionally on the real observables $t_1< \ldots < t_N$ and $\frak s_1, \ldots, \frak s_N$ under the assumption that the unobservable variables are distributed according to the model with given values $\widehat{\Lambda^b}$, $\widehat{\Lambda^a}$ and $\widehat{Q}$ of $\Lambda^b$, $\Lambda^a$  and $Q$, and, then, carrying out a maximization of the resulting expression over the diagonal coefficients of $\Lambda^b$ and $\Lambda^a$ and the non-diagonal coefficients of $Q$ to update the values of $\widehat{\Lambda^b}$, $\widehat{\Lambda^a}$ and $\widehat{Q}$.\\

Ignoring the first term which contributes almost nothing, we easily see that the EM algorithm boils down to the following updates:  
$$ \widehat{\Lambda^b}_{j_b,j_b} \leftarrow \frac{\sum_{j_a=1}^{m_a}\mathbb{E}_{\widehat{\Lambda^b}, \widehat{\Lambda^a}, \widehat{Q}, t_1, \ldots t_N, \frak s_1, \ldots \frak s_N}\left[\tilde{n}_{(j_b,j_a)}^{b}\right]}{\sum_{j_a=1}^{m_a}\mathbb{E}_{\widehat{\Lambda^b}, \widehat{\Lambda^a}, \widehat{Q}, t_1, \ldots t_N, \frak s_1, \ldots \frak s_N}\left[\tilde{T}^{(j_b,j_a)}\right]}\quad \text{ for } 1 \le j_b \le m_b,$$
$$ \widehat{\Lambda^a}_{j_a,j_a} \leftarrow \frac{\sum_{j_b=1}^{m_b}\mathbb{E}_{\widehat{\Lambda^b}, \widehat{\Lambda^a}, \widehat{Q}, t_1, \ldots t_N, \frak s_1, \ldots \frak s_N}\left[\tilde{n}_{(j_b,j_a)}^{a}\right]}{\sum_{j_b=1}^{m_b}\mathbb{E}_{\widehat{\Lambda^b}, \widehat{\Lambda^a}, \widehat{Q}, t_1, \ldots t_N, \frak s_1, \ldots \frak s_N}\left[\tilde{T}^{(j_b,j_a)}\right]}\quad  \text{ for } 1 \le j_a \le m_a,$$
and, for $1 \le j_b,k_b \le m_b$ and $1 \le j_a,k_a \le m_a$ with $(j_b, j_a) \neq (k_b,k_a)$,
$$\widehat{Q}_{(j_b-1)m_a+j_a,(k_b-1)m_a+k_a} \leftarrow \frac{\mathbb{E}_{\widehat{\Lambda^b}, \widehat{\Lambda^a}, \widehat{Q}, t_1, \ldots t_N, \frak s_1, \ldots \frak s_N}\left[\tilde{n}^{(j_b,j_a),(k_b,k_a)}\right]}{\mathbb{E}_{\widehat{\Lambda^b}, \widehat{\Lambda^a}, \widehat{Q}, t_1, \ldots t_N, \frak s_1, \ldots \frak s_N}\left[\tilde{T}^{(j_b,j_a)}\right]}.$$ 

Assuming that the initial intensity is distributed according to a distribution represented by a column vector~$\pi_0$, we get that the conditional expectation of the number of RFQs at the bid while the intensity process is equal to $(\lambda^{j_b,b},\lambda^{j_a,a})$ is\footnote{We write $\widetilde{\widehat{\Lambda^b}} = \widehat{\Lambda^b} \otimes I_{m_a}$ and $\widetilde{\widehat{\Lambda^a}} =  I_{m_b} \otimes \widehat{\Lambda^a}$.}
\newpage
\begin{eqnarray*}
&&\mathbb{E}_{\widehat{\Lambda^b}, \widehat{\Lambda^a}, \widehat{Q}, t_1, \ldots t_N, \frak s_1, \ldots \frak s_N}\left[\tilde{n}_{(j_b,j_a)}^{b}\right]\\
&=&\sum_{r=1}^N 1_{\frak s_r = b}  \mathbb{E}_{\widehat{\Lambda^b}, \widehat{\Lambda^a}, \widehat{Q}, t_1, \ldots t_N, \frak s_1, \ldots \frak s_N}\left[1_{\lambda^b_{t_r} = \lambda^{j_b,b}}\right]\\
&=& \frac{1}{\pi_0'\left(\prod_{n=1}^N \exp((\widehat{Q} - \widetilde{\widehat{\Lambda^b}} - \widetilde{\widehat{\Lambda^a}})(t_n-t_{n-1})) \widetilde{\widehat{\Lambda^{\frak s_n}}}\right) e}\\
&&\quad \times  \sum_{r=1}^N 1_{\frak s_r = b} \pi_0'\left(\prod_{n=1}^{r} \exp((\widehat{Q} - \widetilde{\widehat{\Lambda^b}} - \widetilde{\widehat{\Lambda^a}})(t_n-t_{n-1})) \widetilde{\widehat{\Lambda^{\frak s_n}}}\right)e^{(j_b-1)m_a+j_a}\\
&& \quad \times {e^{(j_b-1)m_a+j_a}}'\left(\prod_{n=r+1}^{N} \exp((\widehat{Q} - \widetilde{\widehat{\Lambda^b}} - \widetilde{\widehat{\Lambda^a}})(t_n-t_{n-1})) \widetilde{\widehat{\Lambda^{\frak s_n}}}\right)e.
\end{eqnarray*}
Similarly, we have
\begin{eqnarray*}
&&\mathbb{E}_{\widehat{\Lambda^b}, \widehat{\Lambda^a}, \widehat{Q}, t_1, \ldots t_N, \frak s_1, \ldots \frak s_N}\left[\tilde{n}_{(j_b,j_a)}^{a}\right]\\
&=&\sum_{r=1}^N 1_{\frak s_r = a}  \mathbb{E}_{\widehat{\Lambda^b}, \widehat{\Lambda^a}, \widehat{Q}, t_1, \ldots t_N, \frak s_1, \ldots \frak s_N}\left[1_{\lambda^b_{t_r} = \lambda^{j_b,b}}\right]\\
&=& \frac{1}{\pi_0'\left(\prod_{n=1}^N \exp((\widehat{Q} - \widetilde{\widehat{\Lambda^b}} - \widetilde{\widehat{\Lambda^a}})(t_n-t_{n-1})) \widetilde{\widehat{\Lambda^{\frak s_n}}}\right) e}\\
&&\quad \times  \sum_{r=1}^N 1_{\frak s_r = a} \pi_0'\left(\prod_{n=1}^{r} \exp((\widehat{Q} - \widetilde{\widehat{\Lambda^b}} - \widetilde{\widehat{\Lambda^a}})(t_n-t_{n-1})) \widetilde{\widehat{\Lambda^{\frak s_n}}}\right)e^{(j_b-1)m_a+j_a}\\
&& \quad \times {e^{(j_b-1)m_a+j_a}}'\left(\prod_{n=r+1}^{N} \exp((\widehat{Q} - \widetilde{\widehat{\Lambda^b}} - \widetilde{\widehat{\Lambda^a}})(t_n-t_{n-1})) \widetilde{\widehat{\Lambda^{\frak s_n}}}\right)e.
\end{eqnarray*}

Regarding the time spent in $(\lambda^{j_b,b}, \lambda^{j_a,a})$, we get 
\begin{eqnarray*}
&&\mathbb{E}_{\widehat{\Lambda^b}, \widehat{\Lambda^a}, \widehat{Q}, t_1, \ldots t_N, \frak s_1, \ldots \frak s_N}\left[\tilde{T}^{(j_b,j_a)}\right]\\
&=& \int_0^{t_N} \mathbb{E}_{\widehat{\Lambda^b}, \widehat{\Lambda^a}, \widehat{Q}, t_1, \ldots t_N, \frak s_1, \ldots \frak s_N}\left[1_{(\lambda^b_t, \lambda^a_t) = (\lambda^{j_b,b},\lambda^{j_a,a})}\right] dt\\
&=& \frac{1}{\pi_0'\left(\prod_{n=1}^N \exp((\widehat{Q} - \widetilde{\widehat{\Lambda^b}} - \widetilde{\widehat{\Lambda^a}})(t_n-t_{n-1})) \widetilde{\widehat{\Lambda^{\frak s_n}}}\right) e}\\
&&\quad \times  \int_0^{t_N} \pi_0'\left(\prod_{n=1}^{n(t)} \exp((\widehat{Q} - \widetilde{\widehat{\Lambda^b}} - \widetilde{\widehat{\Lambda^a}})(t_n-t_{n-1})) \widetilde{\widehat{\Lambda^{\frak s_n}}}\right)\\
&& \quad\quad\quad \times \exp((\widehat{Q} - \widetilde{\widehat{\Lambda^b}} - \widetilde{\widehat{\Lambda^a}})(t-t_{n(t)}))e^{(j_b-1)m_a+j_a}{e^{(j_b-1)m_a+j_a}}' \exp((\widehat{Q} - \widetilde{\widehat{\Lambda^b}} - \widetilde{\widehat{\Lambda^a}})(t_{n(t)+1}-t))\\
&&\quad\quad\quad \times\left(\prod_{n=n(t)+2}^N \exp((\widehat{Q} - \widetilde{\widehat{\Lambda^b}} - \widetilde{\widehat{\Lambda^a}})(t_n-t_{n-1})) \widetilde{\widehat{\Lambda^{\frak s_n}}}\right)e\ dt
\end{eqnarray*}
where $n(t) = \max\{n, t_n < t\}$.\\
\newpage
This also writes
\begin{eqnarray*}
&&\mathbb{E}_{\widehat{\Lambda^b}, \widehat{\Lambda^a}, \widehat{Q}, t_1, \ldots t_N, \frak s_1, \ldots \frak s_N}\left[\tilde{T}^{(j_b,j_a)}\right]\\
&=& \frac{1}{\pi_0'\left(\prod_{n=1}^N \exp((\widehat{Q} - \widetilde{\widehat{\Lambda^b}} - \widetilde{\widehat{\Lambda^a}})(t_n-t_{n-1})) \widetilde{\widehat{\Lambda^{\frak s_n}}}\right) e}\\
&&\quad \times  \sum_{r=1}^N \left(\pi_0'\left(\prod_{n=1}^{r-1} \exp((\widehat{Q} - \widetilde{\widehat{\Lambda^b}} - \widetilde{\widehat{\Lambda^a}})(t_n-t_{n-1})) \widetilde{\widehat{\Lambda^{\frak s_n}}}\right)\right.\\
&& \quad \quad\quad\times \int_{t_{r-1}}^{t_r}\exp((\widehat{Q} - \widetilde{\widehat{\Lambda^b}} - \widetilde{\widehat{\Lambda^a}})(t-t_{r-1}))e^{(j_b-1)m_a+j_a}{e^{(j_b-1)m_a+j_a}}' \exp((\widehat{Q} - \widetilde{\widehat{\Lambda^b}} - \widetilde{\widehat{\Lambda^a}})(t_{r}-t))dt\\
&&\quad\quad\quad\times\left.\left(\prod_{n=r+1}^N \exp((\widehat{Q} - \widetilde{\widehat{\Lambda^b}} - \widetilde{\widehat{\Lambda^a}})(t_n-t_{n-1})) \widetilde{\widehat{\Lambda^{\frak s_n}}}\right)e\right).
\end{eqnarray*}

Using a similar reasoning, we have
\begin{eqnarray*}
&&\mathbb{E}_{\widehat{\Lambda^b}, \widehat{\Lambda^a}, \widehat{Q}, t_1, \ldots t_N, \frak s_1, \ldots \frak s_N}\left[\tilde{n}^{(j_b,j_a),(k_b,k_a)}\right]\\
&=& \frac{\widehat{Q}_{(j_b-1)m_a+j_a,(k_b-1)m_a+k_a}}{\pi_0'\left(\prod_{n=1}^N \exp((\widehat{Q} - \widetilde{\widehat{\Lambda^b}} - \widetilde{\widehat{\Lambda^a}})(t_n-t_{n-1})) \widetilde{\widehat{\Lambda^{\frak s_n}}}\right) e}\\
&&\quad \times  \int_0^{t_N} \pi_0'\left(\prod_{n=1}^{n(t)} \exp((\widehat{Q} - \widetilde{\widehat{\Lambda^b}} - \widetilde{\widehat{\Lambda^a}})(t_n-t_{n-1})) \widetilde{\widehat{\Lambda^{\frak s_n}}}\right)\\
&& \quad\quad\quad \times \exp((\widehat{Q} - \widetilde{\widehat{\Lambda^b}} - \widetilde{\widehat{\Lambda^a}})(t-t_{n(t)}))e^{(j_b-1)m_a+j_a}{e^{(k_b-1)m_a+k_a}}' \exp((\widehat{Q} - \widetilde{\widehat{\Lambda^b}} - \widetilde{\widehat{\Lambda^a}})(t_{n(t)+1}-t))\\
&&\quad\quad\quad \times\left(\prod_{n=n(t)+2}^N \exp((\widehat{Q} - \widetilde{\widehat{\Lambda^b}} - \widetilde{\widehat{\Lambda^a}})(t_n-t_{n-1})) \widetilde{\widehat{\Lambda^{\frak s_n}}}\right)e\ dt\\
&=& \frac{\widehat{Q}_{(j_b-1)m_a+j_a,(k_b-1)m_a+k_a}}{\pi_0'\left(\prod_{n=1}^N \exp((\widehat{Q} - \widetilde{\widehat{\Lambda^b}} - \widetilde{\widehat{\Lambda^a}})(t_n-t_{n-1})) \widetilde{\widehat{\Lambda^{\frak s_n}}}\right) e}\\
&&\quad \times  \sum_{r=1}^N \left(\pi_0'\left(\prod_{n=1}^{r-1} \exp((\widehat{Q} - \widetilde{\widehat{\Lambda^b}} - \widetilde{\widehat{\Lambda^a}})(t_n-t_{n-1})) \widetilde{\widehat{\Lambda^{\frak s_n}}}\right)\right.\\
&& \quad \quad\quad\times \int_{t_{r-1}}^{t_r}\exp((\widehat{Q} - \widetilde{\widehat{\Lambda^b}} - \widetilde{\widehat{\Lambda^a}})(t-t_{r-1}))e^{(j_b-1)m_a+j_a}{e^{(k_b-1)m_a+k_a}}' \exp((\widehat{Q} - \widetilde{\widehat{\Lambda^b}} - \widetilde{\widehat{\Lambda^a}})(t_{r}-t))dt\\
&&\quad\quad\quad\times\left.\left(\prod_{n=r+1}^N \exp((\widehat{Q} - \widetilde{\widehat{\Lambda^b}} - \widetilde{\widehat{\Lambda^a}})(t_n-t_{n-1})) \widetilde{\widehat{\Lambda^{\frak s_n}}}\right)e\right).
\end{eqnarray*}

These quantities can be computed iteratively and it is noteworthy that we do not need to compute the denominators as they cancel out when we update $\widehat{\Lambda^b}$, $\widehat{\Lambda^a}$ and $ \widehat{Q}$. It must also be noted that the only computational difficulty lies in finding a scaling factor to avoid ending up with very low or very high values.

\subsubsection{Estimating the current state}
\label{current_est}
Once an estimation of the parameters has been carried out, it is possible to estimate the state of the intensity processes at any point in time $t$. If indeed we consider a prior probability distribution for the initial value $\lambda_0 = (\lambda^b_0, \lambda^a_0)$ of intensity process represented by a column vector $\pi_0$, then, given a sequence of observed RFQs times $0=t_0<t_1< \ldots < t_n (\le t)$ prior to time $t$ along with their associated sides $\frak s_1, \ldots \frak s_n$, the a posteriori distribution $\pi_t$ of $(\lambda^b_t, \lambda^a_t)$ writes
$$(\pi_t)_{(j_b-1)m_a+j_a} \propto \pi_0'\left(\prod_{r=1}^n \exp((Q - \tilde{\Lambda}^b - \tilde{\Lambda}^a)(t_r-t_{r-1})) \tilde{\Lambda}^{\frak s_r}\right)\exp((Q - \tilde{\Lambda}^b - \tilde{\Lambda}^a)(t-t_n))e^{(j_b-1)m_a+j_a}$$
\textit{i.e.}
\begin{equation}
\pi_t = \frac{\pi_0'\left(\prod_{r=1}^n \exp((Q - \tilde{\Lambda}^b - \tilde{\Lambda}^a)(t_r-t_{r-1})) \tilde{\Lambda}^{\frak s_r}\right)\exp((Q - \tilde{\Lambda}^b - \tilde{\Lambda}^a)(t-t_n))e^{(j_b-1)m_a+j_a}}{\pi_0'\left(\prod_{r=1}^n \exp((Q - \tilde{\Lambda}^b - \tilde{\Lambda}^a)(t_r-t_{r-1})) \tilde{\Lambda}^{\frak s_r}\right)\exp((Q - \tilde{\Lambda}^b - \tilde{\Lambda}^a)(t-t_n))e}. \label{apost2}
\end{equation}

\section{New notions of price}

\subsection{A micro-price for RFQ markets}

\subsubsection{Definition}

In \cite{stoikov_micro-price_2018}, Stoikov introduced the notion of micro-price for an asset traded through a limit order book. It is defined as the asymptotic value of the expected mid-price, given all the information available (in the limit order book).\\

It is reasonable to extend the ideas introduced in \cite{stoikov_micro-price_2018} to RFQ markets through the use of our model for RFQ arrival. If we consider a reference price process,\footnote{In the case of corporate bonds, it can be CBBT, CP+, or another composite.} \((S_t)_t\), it is commonplace to assume a Brownian dynamics \(dS_t = \sigma dW_t\). However, if we know the current state of liquidity, it makes more sense to consider a dynamics of the form
$$
dS_t = \sigma dW_t + \kappa(\lambda^a_t - \lambda^b_t)dt
$$
where \(\kappa\) is a nonnegative constant. Then, the micro-price at time \(t\) is naturally defined by 
$$
S^{\text{micro}}_t = \lim_{T \to +\infty}\mathbb{E}[S_{T} \mid S_t, \lambda^b_t, \lambda^a_t] = S_t + \kappa \lim_{T \to +\infty} \mathbb{E}\left[\left. \int_t^{T} (\lambda^a_s - \lambda^b_s) \, ds \right| \lambda^b_t, \lambda^a_t \right]
$$
if that limit exists.\footnote{For bonds, Brownian dynamics can only be valid in the short run. Nevertheless, we consider a long-term limit to define the micro-price. This may seem problematic at first sight, but in our model, the long-term limit corresponds to a time horizon equivalent to that of the return to a symmetric state of liquidity, which is typically short.}

\subsubsection{Mathematical analysis}

To study this notion, let us define
$$v_T(t, \lambda^b, \lambda^a) = \mathbb E\left[\left. \int_t^T(\lambda^a_s - \lambda^b_s)ds \right| \lambda^b_t = \lambda^b, \lambda^a_t = \lambda^a \right].$$

If we write $v_T(t)$ the vector with coordinates $(v_T(t, \lambda^{b,j_b}, \lambda^{a,j_a}))_{1 \le j_b \le m_b, 1 \le j_a \le m_a }$ in lexicographic order, then we have that $v_T$ solves
$$\frac d{dt}v_T(t) + I_{m_b} \otimes \lambda_{\text{vec}}^a - \lambda_{\text{vec}}^b \otimes I_{m_a} + Qv_T(t) = 0 \quad \text{and} \quad v_T(T) = 0$$
where $\lambda_{\text{vec}}^b = (\lambda^{b,1}, \ldots, \lambda^{b,m_b})'$ and $\lambda_{\text{vec}}^a = (\lambda^{a,1}, \ldots, \lambda^{a,m_a})'$, \textit{i.e.}
$$v_T(t) = \int_t^T \exp(Q(s-t)) (I_{m_b} \otimes \lambda_{\text{vec}}^a - \lambda_{\text{vec}}^b \otimes I_{m_a}) ds.$$

Let us now assume as in Appendix \ref{MMPPex} that $(\lambda^b_s, \lambda^a_s)_s$ and $(\lambda^a_s, \lambda^b_s)_s$ have the same distribution. Then, it is straightforward to see that $v_T(t,\lambda, \lambda) = 0$ for all $\lambda \in \{\lambda^{1}, \ldots, \lambda^{m}\}$. Moreover, writing $\lambda_{\text{vec}} = (\lambda^{1}, \ldots, \lambda^{m})'$, we have that $I_{m_b} \otimes \lambda_{\text{vec}}^a - \lambda_{\text{vec}}^b \otimes I_{m_a} = I_{m} \otimes \lambda_{\text{vec}} - \lambda_{\text{vec}} \otimes I_{m}$ has all coordinates corresponding to symmetric states equal to~$0$. Therefore, if we denote by the superscript ${}^{\text{ns}}$ vectors and matrices where all symmetric states have been dropped, we have:
$$v^{\text{ns}}_T(t) = \int_t^T \exp(Q^{\text{ns}}(s-t)) (I_{m} \otimes \lambda_{\text{vec}} - \lambda_{\text{vec}} \otimes I_{m})^{\text{ns}} ds.$$

If we add the assumption that the matrix $Q$ is such that, in each asymmetric state, the intensity associated with returning to at least one symmetric state is positive, then $Q^{\text{ns}}$ is a strictly diagonally-dominant matrix with negative diagonal terms and we have therefore, from Gershgorin circle theorem, that $Q^{\text{ns}}$ is invertible and that $\lim_{T \to +\infty}\exp(Q^{\text{ns}}T) =0$. We conclude that  
\begin{eqnarray}\label{vas}
v^{\text{ns}}_T(t) &=& (Q^{\text{ns}})^{-1} \exp(Q^{\text{ns}}(T-t)) (I_{m} \otimes \lambda_{\text{vec}} - \lambda_{\text{vec}} \otimes I_{m})^{\text{ns}} - (Q^{\text{ns}})^{-1} (I_{m} \otimes \lambda_{\text{vec}} - \lambda_{\text{vec}} \otimes I_{m})^{\text{ns}}\nonumber \\
&&\to_{T \to +\infty} - (Q^{\text{ns}})^{-1}(I_{m} \otimes \lambda_{\text{vec}} - \lambda_{\text{vec}} \otimes I_{m})^{\text{ns}} .\end{eqnarray}

Under the above assumptions, we can therefore define $v(\lambda^b, \lambda^a) = \lim_{T \to +\infty} v_T(t, \lambda^b, \lambda^a)$ and write  $$S^{\text{micro}}_t = S_t + \kappa v(\lambda^b_t, \lambda^a_t).$$

Of course, in practice, one never knows the current state of liquidity and rather uses a probability distribution~$\pi$ over the states. Then, one gets at time $t$ a micro-price with mean \begin{equation}
\label{mpestim}
\bar{S}^{\text{micro}}_t = S_t + \kappa \sum_{1\le j_b \le m, 1\le j_a \le m} \pi^{j_b, j_a} v(\lambda^{j_b}, \lambda^{j_a})
\end{equation} and standard deviation\footnote{This standard deviation only quantifies the uncertainty linked to the estimation of the current market liquidity state for given values of the model parameters.} given by
$$\kappa\sqrt{\sum_{1\le j_b \le m, 1\le j_a \le m} \pi^{j_b, j_a} v(\lambda^{j_b}, \lambda^{j_a})^2 - \left(\sum_{1\le j_b \le m, 1\le j_a \le m} \pi^{j_b, j_a} v(\lambda^{j_b}, \lambda^{j_a})\right)^2 }.$$

\subsubsection{Main remarks on our assumptions to obtain a micro-price}

To be able to define the notion of micro-price and ensure that the limit exists, we imposed three structural assumptions to our model for the flow of RFQs. We indeed imposed that:
\begin{itemize}
\item the set of possible intensities is shared across the bid and the ask;
\item the transition matrix $Q$ is both that of $(\lambda^b_t, \lambda^a_t)$ and $(\lambda^a_t, \lambda^b_t)$ -- in the case of two liquidity states, high and low, this means that the chance of any transition leading to or from an unbalanced state does not depend on the side of the imbalance;
\item any unbalanced state has a chance to be followed by at least one balanced state.
\end{itemize}

These assumptions are quite light and natural. What is more questionable is the linearity assumption in the drift. It is, however, important to keep in mind that intensities are not observed directly; only the probability of being in the different states can be estimated. This results in noisy estimates of \(\kappa\), as we shall see. Parsimony clearly guided our modeling choice.

\subsection{A fair transfer price} 
 
\subsubsection{From a market making model to a fair transfer price}\label{MMmodel}

We now want to go beyond the notion of micro-price and use ideas coming from the OTC market making literature to define a fair transfer price.\\

We consider a theoretical market maker receiving RFQs to buy and sell an asset (for a given unique size $z$ in this model). We model the number of RFQs at the bid and at the ask by a bidimensional MMPP as above and do not impose here, for the sake of generality, our symmetry assumptions that guaranteed the well-posedness of the definition of micro-price.\\

We consider a reference price process $(S_t)_t$ and we assume that
$$dS_t = \sigma dW_t + \kappa(\lambda^a_t - \lambda^b_t)dt.$$

Upon receiving at time $t$ an RFQ at the bid (resp. at the ask), the market maker answers a price $S_t^b = S_t - \delta_t^b$ (resp. $S_t^a = S_t + \delta_t^a$) and this leads to a trade with probability $f^b(S_t-S_t^b) = f^b(\delta_t^b)$ (resp. $f^a(S_t^a-S_t) = f^a(\delta_t^a)$) where $f^b$ (resp. $f^a$) is a decreasing function from $1$ to $0$ (sometimes called S-function or S-curve). The inventory process $(q_t)_t$ of the market maker evolves subsequently as
$$dq_t = zdN^b_t - zdN^a_t,$$
where $z>0$ is the (constant) transaction size. The cash process $(X_t)_t$ evolves therefore as
$$dX_t = zS^a_t dN^a_t - zS^b_t dN^b_t = -S_t dq_t + z\delta_t^b dN_t^b + z\delta_t^a dN_t^a$$
and the PnL process $(\text{PnL}_t)_t = (X_t+q_t S_t)_t$ as
$$d\text{PnL}_t = z\delta_t^b dN_t^b + z\delta_t^a dN_t^a + \sigma q_t dW_t + \kappa(\lambda_t^a - \lambda_t^b)q_t dt.$$

A market maker wishing to capture the bid-ask spread while mitigating the risk (see \cite{cartea2015algorithmic,cartea2014buy, gueant2016financial}) typically maximizes, over the set of predictable processes $(\delta^b_t)_t$ and $(\delta^a_t)_t$, the objective function 
\begin{align*}
\mathbb E \left[\int_0^T \left(z\lambda_t^b \delta_t^b f^b(\delta_t^b) + z\lambda_t^a \delta_t^a f^a(\delta_t^a) + \kappa(\lambda_t^a - \lambda_t^b)q_t - \frac \gamma 2  \sigma^2 q_t^2 \right)dt\right]
\end{align*}
for a given risk aversion parameter $\gamma >0$.\\

Assuming that the theoretical market maker is able to identify in which state $(\lambda^{j_b,b}, \lambda^{j_a,a})$ the market is at any point in time, the value functions $(\theta^{j_b, j_a})_{1\le j_b \le m_b, 1\le j_a \le m_a}$ satisfy the following system of Hamilton-Jacobi-Bellman (HJB) equations:
\begin{align*}
    & \partial_t \theta^{j_b, j_a}(t,q) + \kappa(\lambda^{j_a,a} - \lambda^{j_b,b}  )q - \frac 12 \gamma \sigma^2 q^2 + \sum_{1\le k_b \le m_b, 1\le k_a \le m_a}  Q_{(j_b-1) m_a+j_a, (k_b-1) m_a+k_a} \theta^{k_b, k_a}(t,q) \\
    &+  z \lambda^{j_b,b} H^{b} \left( \frac{\theta^{j_b, j_a}(t,q) - \theta^{j_b, j_a}(t,q+z)}{z} \right) + z \lambda^{j_a,a} H^{a} \left( \frac{\theta^{j_b, j_a}(t,q) - \theta^{j_b, j_a}(t,q-z)}{z} \right) = 0 
\end{align*}with terminal condition $\theta^{j_b, j_a}(T,q) = 0$, where
$H^{b/a}(p) = \underset{\delta \in \mathbb R}{\sup} f^{b/a}(\delta)(\delta-p).$

Under mild assumptions on the functions $f^b$ and $f^a$ (see for instance \cite{bergault2021size}), the optimal bid and ask quotes of the market maker if the current state of the market is $(\lambda^{j_b,b}, \lambda^{j_a,a})$ write 
\[
\delta^{b,i,\star}_t = \bar\delta^{b}\left( \frac{\theta^{j_b, j_a}(t,q_{t-}) - \theta^{j_b, j_a}(t,q_{t-}+z)}{z}\right) \quad \text{and} \quad
\delta^{a,i,\star}_t = \bar\delta^{a}\left( \frac{\theta^{j_b, j_a}(t,q_{t-}) - \theta^{j_b, j_a}(t,q_{t-}-z)}{z}\right),
\]
where
\begin{align*}
    \bar \delta^b(p) = {f^b}^{-1} \left( - {H^b}'(p) \right) \quad \text{and} \quad  \bar \delta^a(p) = {f^a}^{-1} \left( - {H^a}'(p) \right).
\end{align*}
If the matrix $Q$ is irreducible, an ergodic limit exists, \textit{i.e.} there exists a constant $c$ such that we have $\lim_{T \to \infty}\theta^{j_b, j_a}(t,q) - c(T-t) = \theta^{j_b, j_a}_{\infty}(q)$ (see~\cite{gueant2020optimal} for a general framework), and we can define a time-independent notion of skew by $$\text{skew}^{j_b, j_a}_\infty = \bar\delta^{a}\left(\frac{\theta^{j_b, j_a}_\infty(0) - \theta^{j_b, j_a}_\infty(-z)}{z}\right) - \bar\delta^{b}\left(\frac{\theta^{j_b, j_a}_\infty(0) - \theta^{j_b, j_a}_\infty(z)}{z}\right).$$

This skew ``projects'' the asymmetry of the market liquidity into the price space because of the market maker's need to quote asymmetrically in order to account for that asymmetric liquidity (even in the absence of inventory).\\

The notion of Fair Transfer Price we propose (from now on FTP) is then defined as the mid-price of a market maker with infinite horizon and no inventory, \textit{i.e.} it is defined at time $t$ by:
$$S^{\text{FTP}}_t = S_t + \frac 12 \text{skew}_\infty.$$
It corresponds to the average between the price answered to a buyer and the price answered to a seller by a theoretical market maker with no inventory, if they were requested -- hence the dimension of fairness.\\

Of course, in practice, one never knows the current state of liquidity and rather uses a probability distribution~$\pi$ over the states. One gets at time $t$ an FTP with mean 
\begin{align}\label{SFTP}
    \bar{S}^{\text{FTP}}_t = S_t + \frac 12 \sum_{1\le j_b \le m_b, 1\le j_a \le m_a} \pi^{j_b, j_a} \text{skew}^{j_b, j_a}_\infty
\end{align}
and standard deviation given by
$$\frac 12\sqrt{\sum_{1\le j_b \le m_b, 1\le j_a \le m_a} \pi^{j_b, j_a} \left(\text{skew}^{j_b, j_a}_\infty\right)^2 - \left(\sum_{1\le j_b \le m_b, 1\le j_a \le m_a} \pi^{j_b, j_a} \text{skew}^{j_b, j_a}_\infty\right)^2 }.$$

\subsubsection{Main remarks on FTP}

It is a priori hard to relate our concept of FTP to those used in the case of LOBs. However, since top-of-book volumes in LOBs are inversely related to the appetite of liquidity takers (the clients in a dealer market), FTP shares many characteristics with the weighted mid-prices of LOBs. Bid/ask imbalances in LOBs are indeed comparable to bid/ask asymmetries in client flows, though in the opposite direction. A very high volume at the bid (relative to the ask) in an LOB is similar to a situation in a dealer market where clients are more willing to buy than to sell. In this context, a market maker should skew their quotes towards the right, pushing the FTP upwards. This is in line with a weighted mid-price above the mid-price in an LOB.\\

At first sight, the notion of FTP depends strongly on the reference price chosen to build the market making model. This dependence is real but it is not the serious caveat it might seem. If indeed we replace \((S_t)_t\) by \((S_t + \xi)_t\), then the functions \(f^b\) and \(f^a\) should be shifted accordingly in the estimation procedure if we assume that trading decisions depend on the absolute values (as opposed to relative) of proposed prices. Subsequently, value functions should be translated by a term \(\xi q\), and it is easy to see that the FTP would be unchanged since the functions \(\bar{\delta}^b\) and \(\bar{\delta}^a\) are themselves translated by \(\pm \xi\). Of course, this invariance is limited to constant shifts, but it shows that differences between (relevant) reference prices should be partially or entirely compensated by their impact on the definition/estimation of \(f^b\) and \(f^a\).\\

In the definition of FTP, the transaction size \(z\) inputted in the market making model plays a role, although it might seem arbitrary. This transaction size \(z\) could be the reference size for which market makers stream prices on electronic platforms. One can naturally generalize the concept to consider any size.\\

Another important point is that the FTP depends on the risk aversion parameter \(\gamma\) inputted into the objective function. This might seem problematic since the parameter \(\gamma\) can be chosen arbitrarily. However, the choice of \(\gamma\) leaves one degree of freedom, which is rather an opportunity. In Section~4, we use \(\gamma\) to calibrate the model to observed bid-ask spreads.\\

Many improvements can be made to the market making model in line with what exists in the literature, such as considering several transaction sizes (see \cite{bergault2021size}), taking account of client tiering (see \cite{barzykin2021market}), using the possibility to externalize part of the flow (see \cite{barzykin2021market, barzykin2023algorithmic}), replacing the quadratic running penalty -- linked to the objective function proposed in \cite{cartea2014buy} -- with a more complicated one, etc.\footnote{The introduction of asymmetric inventory costs, linked to repo rates for instance, is however not recommended if the goal is to define a fair price between two parties.} One can also decide to use a multi-asset market making model instead of a single-asset one (see, for instance, \cite{barzykin2022dealing}, \cite{bergault_closed-form_2021}, and \cite{gueant2017optimal}). In fact, the notion of FTP is versatile, and one can choose the market making model they prefer. It must be noted that the role of the market making model here is not to provide quotes that will be used inside a trading algorithm, but rather to project information regarding liquidity levels, liquidity imbalances, and volatility into the price space. In particular, there is no real problem in keeping the market making model relatively simple (as long as liquidity dynamics are taken into account), especially since one can rely on the degree of freedom provided by the risk aversion parameter \(\gamma\) to match a desired target (see Section~\ref{sec:num} below).

\section{From theory to practice}
\label{sec:num}
\subsection{Introduction}

In the above sections, we have extended the notion of micro-price and defined the new concept of Fair Transfer Price (FTP). To use these notions in practice, we need to estimate several parameters.\\

First, we need to estimate the parameters of the bidimensional MMPP. In Section~2, we detailed an estimation procedure based on an EM algorithm (extensions are presented in Appendix~\ref{annexe}). Then, to compute the micro-price and/or estimate the dynamics of the reference price in the market making model, we need to estimate the constant \(\kappa\). This is typically done with a linear regression of price moves on terms of the form \(\sum_{1\le j_b \le m, 1\le j_a \le m} \pi^{j_b, j_a} v(\lambda^{j_b}, \lambda^{j_a})\), following Eq.~\eqref{mpestim}. When it comes to using FTP, the volatility parameter \(\sigma\) is also necessary, and classical estimators can be used for that purpose. One also needs to estimate the parameters used in modeling the conversion of an RFQ into a trade, \textit{i.e.}, the parameters of \(f^b\) and \(f^a\) once a parametric functional form has been chosen. \(f^b\) and \(f^a\) are typically chosen to be logistic, and the estimation procedure boils down to logistic regressions. In addition to the estimation of parameters, the use of FTP requires choosing a risk aversion parameter and solving an HJB equation to get optimal quotes.\\

In what follows, we illustrate our approach and the different concepts of price on corporate bond data. For that purpose, we use an anonymized dataset of RFQs on high-yield corporate bonds kindly provided by J.P.~Morgan. It contains, for each RFQ, the date and time of the request, the bond requested, the direction of the request (buy or sell), the notional (odd lots have been removed from the dataset), the price answered to the client, the current market (in fact composite) prices at the bid and at the ask, and the status -- \textit{i.e.}, whether the price was accepted by the client or not. Because some requests are only sent by clients to gather information, we focused on RFQs that led to a trade with J.P.~Morgan or with another dealer (this piece of information, but of course not the identity of the other dealer, is known as we focus on RFQs sent through multi-dealer-to-client platforms). Our dataset contains bonds from four different sectors.\footnote{For confidentiality reasons, we do not document the list of bonds and sectors.} It covers more than half a year of trading, over the post-COVID period.\footnote{For confidentiality reasons, we do not document the exact period of time. Throughout the paper, the unit for times is in days since the beginning of the period, excluding weekends. Nights have also been excluded so that the beginning of the next trading day follows the end of the current one -- trading hours have been set from 7am to 5pm.}

\subsection{Estimation of the parameters of the bidimensional Markov-modulated Poisson process}

For the estimation of the parameters of the bidimensional Markov-modulated Poisson process, we consider the multi-asset extension presented in Appendix \ref{MMPPmulti} to carry out the process at the sector level. To illustrate our notion of micro-price, we also rely on the exchangeability assumption detailed in Appendix \ref{MMPPex}.\\

\begin{figure}[h!]
\centering
\includegraphics[width=\textwidth]{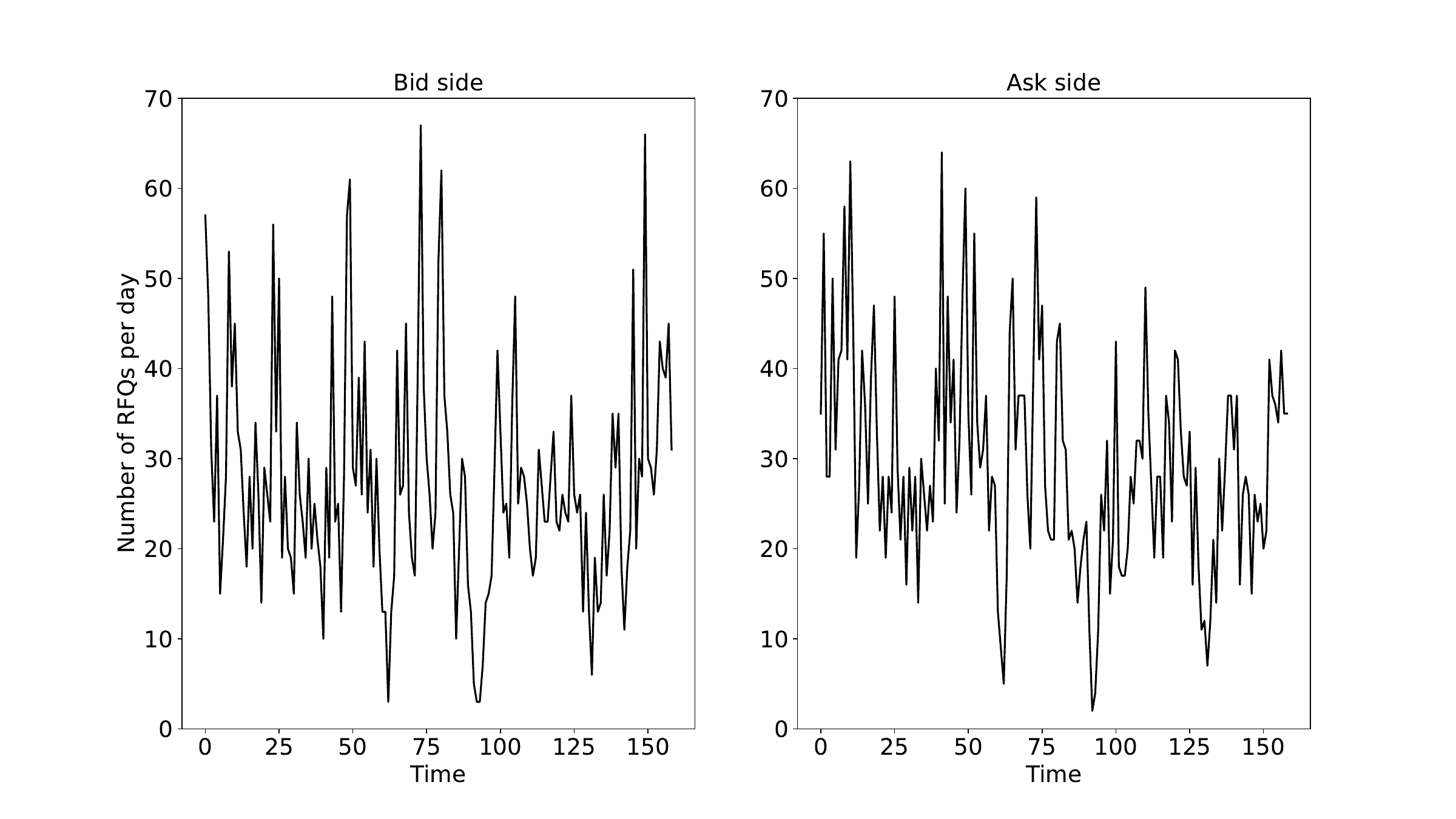}\\
\caption{Number of RFQs per day at the bid and at the ask for Sector 1.}
\label{rolling_1}
\end{figure}

\begin{figure}[h!]
\centering
\includegraphics[width=\textwidth]{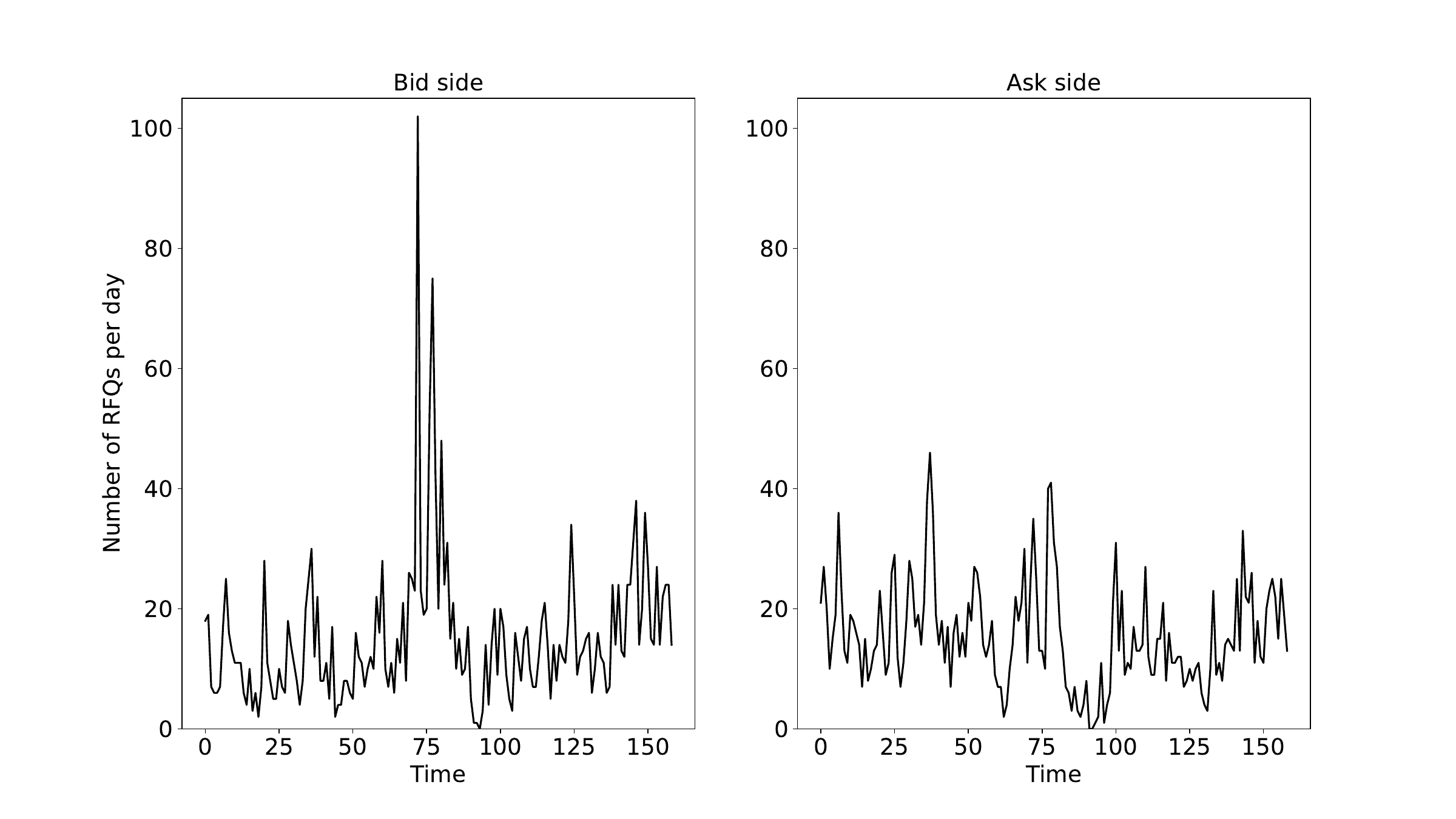}\\
\caption{Number of RFQs per day at the bid and at the ask for Sector 2.}
\label{rolling_2}
\vspace{1.75cm}
\end{figure}

\begin{figure}[h!]
\centering
\includegraphics[width=\textwidth]{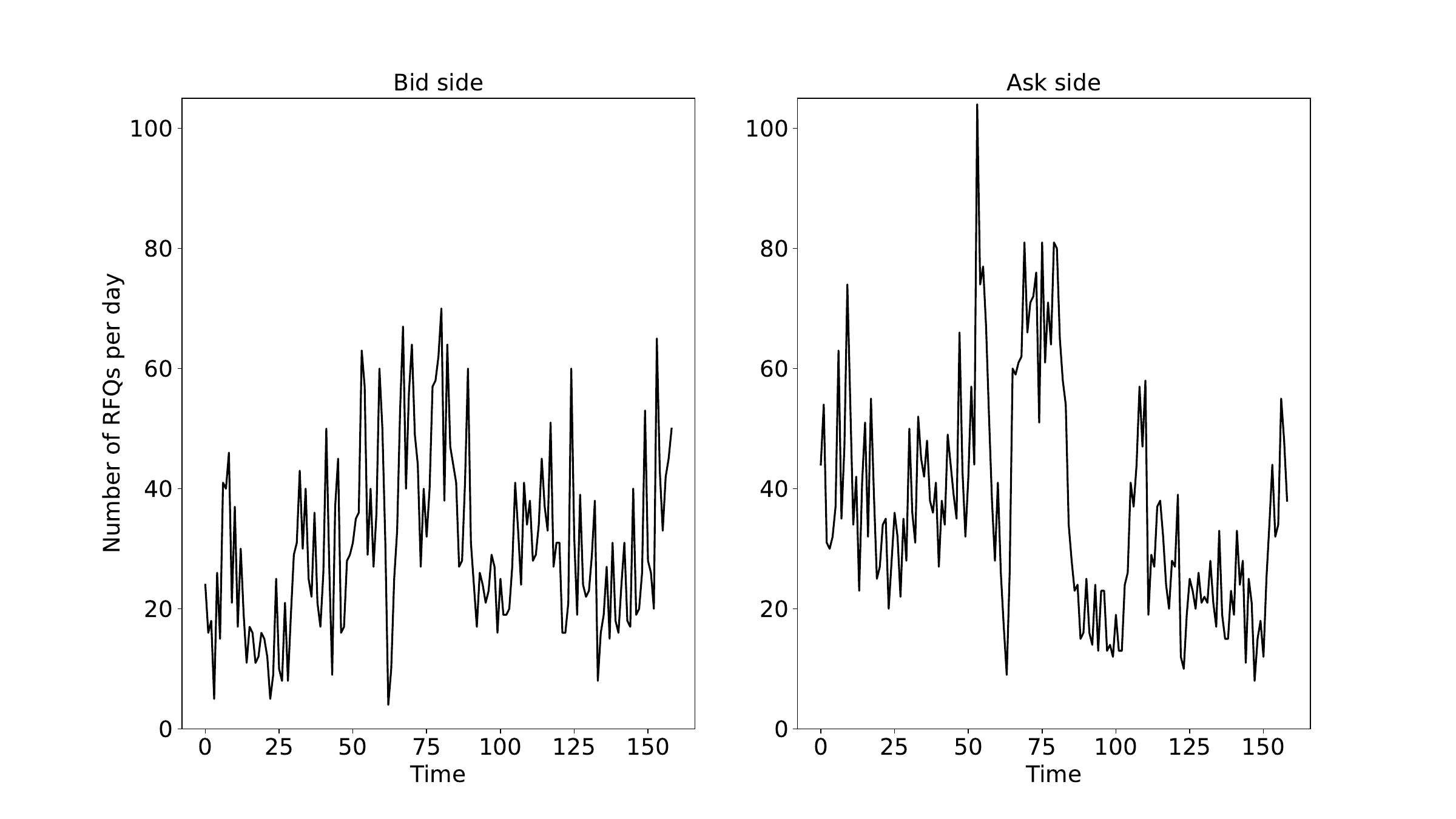}\\
\caption{Number of RFQs per day at the bid and at the ask for Sector 3.}
\label{rolling_3}
\end{figure}

\begin{figure}[h!]
\centering
\includegraphics[width=0.97\textwidth]{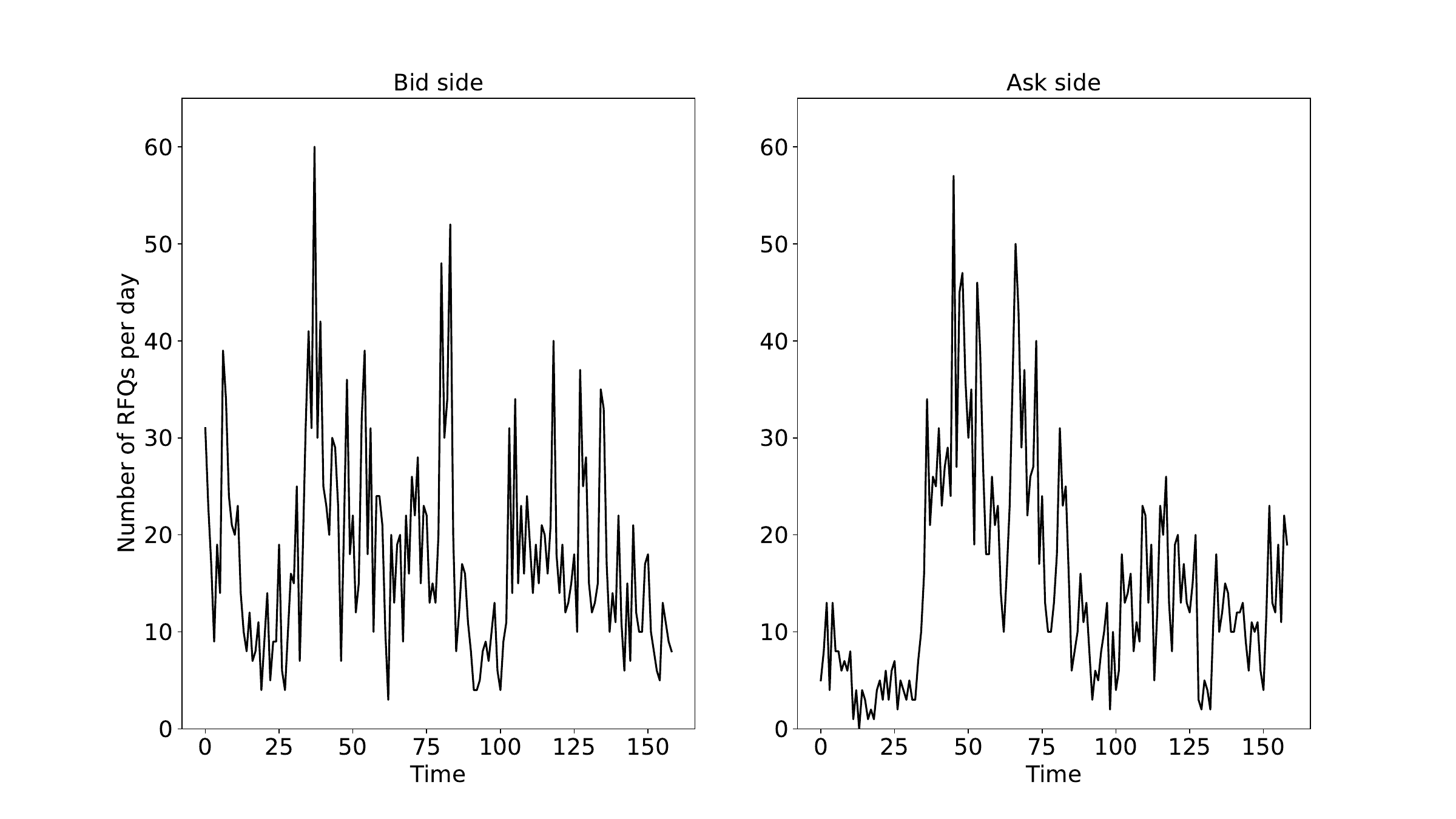}\\
\caption{Number of RFQs per day  at the bid and at the ask for Sector 4.}
\label{rolling_4}
\vspace{1.5cm}
\end{figure}

In the EM algorithm corresponding to the extension presented in Appendix \ref{MMPPex}, one must choose the number \(m\) of intensities, set their initial values, and those of the coefficients of the transition matrix. To obtain a first naive estimation of the intensities of the bidimensional Markov-modulated Poisson process for each sector, we started by computing the number of RFQs per day at the bid and at the ask. The results are plotted in Figures~\ref{rolling_1}, \ref{rolling_2}, \ref{rolling_3}, and \ref{rolling_4}. We clearly see that liquidity is volatile and that upward or downward bumps may be simultaneous across bid and ask (see, for instance, what happens around \(t=60\) and \(t=90\) in Figure~\ref{rolling_1}), but also asymmetric with one side seeing a rise or a decrease in liquidity while the other does not (see, for instance, what happens around \(t=75\) in Figure~\ref{rolling_2}).\\

For each sector, we decided to consider two intensities (\(m = 2\)), and initialized them using the average over the bid and ask sides of the 10th percentile (for the low liquidity state) and the 90th percentile (for the high liquidity state) of the distribution of the number of RFQs per day. To initialize the matrix \(Q\), we used very naive values corresponding to independence between the intensities at the bid and the ask and transition rates from low to high and high to low equal to \(1\) (per day).\\

We ran the EM algorithm over our database of RFQs, sector by sector (using the technique described in Appendix \ref{MMPPmulti} on the multi-asset extension). To normalize likelihoods, we used classical regression techniques. We noticed convergence of the values of \(\lambda^1\) and \(\lambda^2\) and the coefficients of the matrix \(Q\) after approximately 50 steps. The resulting parameters of the bidimensional MMPP are reported in Table \ref{tableMMPP}. We clearly see that, for the four sectors, the algorithm manages to separate low liquidity from high liquidity. We also see that the transition matrices are different across sectors: high transition rates and a relatively high probability of jumping from an imbalanced state to the opposite imbalanced state in the case of Sector 1, a very stable (resp. unstable) low/low-liquidity (resp. high/high-liquidity) state in the case of Sector 2, and low transition rates for Sector~4.\\

\begin{figure}[h!]
\centering
\includegraphics[width=\textwidth]{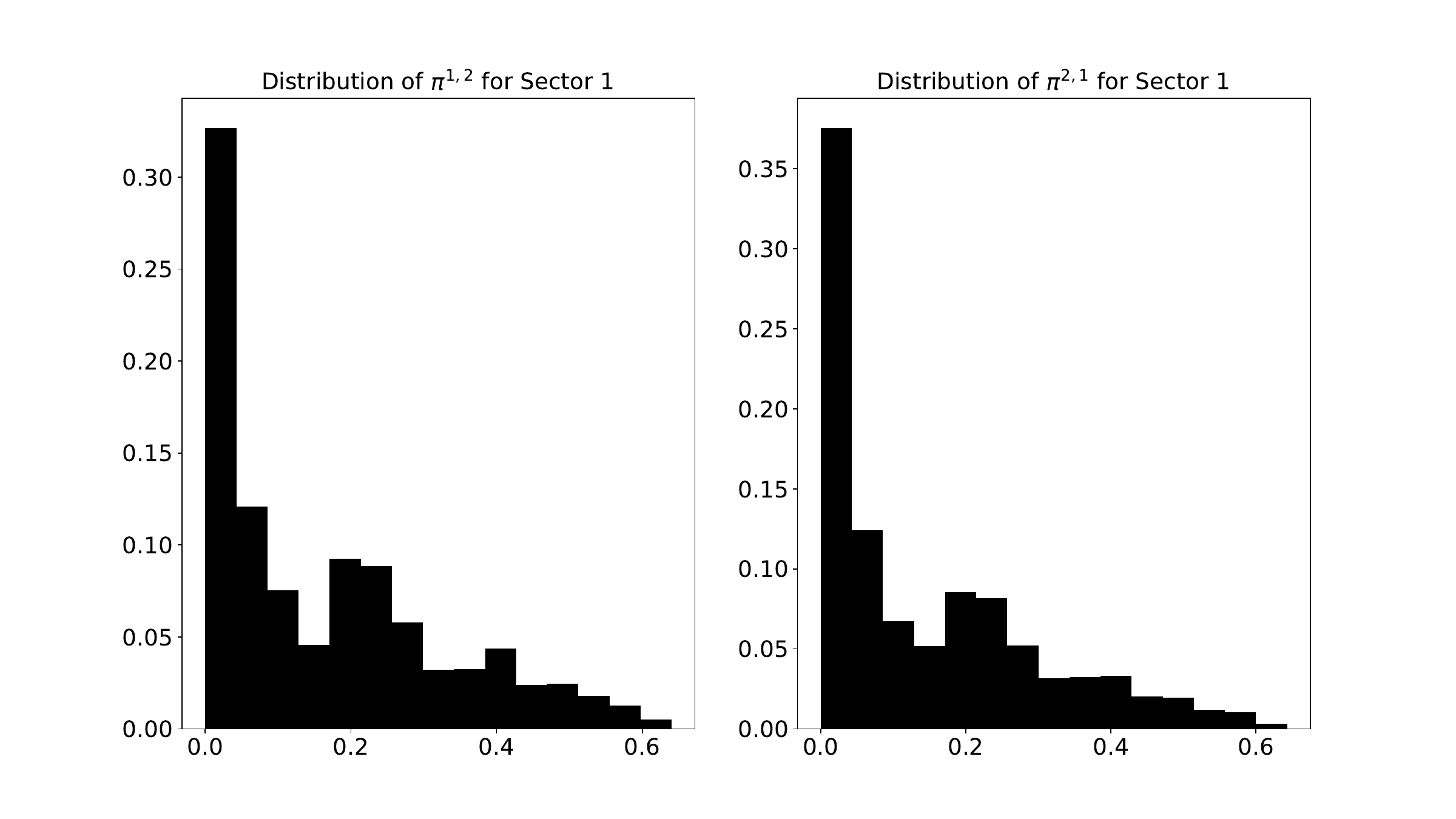}\\
\caption{Distribution of the values taken by $\pi^{1,2}$ and $\pi^{2,1}$ for Sector 1.}
\label{hist_1}
\end{figure}

\begin{figure}[h!]
\centering
\includegraphics[width=\textwidth]{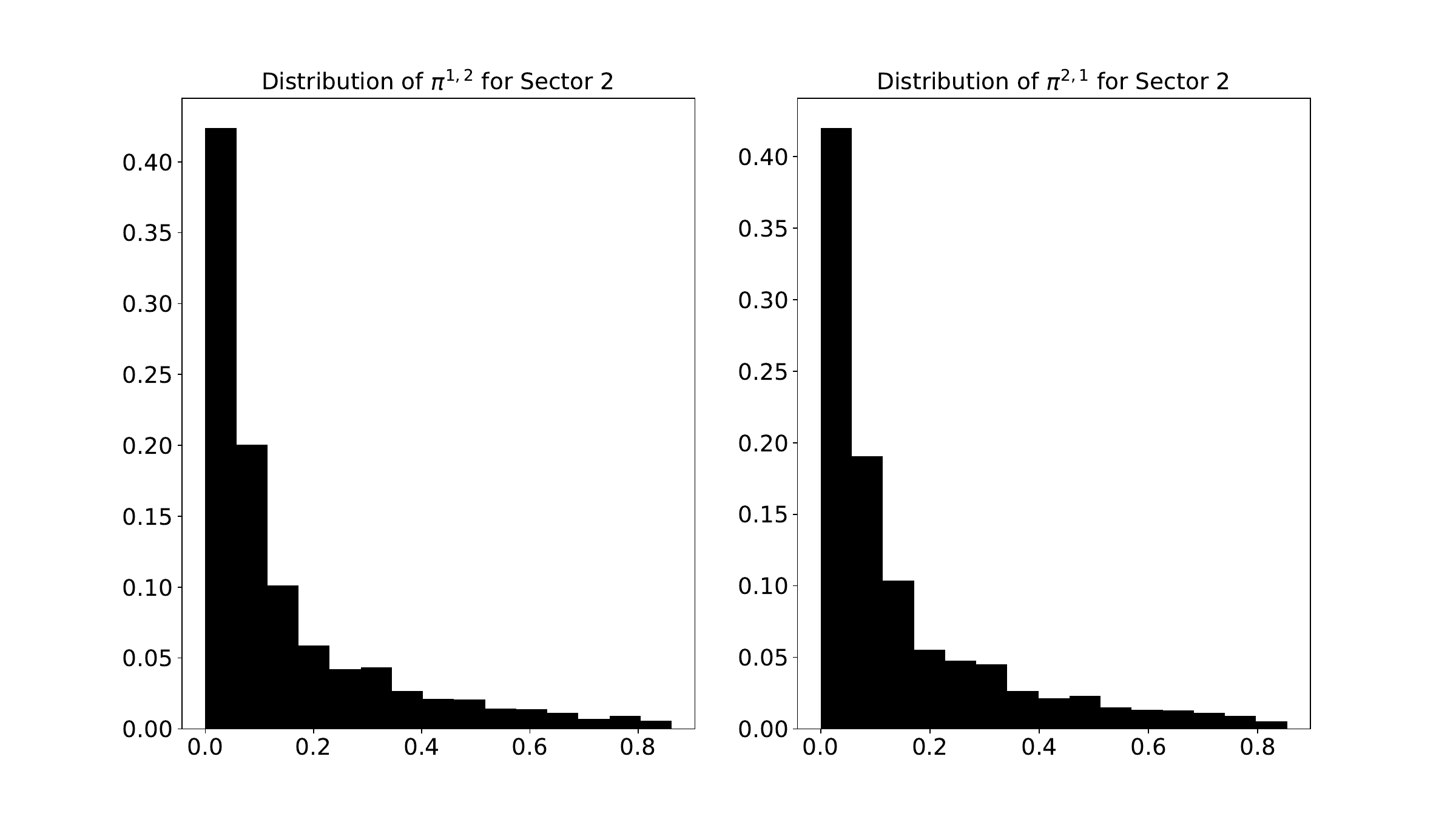}\\
\caption{Distribution of the values taken by $\pi^{1,2}$ and $\pi^{2,1}$ for Sector 2.}
\label{hist_2}
\vspace{15mm}
\end{figure}

\begin{figure}[h!]
\centering
\includegraphics[width=\textwidth]{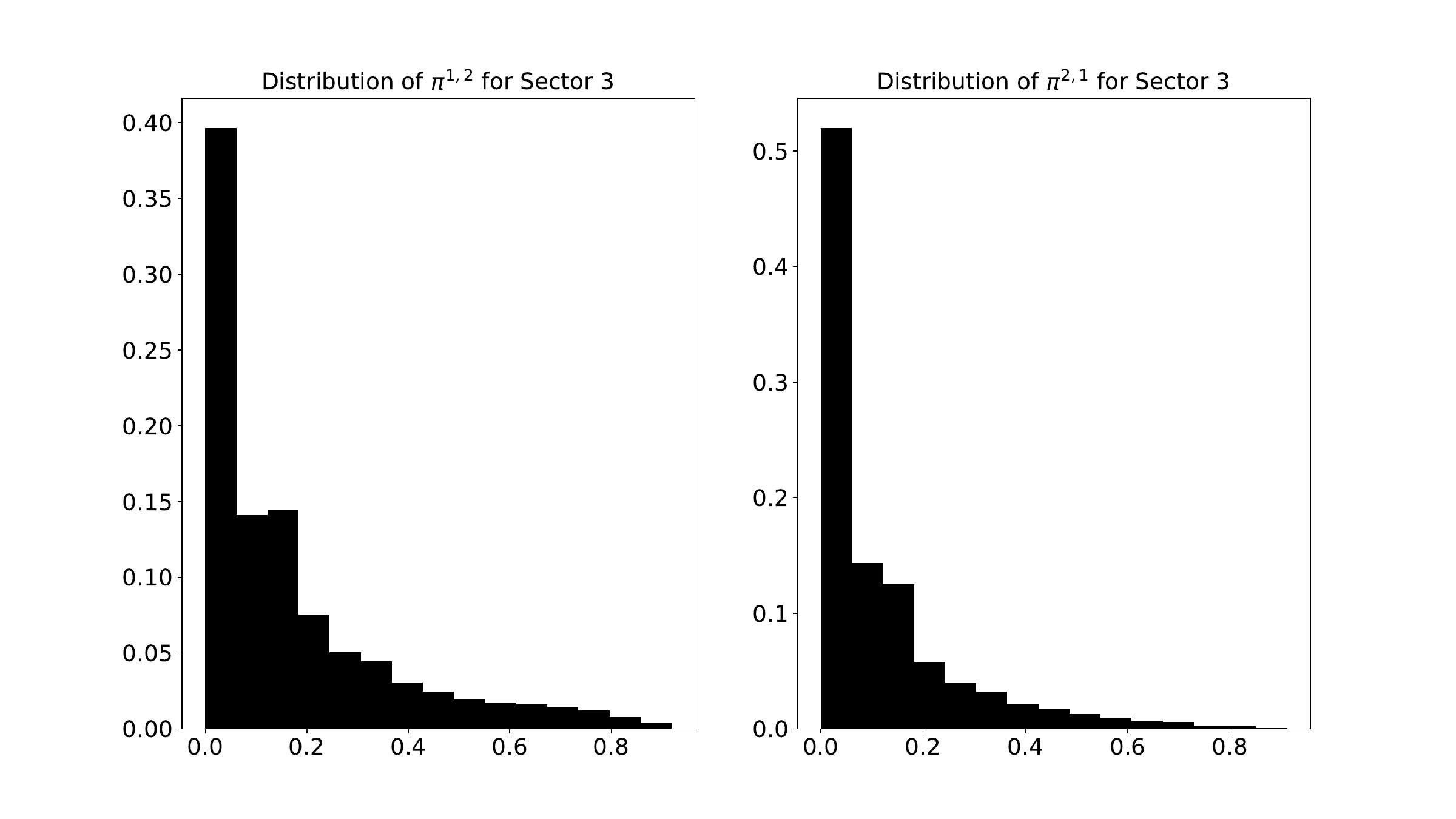}\\
\caption{Distribution of the values taken by $\pi^{1,2}$ and $\pi^{2,1}$ for Sector 3.}
\label{hist_3}
\end{figure}

\begin{figure}[h!]
\centering
\includegraphics[width=\textwidth]{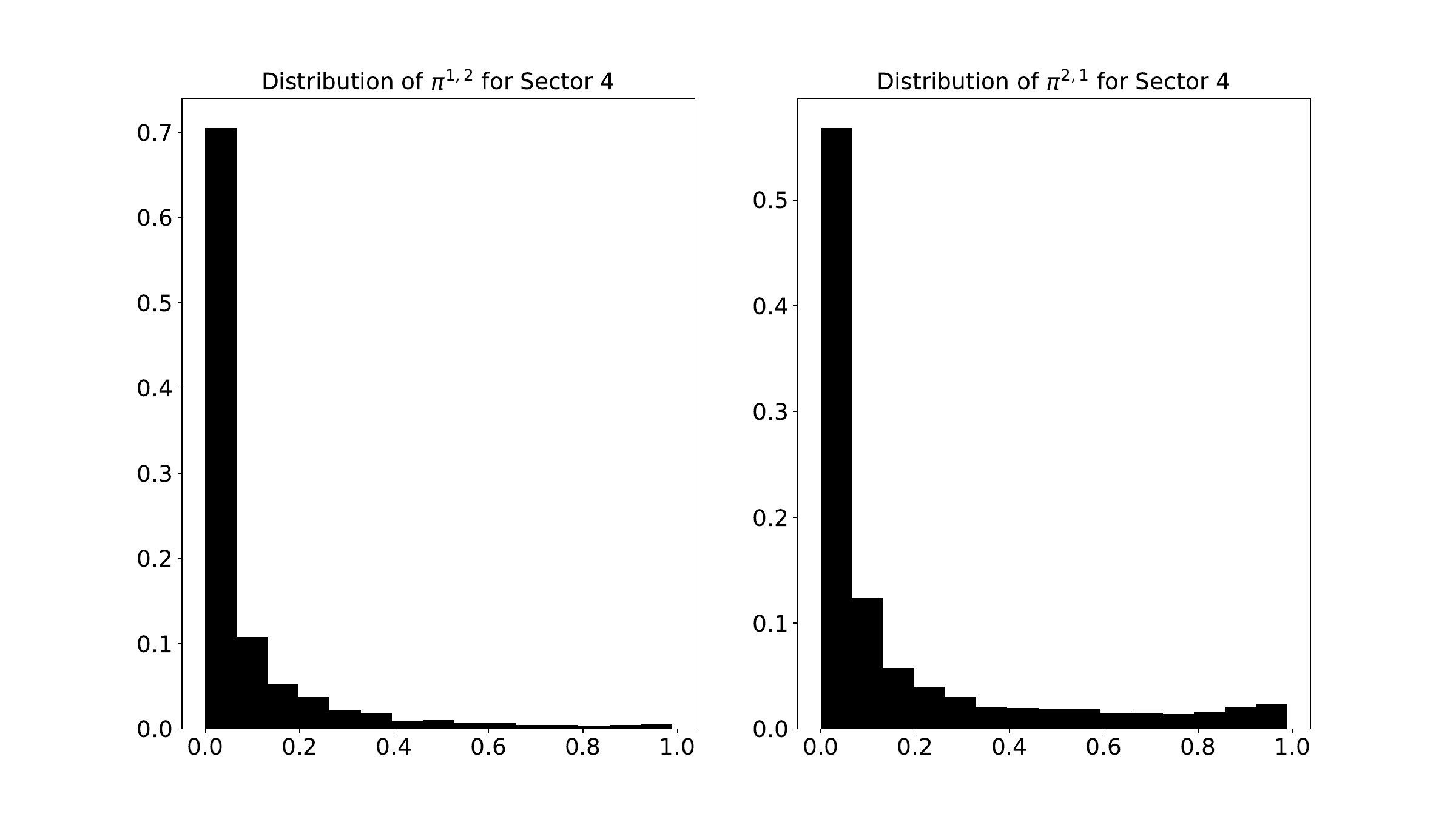}\\
\caption{Distribution of the values taken by $\pi^{1,2}$ and $\pi^{2,1}$ for Sector 4.}
\label{hist_4}
\vspace{15mm}
\end{figure}

Once the parameters of the bidimensional MMPP have been estimated, we can evaluate at each point in time the probability of being in each state (see Section \ref{current_est}). In Figures~\ref{hist_1}, \ref{hist_2}, \ref{hist_3}, and \ref{hist_4}, we document the distribution of the values taken by the probability \(\pi^{1,2}\) (resp. \(\pi^{2,1}\)) of being in a low/high-liquidity (resp. high/low-liquidity) state. We clearly see that high values of these probabilities are quite rare: it is hard to be certain that a disequilibrium in RFQs indeed corresponds to an underlying asymmetric regime.\\



\begin{table}[h!]
\begin{center}
\begin{tabular}{c | c c c } 
 \hline
Sector &  $\lambda^{1}$ & $\lambda^{2}$ & $Q$\\ [2ex] 
 \hline
Sector 1 & 10.83 & 73.03 & $\begin{pmatrix} 
-14.01 &   4.37 & 4.37  &5.27\\
 19.32 & -60.91 & 12.54  &29.05\\
 19.32 & 12.54 &-60.91  &29.05\\
  23.67 & 15.00  &15.00 &-53.67
\end{pmatrix}$ \\ [7ex]  

Sector 2 & 8.44 & 58.28 & $\begin{pmatrix} -4.55 &  1.00 & 1.00 & 2.55\\
  18.53 & -28.31 &  0.13  & 9.65\\
  18.53 & 0.13 &  -28.31 &  9.65\\
 14.77 & 16.73 & 16.73 & -48.23 \end{pmatrix}$ \\ [7ex] 

 Sector 3 & 15.73 & 81.78 & $\begin{pmatrix} -9.98 &  2.79& 2.79 & 4.40\\
  20.53 & -23.73 &  0.02  & 3.18\\
  20.53 & 0.02 &  -23.73 &  3.18\\
 9.87 & 4.17 & 4.17 & -18.21 \end{pmatrix}$ \\ [7ex] 

 Sector 4 & 7.33 & 28.32 & $\begin{pmatrix} -1.67 &  0.48 & 0.48 & 0.71 \\
  1.92 & -2.02 &  0.00  & 0.10\\
  1.92 & 0.00 &  -2.02 &  0.10\\
 0.84 & 0.11 & 0.11 & -1.06 \end{pmatrix}$ \\ [7ex] 
 \hline
\end{tabular}
\end{center}
\caption {Estimated parameters of the bidimensional Markov-modulated Poisson process for the four sectors (the intensities are given in $\text{day}^{-1}$).}
\label{tableMMPP}
\end{table}

\vspace{-4mm}



\subsection{Micro-price}

\subsubsection{Estimation of $\kappa$}

In order to illustrate our concept of micro-price, we need first to estimate the parameter $\kappa$ in the dynamics of bond prices. For each sector, we consider four bonds amongst those available in the database. We first compute the functions $v$ given by Eq. \eqref{vas} and then use Eq. \eqref{mpestim} to perform a linear regression and estimate~$\kappa$ for each bond. We also compute the arithmetic volatility $\sigma$ and the weight $\beta$ associated with each bond (see Appendix~\ref{MMPPmulti}). The results are reported in Table~\ref{params}.\\

\begin{table}[h!]
\vspace{5mm}
\begin{center}
\begin{tabular}{c c | c c c } 
 \hline
Sector & Bond & $\beta$ & $\kappa$ (stdev) & $\sigma$\\ [1ex] 
 \hline
1 & 1 & 0.10 & 2.29 (0.55) & 18.39  \\ [0.6ex]  
 & 2 & 0.10 & 0.25 (0.49) & 15.43  \\ [0.6ex] 
 & 3 & 0.06 & 2.83 (1.66) & 22.55  \\ [0.6ex]  
 & 4 & 0.05 & 0.33 (2.23) & 19.75  \\ [0.6ex]
\hline
2 & 1 & 0.19 & 0.57 (0.19) & 13.75   \\ [0.6ex]  
 & 2 & 0.14 & 0.90 (0.22) & 16.05  \\ [0.6ex] 
 & 3 & 0.11 & 0.65 (0.16) & 9.80  \\ [0.6ex]  
 & 4 & 0.10 & 0.86 (0.68) & 20.36  \\ [0.6ex]
\hline
3 & 1 & 0.11 & 0.61 (0.34) & 9.93 \\ [0.6ex]  
 & 2 & 0.09 & 0.05 (0.16) & 18.41   \\ [0.6ex] 
 & 3 & 0.06 & 0.11 (0.08) & 12.23   \\ [0.6ex]  
 & 4 & 0.05 & 0.08 (0.11) & 18.68 \\ [0.6ex]
\hline
4 & 1 & 0.21 & 0.04 (0.02) & 13.00  \\ [0.6ex]  
 & 2 & 0.12 & 0.01 (0.01) & 24.09  \\ [0.6ex] 
 & 3 & 0.12 & 0.08 (0.04) & 16.91   \\ [0.6ex]  
 & 4 & 0.07 & 0.09 (0.05) & 12.67  \\ [0.6ex]
 \hline
\end{tabular}
\end{center}
\caption {Estimations of $\beta$, $\kappa$ (in $\$$) and $\sigma$ (in $\$ \cdot \text{day}^{-1/2}$).}
\label{params}
\vspace{1cm}
\end{table}

The estimated values for $\kappa$ are not all significantly different for $0$ (given the standard deviations reported), but it is nevertheless interesting to notice that the figures are positive for all bonds. This tends to prove that imbalance in the flow of RFQs has a consistant predictive power on the variation of the price, hence the interest of the concept of micro-price.

\subsubsection{Micro-price in practice}

In Table \ref{micro}, we took the last composite mid-price and bid-ask spread in the dataset for each of the 16 bonds we focus on, and computed the corresponding micro-price when we are \(100\%\) sure that the market is imbalanced, one way or the other.\\

Of course, and as confirmed by the above histograms, one can seldom be certain to be in any of the two imbalanced states. In practice, micro-prices must therefore be computed as expectations over the different possible states, \textit{i.e.}, as functions of the current estimates of the probabilities of being in each state. In particular, the micro-prices exhibited in Table \ref{micro} correspond to theoretical bounds for the micro-prices that would be used in practice.

\begin{table}[h!]
\begin{center}
\begin{tabular}{c c | c c c c c c } 
 \hline
Sector & Bond & Mid-price  & Bid price & Ask price & Micro-price $\pi^{2,1}\!\!\!=\!\!1$ & Micro-price $\pi^{1,2\!}\!\!=\!\!1$\\ [1ex] 
 \hline
1 &1 & 103.593  & 103.098 & 104.088 & 101.652 & 105.534\\ [0.6ex]  
&2 & 97.107  & 96.614  & 97.600 & 96.892 & 97.322 \\ [0.6ex] 
&3 & 99.146  & 98.631 & 99.661 & 96.752 & 101.541\\ [0.6ex]  
&4 & 94.187  & 93.049  & 95.325 & 93.909 & 94.465\\ [0.6ex]
\hline
2 &1 & 99.823  & 99.291 & 100.355 & 98.819 & 100.827 \\ [0.6ex]  
&2 & 99.270  & 98.603  & 99.936 & 97.700 & 100.840 \\ [0.6ex] 
&3 & 99.649  & 98.815 & 100.483 & 98.513 & 100.784\\ [0.6ex]  
&4 & 98.903  & 97.570  & 100.235 & 97.970 & 99.835 \\ [0.6ex]
\hline
3 &1 & 95.338  & 94.674  & 96.001 & 93.634 & 97.041 \\ [0.6ex]  
&2 & 92.394  & 91.860  & 92.927 & 92.252 & 92.535 \\ [0.6ex] 
&3 & 97.137  & 96.484 & 97.790 & 96.819 & 97.455 \\ [0.6ex]  
&4 & 94.839  & 94.220  & 95.458 & 94.810 & 94.867 \\ [0.6ex]
\hline
4 &1 & 102.632  & 102.151 & 103.112 & 102.252 & 103.011 \\ [0.6ex]  
&2 & 104.785  & 104.327  & 105.242 & 104.717 & 104.853 \\ [0.6ex] 
&3 & 104.824  & 104.293 & 105.355 & 103.994 & 105.654 \\ [0.6ex]  
&4 & 108.438  & 107.991  & 108.884 & 107.500 & 109.375 \\ [0.6ex]
 \hline
\end{tabular}
\end{center}
\caption {Micro-prices for the different bonds in imbalanced states.}
\label{micro}
\vspace{1cm}
\end{table}

In what follows, we study how the micro-price evolves depending on the probabilities of the different states of the bidimensional MMPP, for the first bond of each sector. Notice that, in our case, the respective values of \(\pi^{1,1}\) and \(\pi^{2,2}\) have no impact on the micro-price: only \(\pi^{1,2}\) and \(\pi^{2,1}\) matter.\\

\begin{figure}[h!]
  \begin{subfigure}[b]{0.5\linewidth}
    \centering
    \includegraphics[width=0.9\linewidth]{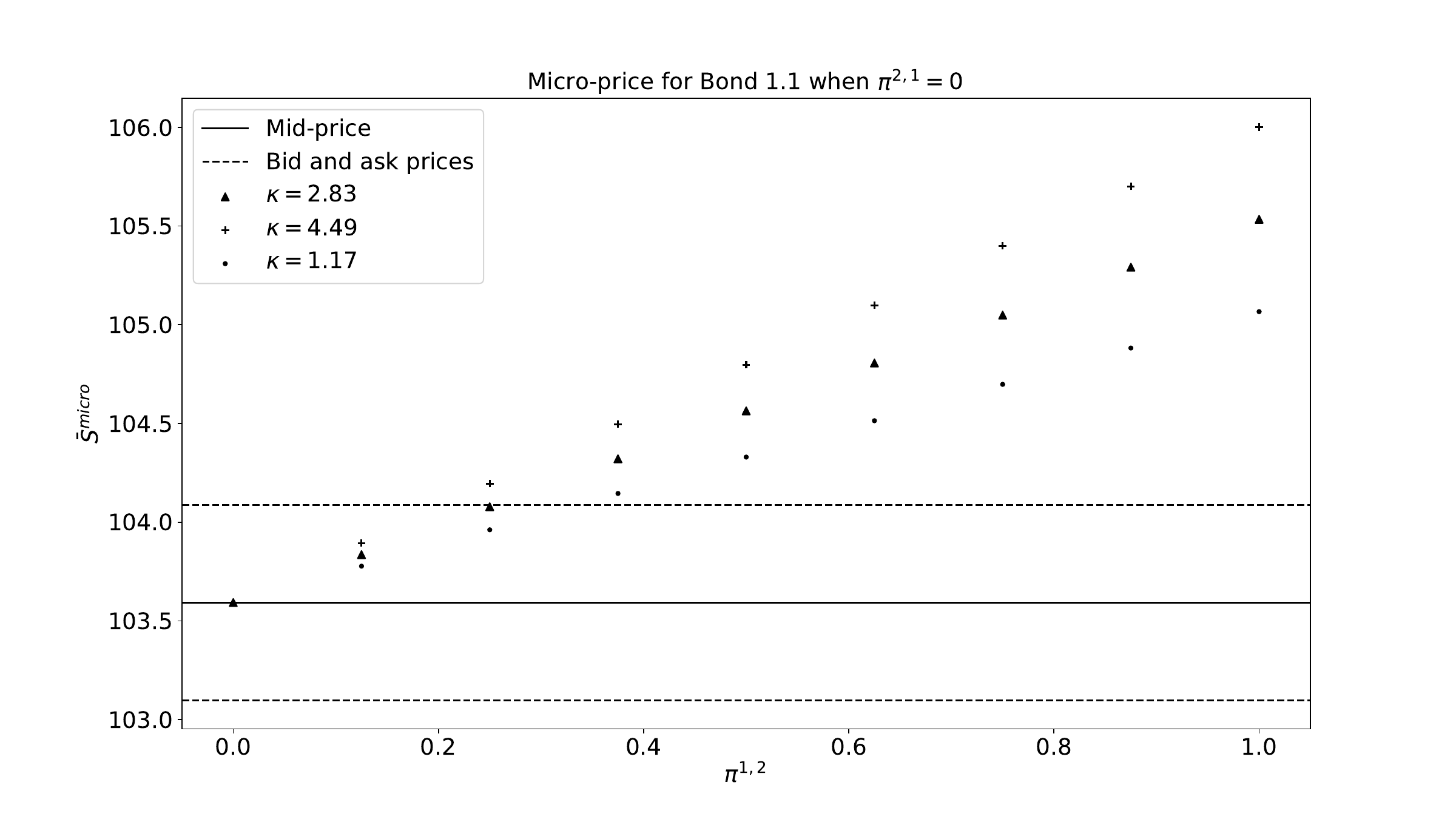} 
    \caption{Bond 1.1} 
    \label{mp_pi10_00:1} 

  \end{subfigure}
  \begin{subfigure}[b]{0.5\linewidth}
    \centering
    \includegraphics[width=0.9\linewidth]{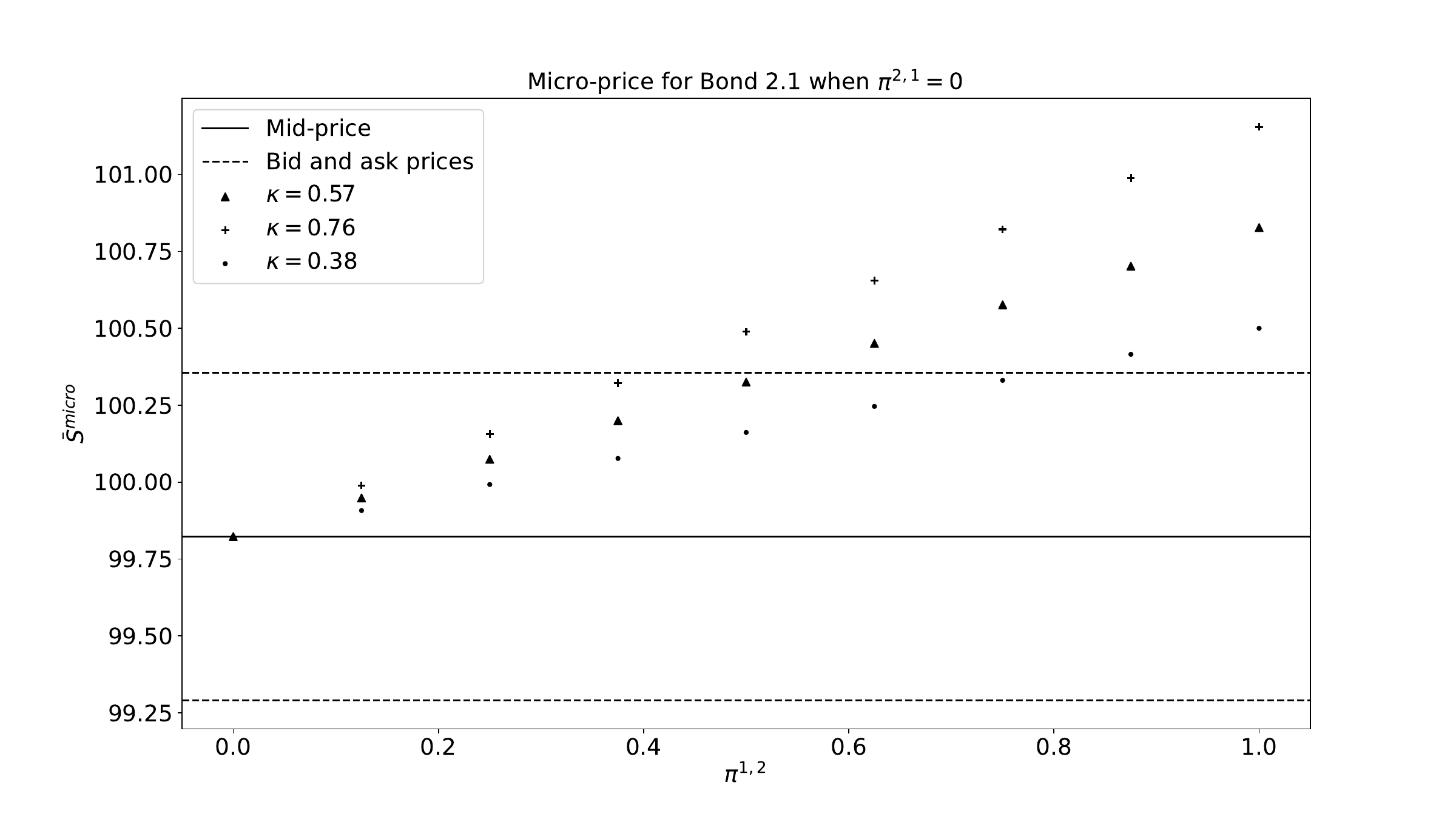} 
    \caption{Bond 2.1} 
    \label{mp_pi10_00:2} 

  \end{subfigure} 
  \begin{subfigure}[b]{0.5\linewidth}
    \centering
    \includegraphics[width=0.9\linewidth]{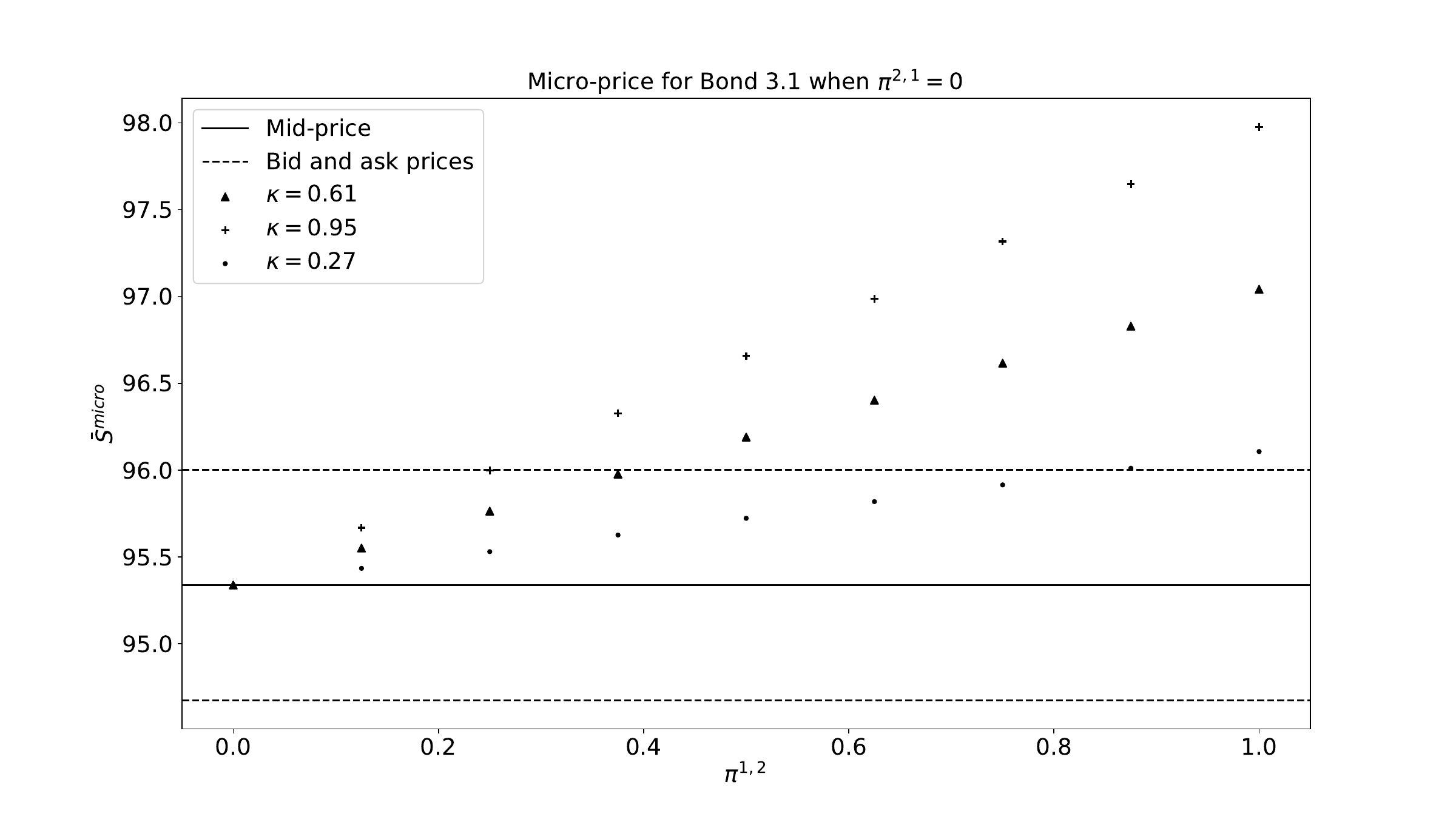} 
    \caption{Bond 3.1} 
    \label{mp_pi10_00:3} 
  \end{subfigure}
  \begin{subfigure}[b]{0.5\linewidth}
    \centering
    \includegraphics[width=0.9\linewidth]{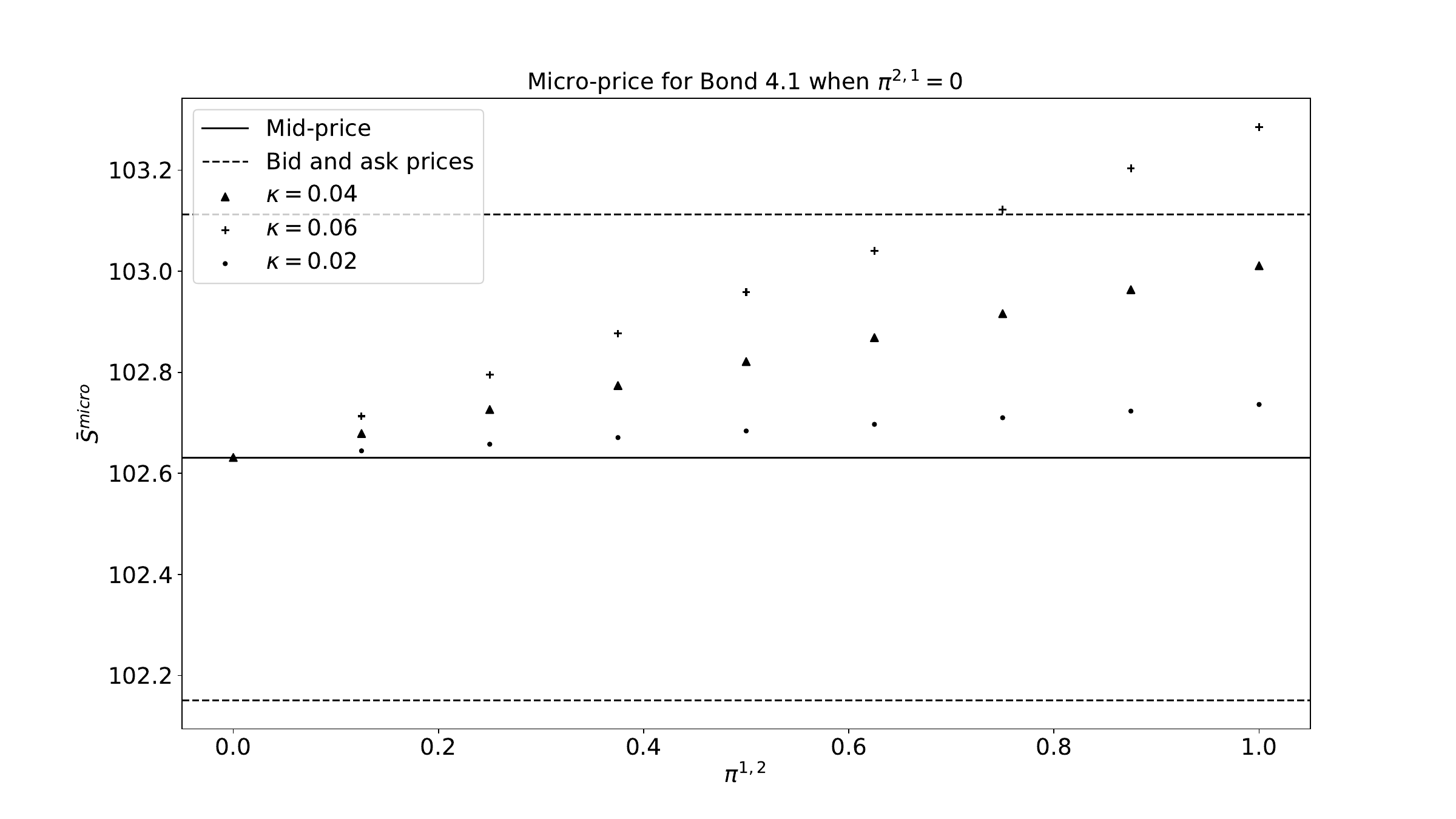} 
    \caption{Bond 4.1} 
    \label{mp_pi10_00:4} 
  \end{subfigure} 
  \caption{Micro-prices for different bonds as a function of $\pi^{1,2}$ when $\pi^{2,1} = 0$.}
  \label{mp_pi10_00}

\end{figure}

\begin{figure}[h!] 
  \begin{subfigure}[b]{0.5\linewidth}
    \centering
    \includegraphics[width=0.9\linewidth]{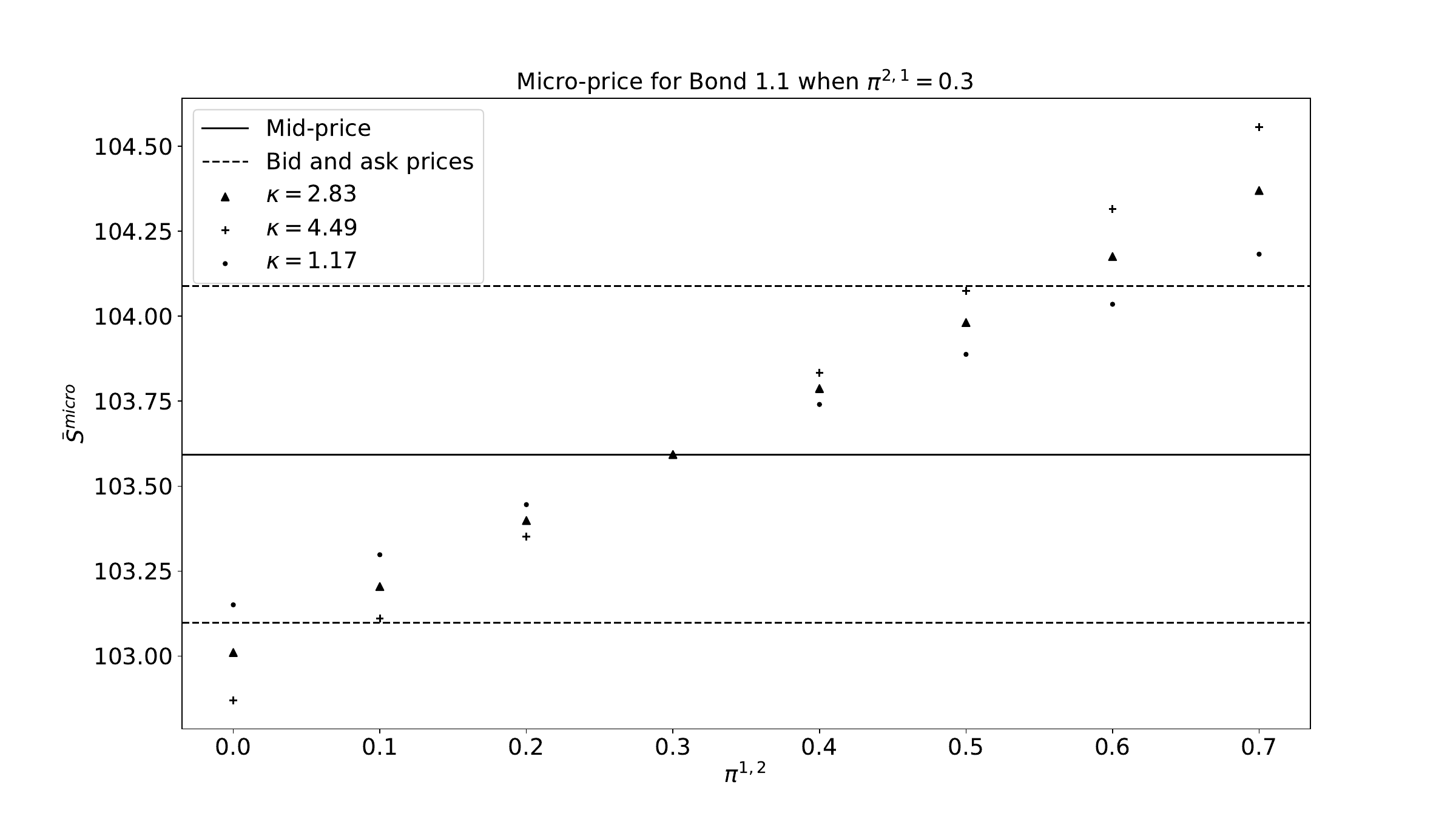} 
    \caption{Bond 1.1} 
    \label{mp_pi10_03:1} 

  \end{subfigure}
  \begin{subfigure}[b]{0.5\linewidth}
    \centering
    \includegraphics[width=0.9\linewidth]{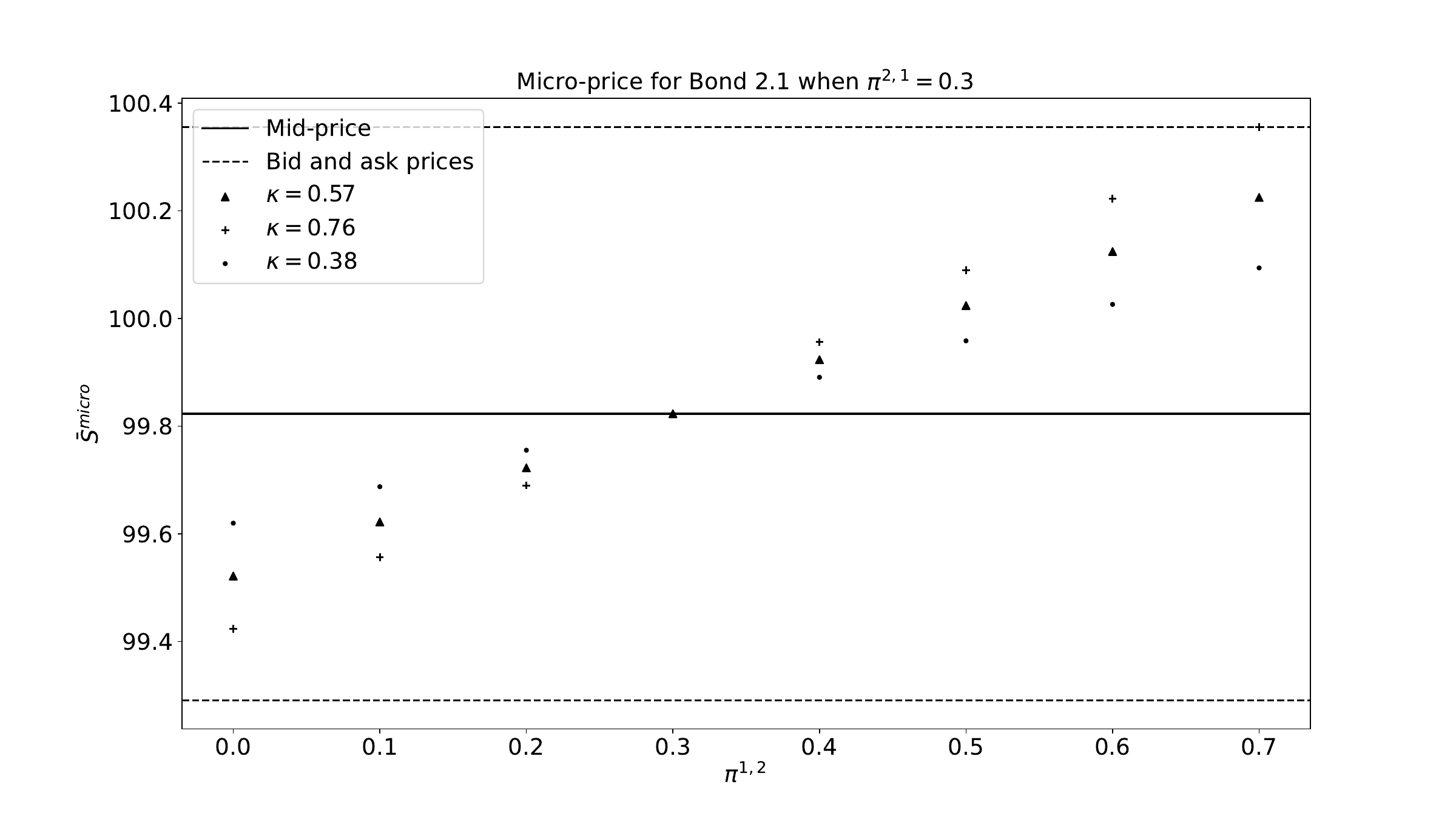} 
    \caption{Bond 2.1} 
    \label{mp_pi10_03:2} 

  \end{subfigure} 
  \begin{subfigure}[b]{0.5\linewidth}
    \centering
    \includegraphics[width=0.9\linewidth]{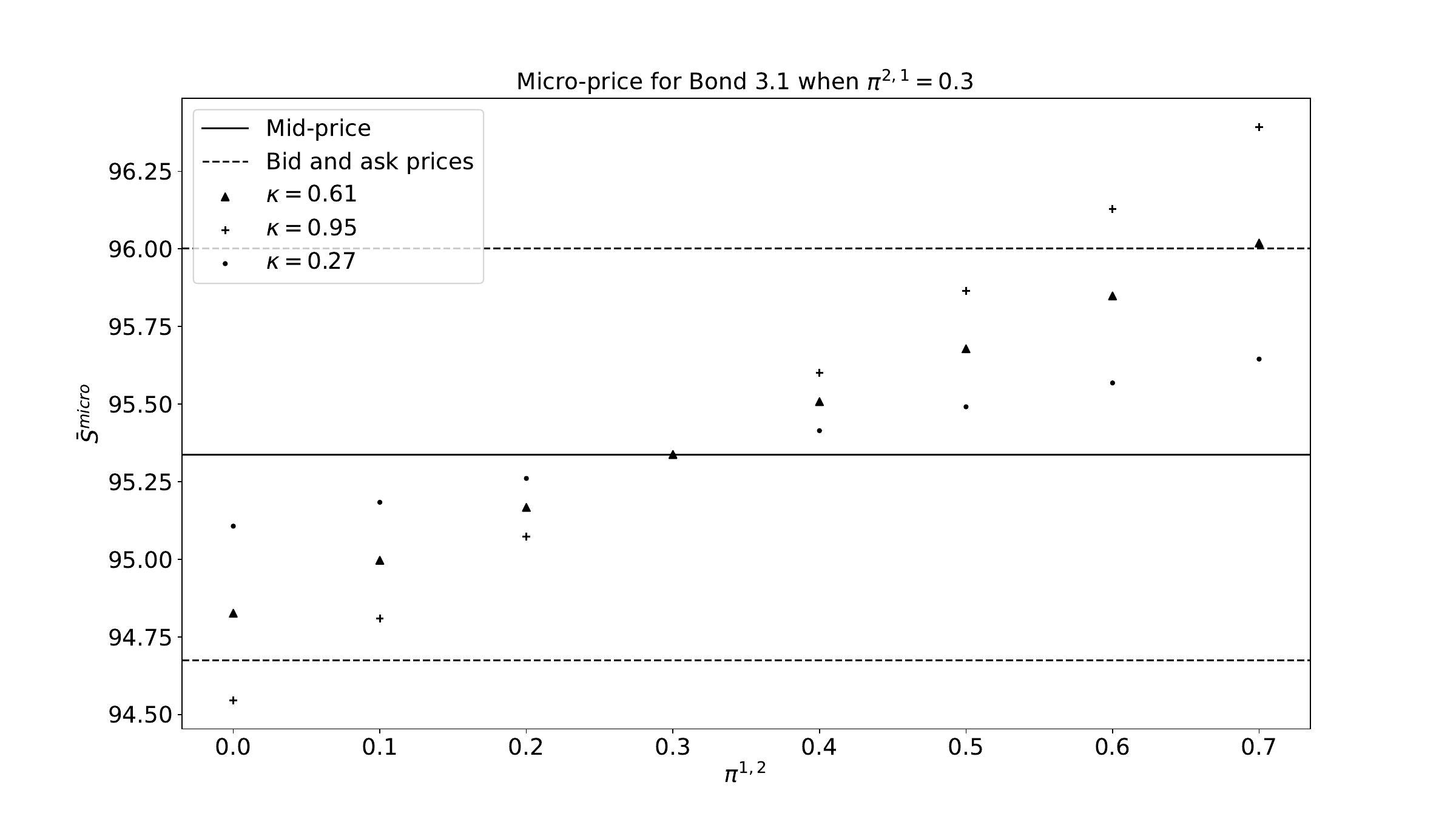} 
    \caption{Bond 3.1} 
    \label{mp_pi10_03:3} 
  \end{subfigure}
  \begin{subfigure}[b]{0.5\linewidth}
    \centering
    \includegraphics[width=0.9\linewidth]{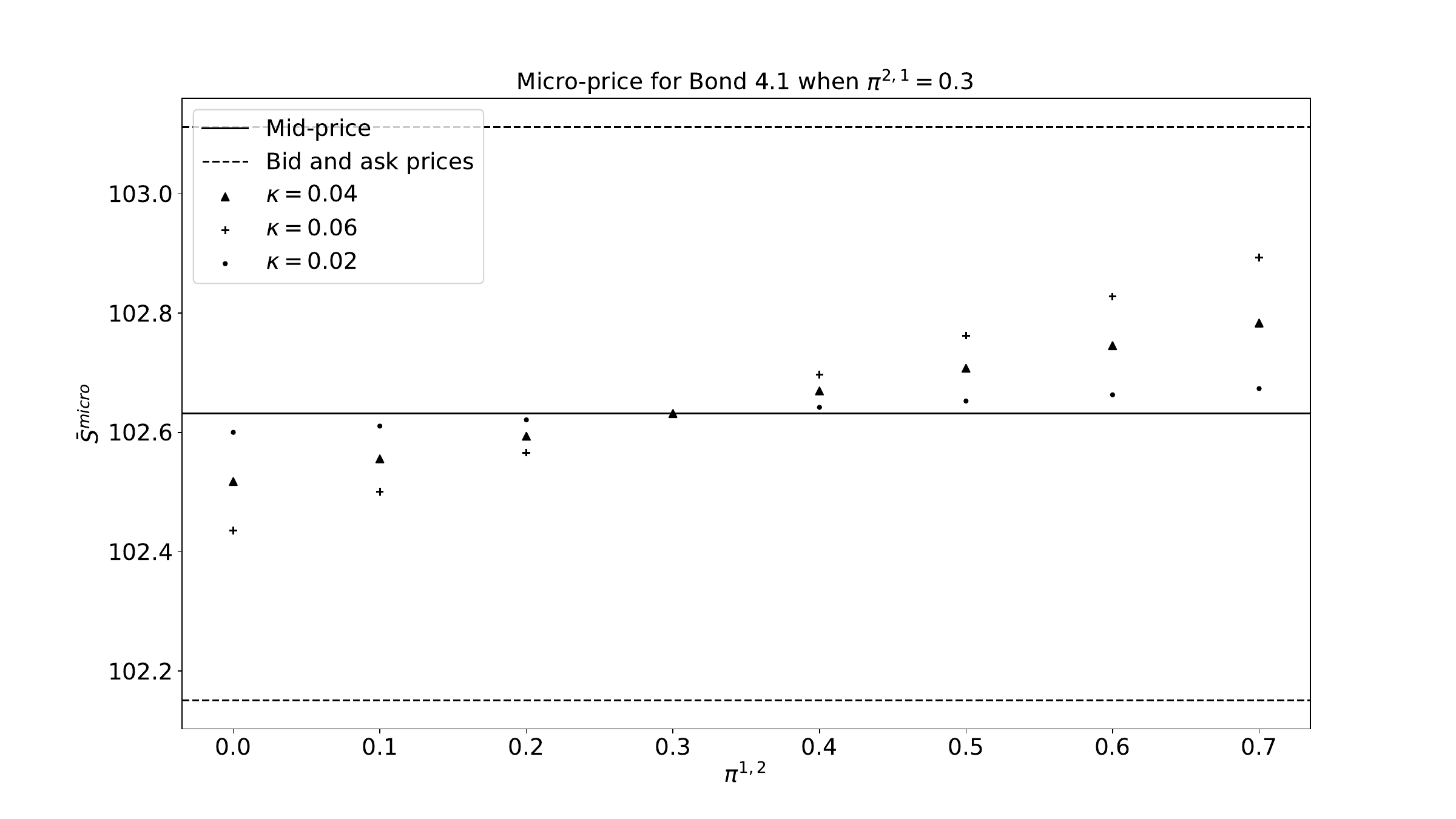} 
    \caption{Bond 4.1} 
    \label{mp_pi10_03:4} 
  \end{subfigure} 
  \caption{Micro-prices for different bonds as a function of $\pi^{1,2}$ when $\pi^{2,1} = 0.3$.}
  \label{mp_pi10_03} 
  \vspace{-5mm}
\end{figure}

In Figure \ref{mp_pi10_00}, we plot the micro-price as a function of \(\pi^{1,2}\) when \(\pi^{2,1} = 0\). Figure \ref{mp_pi10_03} documents similarly the micro-price as a function of \(\pi^{1,2}\) when \(\pi^{2,1} = 0.3\). To study the impact of the uncertainty on the parameter~\(\kappa\), we also plot the micro-prices corresponding to values of \(\kappa\) one standard deviation above and below our estimation. Composite bid-ask spreads are also reported.\\

\begin{figure}[h!]
\centering
\includegraphics[width=\textwidth]{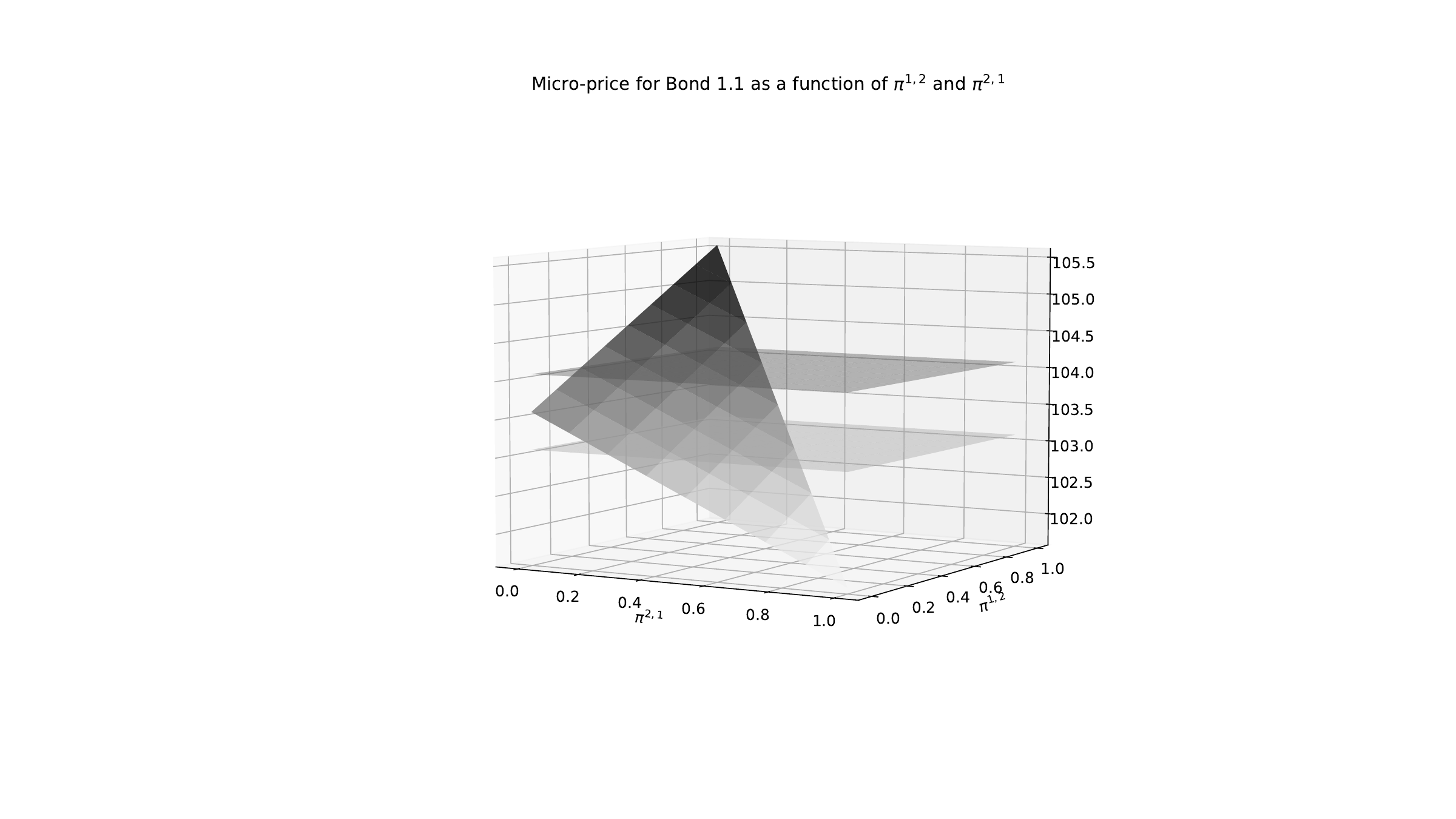}\\
\caption{Micro-prices for Bond 1.1 as a function of $\pi^{1,2}$ and $\pi^{2,1}$.}
\label{mpsurf1}
\end{figure}

\begin{figure}[h!]
\centering
\includegraphics[width=\textwidth]{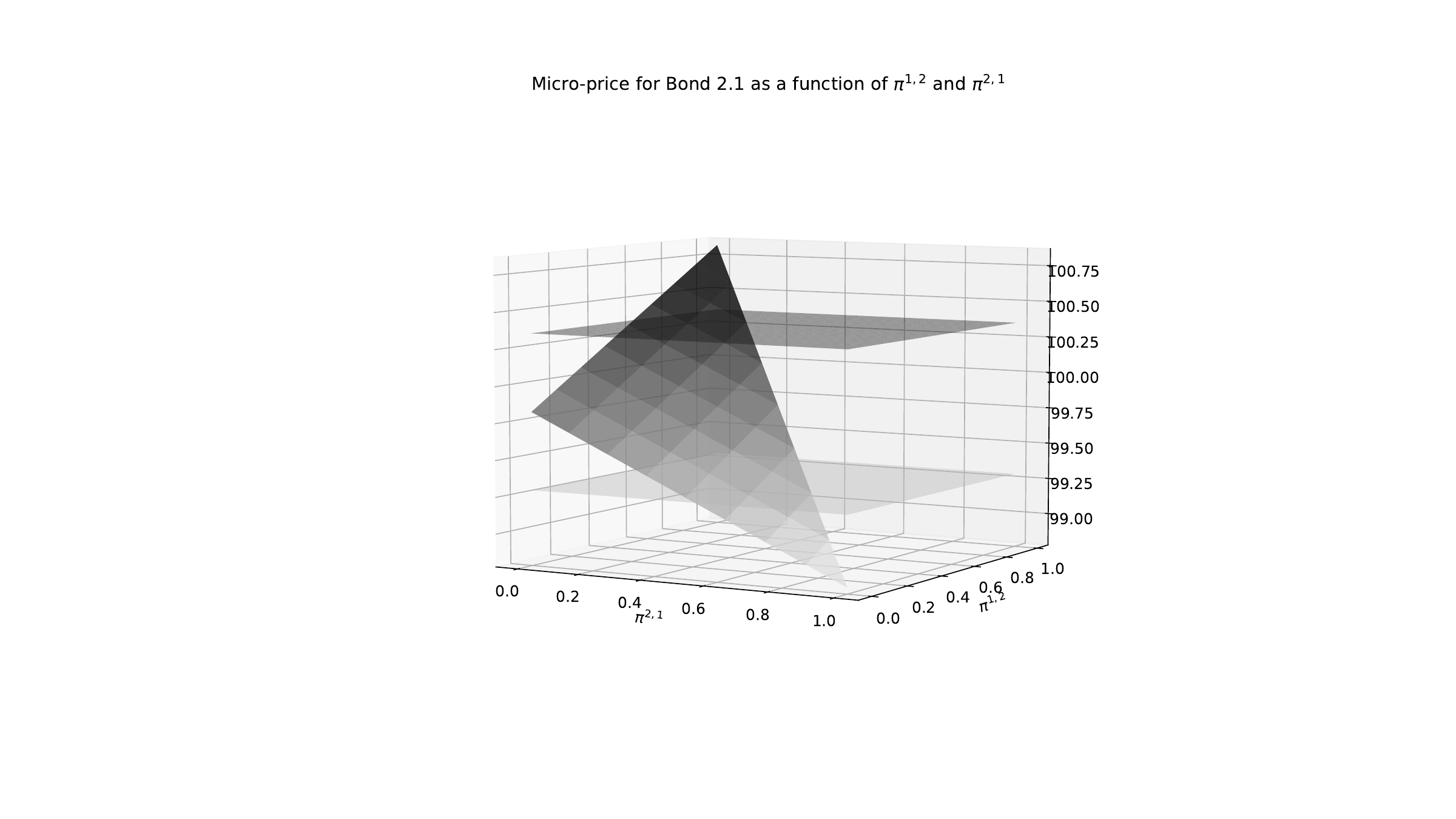}\\
\caption{Micro-prices for Bond 2.1 as a function of $\pi^{1,2}$ and $\pi^{2,1}$.}
\label{mpsurf2}
\end{figure}

\begin{figure}[h!]
\centering
\includegraphics[width=\textwidth]{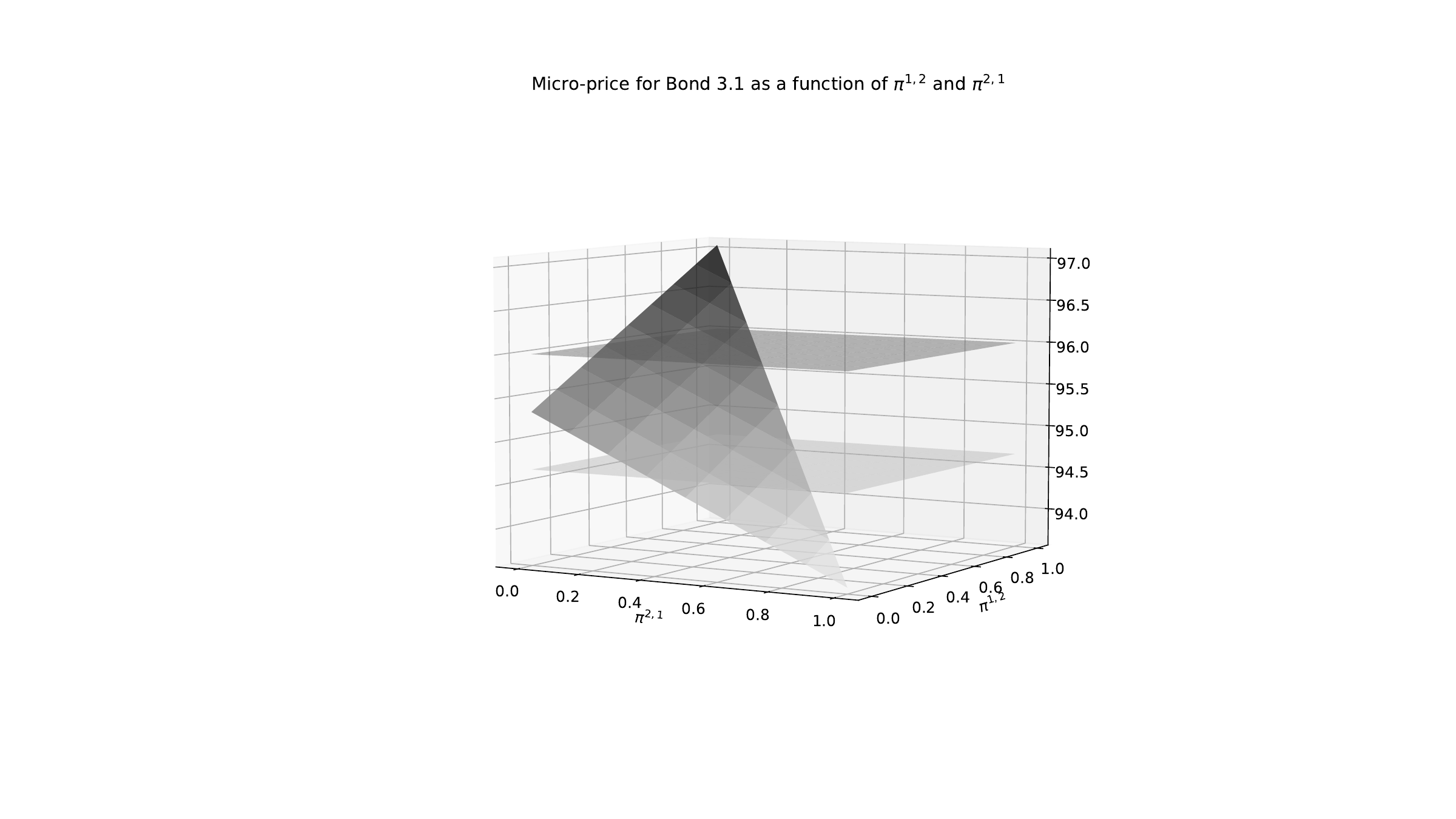}\\
\caption{Micro-prices for Bond 3.1 as a function of $\pi^{1,2}$ and $\pi^{2,1}$.}
\label{mpsurf3}
\end{figure}

\begin{figure}[h!]
\centering
\includegraphics[width=\textwidth]{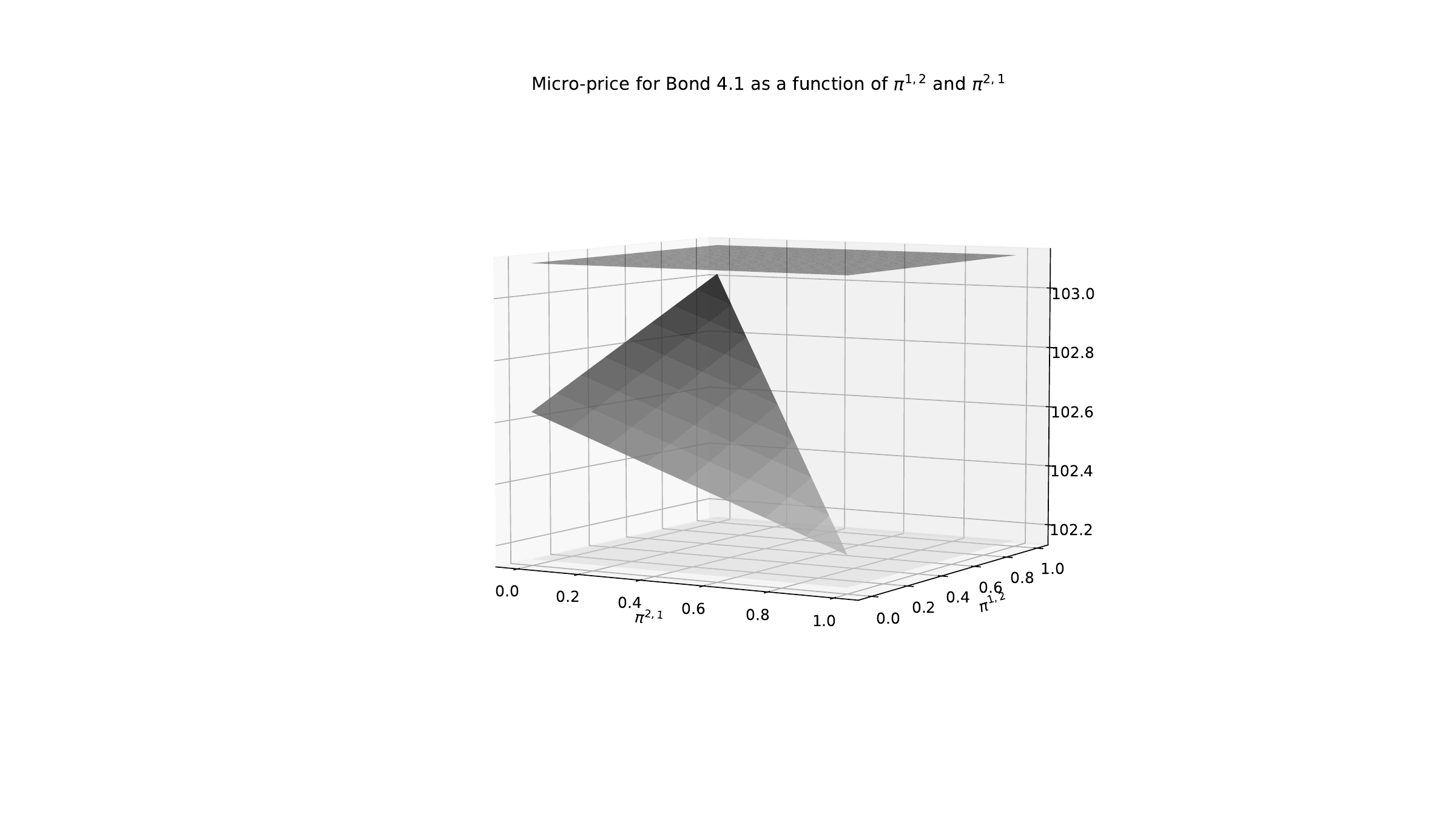}\\

\caption{Micro-prices for Bond 4.1 as a function of $\pi^{1,2}$ and $\pi^{2,1}$.}
\label{mpsurf4}
\end{figure}

Naturally, when \(\pi^{1,2} = \pi^{2,1}\), the micro-price is equal to the mid-price. As expected, we also see that a rise in \(\pi^{1,2}\) leads to an increase in micro-prices. In Figure \ref{mp_pi10_00}, we see that micro-prices are within the composite bid-ask spread for moderate values of \(\pi^{1,2}\) but beyond for large values (except for Bond 4.1). Values beyond the bid-ask spread could be seen as a real trading signal, but it is important to keep in mind that the results obtained for high values of \(\pi^{1,2}\) must be interpreted with caution because the linear regressions have been carried out with only a few high values for \(\pi^{1,2}\) (see the above histograms). We see in Figure \ref{mp_pi10_03} that when \(\pi^{2,1} = 0.3\), most values remain inside the bid-ask spread. In fact, we see in Figures \ref{mpsurf1}, \ref{mpsurf2}, \ref{mpsurf3}, and \ref{mpsurf4} that micro-prices are significantly outside the bid-ask spread for extreme values of \(\pi^{1,2}\) and \(\pi^{2,1}\) only, \textit{i.e.}, when one is really sure that the flow is imbalanced.
\newpage

\newpage

\newpage

\newpage

\subsection{Fair Transfer Price}

Let us now come to the case of Fair Transfer Prices. For that purpose, we need to fit S-curves, choose a risk aversion parameter and solve an HJB equation.

\subsubsection{Estimation of S-curves}

For the functions $f^b$ and $f^a$ defined in Section \ref{MMmodel}, we assumed a logistic form. We noticed no systematic difference between the bid and ask sides. Consequently, we considered
$$f^b(\delta) = f^a(\delta) = \frac{1}{1+e^{\alpha_{\text{logit}} + \beta_{\text{logit}}\frac{\delta}{ \delta^0} }} =:f(\delta),$$
where $\delta^{0}$ is the current composite bid-ask spread of the bond, and the parameters $\alpha_{\text{logit}}$ and $\beta_{\text{logit}}$ are estimated with a logistic regression.\\

\begin{figure}[h!]
\centering
\includegraphics[width=0.96\textwidth]{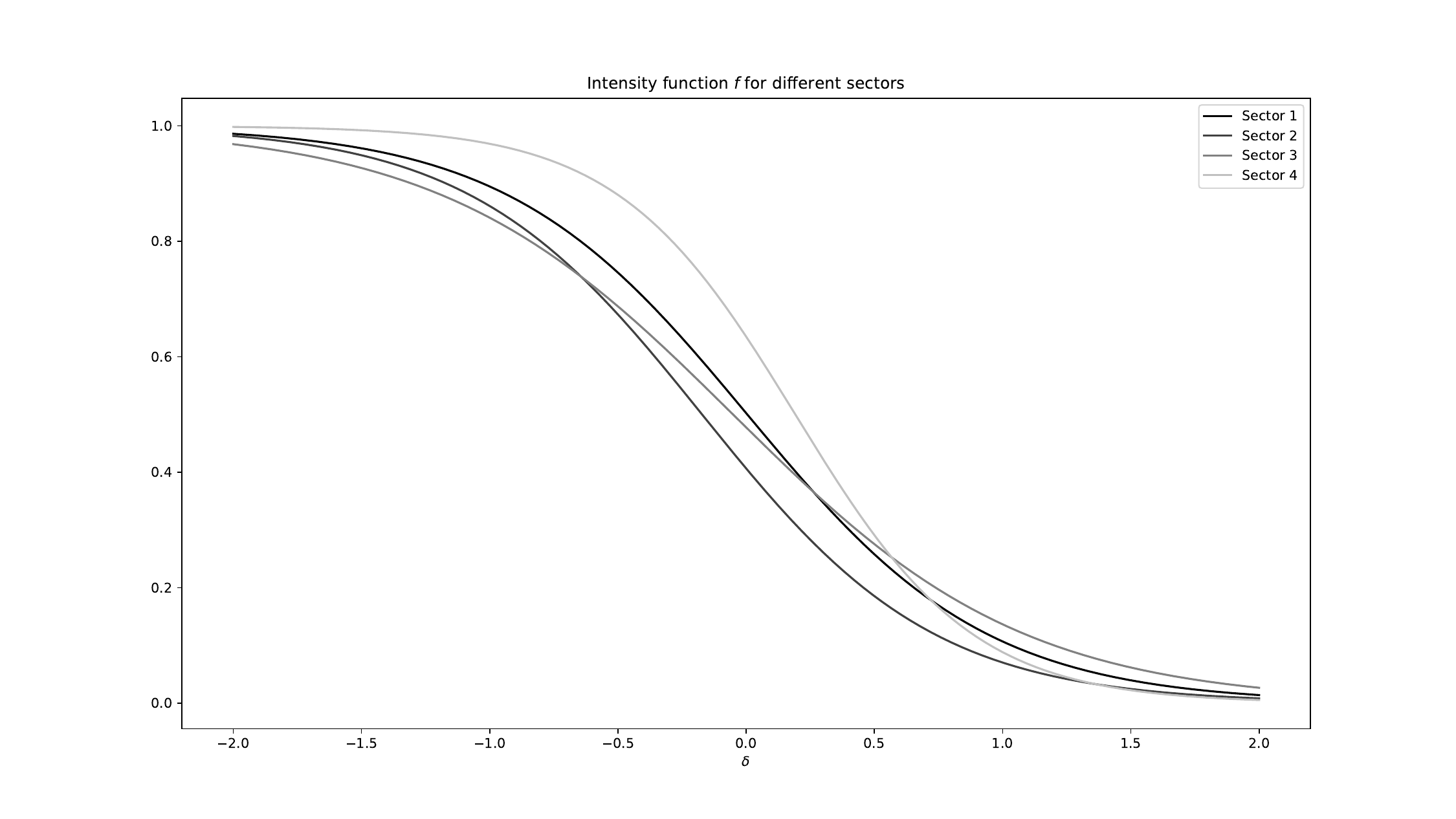}\\
\caption{Intensity function $f$ for the four sectors.}
\label{scurves}
\end{figure}

With this parametrization, the functions appeared to be almost uniform accross sectors (as shown in Figure \ref{scurves}), and we therefore estimated, for the sake of simplicity, a single S-curve using the entire dataset, independently of the sector. We obtained the following (rounded) values: $\alpha_{\text{logit}} = -0.7$ and $\beta_{\text{logit}} = 3.1$.

\subsubsection{Solving HJB equations}

Our concept of FTP relies on the bid and ask quotes of a theoretical market maker who knows the current state of the market. To solve the stochastic optimal control problem of that market maker and obtain the associated quotes, one needs to compute the value functions numerically.\\

Let us recall that, in the model of Section 3, the value functions $(\theta^{j_b, j_a})_{1\le j_b \le m_b, 1\le j_a \le m_a}$ of the market maker satisfy the following system of Hamilton-Jacobi-Bellman (HJB) equations:\footnote{We state the equations in the general case, \textit{i.e.} not in the case of the extension of Appendix \ref{MMPPex}, although our illustrations rely on the exchangeability assumption.}

$$
    \partial_t \theta^{j_b, j_a}(t,q) + \kappa(\lambda^{j_a,a} - \lambda^{j_b,b}  )q - \frac 12 \gamma \sigma^2 q^2 + \sum_{1\le k_b \le m_b, 1\le k_a \le m_a}  Q_{(j_b-1) m_a+j_a, (k_b-1) m_a+k_a} \theta^{k_b, k_a}(t,q)$$
$$+  z \lambda^{j_b,b} H^{b} \left( \frac{\theta^{j_b, j_a}(t,q) - \theta^{j_b, j_a}(t,q+z)}{z} \right) + z \lambda^{j_a,a} H^{a} \left( \frac{\theta^{j_b, j_a}(t,q) - \theta^{j_b, j_a}(t,q-z)}{z} \right) = 0 $$
    with terminal condition $\theta^{j_b, j_a}(T,q) = 0$.\\

We need to compute or approximate numerically the solution of this system of equations in order to compute FTPs. A natural approach is to use a Euler scheme, preferably implicit. In that case, relevant boundary conditions can be chosen by adding risk limits to the inventory of the theoretical market maker, and the equations become
$$\partial_t \theta^{j_b, j_a}(t,q) + \kappa(\lambda^{j_a,a} - \lambda^{j_b,b}  )q - \frac 12 \gamma \sigma^2 q^2 + \sum_{1\le k_b \le m_b, 1\le k_a \le m_a}  Q_{(j_b-1) m_a+j_a, (k_b-1) m_a+k_a} \theta^{k_b, k_a}(t,q)$$
$$+  z \lambda^{j_b,b}\mathds 1_{\{q+z \le \bar{q}\}} H^{b} \left( \frac{\theta^{j_b, j_a}(t,q) - \theta^{j_b, j_a}(t,q+z)}{z} \right) + z \lambda^{j_a,a}\mathds 1_{\{q-z \ge -\bar{q}\}} H^{a} \left( \frac{\theta^{j_b, j_a}(t,q) - \theta^{j_b, j_a}(t,q-z)}{z} \right) = 0$$
with terminal condition $\theta^{j_b, j_a}(T,q) = 0$, where $\bar{q}>0$ corresponds to the risk limit, \textit{i.e.} the market maker refuses any trade that would bring the inventory out of the interval $[-\bar{q}, \bar{q}]$. If $\bar{q}$ is large enough, this has almost no impact on the bid and ask quotes of the market maker at $q=0$ that are used to compute FTPs.\\

Euler schemes can be time-consuming when the number of states \(m_b \times m_a\) is large, or even unfeasible if the one-asset market-making model is replaced by a multi-asset one. Using the same approach as in \cite{bergault_closed-form_2021}, we propose in the following paragraphs a quadratic approximation of the value functions.\\

Let us replace the Hamiltonian functions $H^{b}$ and $H^{a}$ by the quadratic functions
$$\check{H}^{b}: p \mapsto  \alpha^{b}_0 + \alpha^{b}_1 p + \frac 12 \alpha^{b}_2 p^2 \quad \textrm{and} \quad \check{H}^{a}: p \mapsto  \alpha^{a}_0 + \alpha^{a}_1 p + \frac 12 \alpha^{a}_2 p^2.$$

A natural choice for the functions $\check{H}^{b}$ and $\check{H}^{a}$ derives from Taylor expansions around $p=0$. In that case, we have
$$\forall i \in \{0,1,2\},\quad  \alpha^{b}_i = {H^{b}}^{(i)}(0) \quad \textrm{and} \quad \alpha^{a}_i = {H^{a}}^{(i)}(0).$$

For $(j_b, j_a)\in \{1, \ldots, m_b\}\times\{1, \ldots, m_a\}$, we denote by $\check \theta^{j_b, j_a}$ the approximation of $\theta^{j_b,j_a}$ associated with the functions $\check{H}^{b}$ and $\check{H}^{a}$. The functions $(\check \theta^{j_b, j_a})_{1\le j_b \le m_b, 1\le j_a \le m_a}$ verify
\begin{eqnarray*}
0 &=& \partial_t \check{\theta}^{j_b, j_a}(t,q) + \kappa(\lambda^{j_a,a} - \lambda^{j_b,b}  )q - \frac{1}{2} \gamma \sigma^2 q^2\\
&&+     \sum_{1\le k_b \le m_b, 1\le k_a \le m_a}  Q_{(j_b-1) m_a+j_a, (k_b-1) m_a+k_a} \check{\theta}^{k_b, k_a}(t,q) + z\left( \lambda^{j_b,b}\alpha^{b}_0 +  \lambda^{j_a,a}\alpha^{a}_0 \right) \\
  &&+  \left( \lambda^{j_b,b}\alpha^{b}_1 \left(\check{\theta}^{j_b, j_a}(t,q) - \check{\theta}^{j_b, j_a}(t, q + z)\right) + \lambda^{j_a,a}\alpha^{a}_1 \left(\check{\theta}^{j_b, j_a}(t,q) - \check{\theta}^{j_b, j_a}(t, q - z)\right) \right) \\
  &&+ \frac{1}{2z}\left( \lambda^{j_b,b}\alpha^{b}_2 \left(\check{\theta}^{j_b, j_a}(t,q) - \check{\theta}^{j_b, j_a}(t,q + z)\right)^2
                                + \lambda^{j_a,a}\alpha^{a}_2 \left(\check{\theta}^{j_b, j_a}(t,q) - \check{\theta}^{j_b, j_a}(t,q - z)\right)^2 \right),
\end{eqnarray*}
and of course we consider the terminal condition $\check{\theta}^{j_b, j_a}(T,q) = 0$.\\

To write the approximations of the value functions in a simple way, let us introduce for $i \in \{0, 1, 2\}$ and~$k \in \mathbb{N}$
$$\Delta_{i,k}^{b} = \alpha^{b}_i z^k \quad \text{and} \quad \Delta_{i,k}^{a} = \alpha^{a}_i z^k.\vspace{-0.1cm}$$

For $(j_b, j_a)\in \{1, \ldots, m_b\}\times\{1, \ldots, m_a\}$, let us consider three differentiable functions $A_{j_b, j_a}: [0,T] \to \mathbb R$, $B_{j_b, j_a}:
[0,T] \to \mathbb{R}$, and $C_{j_b, j_a}: [0,T] \to \mathbb R$ solutions of the system of
ordinary differential equations\vspace{-0.1cm}
\begin{equation*}
\begin{cases}
\displaystyle
{A_{j_b, j_a}'}(t) = & 2 \left(\lambda^{j_b,b}\Delta^b_{2,1} + \lambda^{j_a,a}\Delta^a_{2,1}\right) A_{j_b, j_a}(t)^2 - \frac{1}{2} \gamma \sigma^2\\& - \sum_{1\le k_b \le m_b, 1\le k_a \le m_a}  Q_{(j_b-1) m_a+j_a, (k_b-1) m_a+k_a}A_{k_b, k_a}(t) \\
{B_{j_b, j_a}'}(t) = &  2  \left(\lambda^{j_b,b}\Delta^b_{1,1} - \lambda^{j_a,a}\Delta^a_{1,1}\right)A_{j_b, j_a}(t) + 2\left(\lambda^{j_b,b}\Delta^b_{2,2} - \lambda^{j_a,a}\Delta^a_{2,2}\right) A_{j_b, j_a}(t)^2 \\& + \kappa(\lambda^{j_a,a} - \lambda^{j_b,b}) + 2 \left(\lambda^{j_b,b}\Delta^b_{2,1} + \lambda^{j_a,a}\Delta^a_{2,1}\right) A_{j_b, j_a}(t)B_{j_b, j_a}(t)\\
& - \sum_{1\le k_b \le m_b, 1\le k_a \le m_a}  Q_{(j_b-1) m_a+j_a, (k_b-1) m_a+k_a}B_{k_b, k_a}(t) \\
{C_{j_b, j_a}'}(t) = & \left(\lambda^{j_b,b}\Delta^b_{0,1} + \lambda^{j_a,a}\Delta^a_{0,1}\right) +\left(\lambda^{j_b,b}\Delta^b_{1,2} + \lambda^{j_a,a}\Delta^a_{1,2}\right) A_{j_b, j_a}(t)\\
&+ \left(\lambda^{j_b,b}\Delta^b_{1,1} - \lambda^{j_a,a}\Delta^a_{1,1}\right) B_{j_b, j_a}(t) + \frac 12 \left(\lambda^{j_b,b}\Delta^b_{2,3} + \lambda^{j_a,a}\Delta^a_{2,3}\right) A_{j_b, j_a}(t)^2\\
& + \frac 12  \left(\lambda^{j_b,b}\Delta^b_{2,1} + \lambda^{j_a,a}\Delta^a_{2,1}\right) B_{j_b, j_a}(t)^2 + \left(\lambda^{j_b,b}\Delta^b_{2,2} - \lambda^{j_a,a}\Delta^a_{2,2}\right) A_{j_b, j_a}(t)B_{j_b, j_a}(t)\\
& - \sum_{1\le k_b \le m_b, 1\le k_a \le m_a}  Q_{(j_b-1) m_a+j_a, (k_b-1) m_a+k_a}C_{k_b, k_a}(t),
\end{cases}
\end{equation*}
with terminal conditions $A_{j_b, j_a}(T) = 0$, $B_{j_b, j_a}(T) = 0$ and $C_{j_b, j_a}(T) = 0$.\\

Then, for all $(j_b, j_a)\in \{1, \ldots, m_b\}\times\{1, \ldots, m_a\}$, we have:\vspace{-0.1cm}
$$\check{\theta}^{j_b, j_a}(t,q) = -q^2  A_{j_b, j_a}(t) - q B_{j_b, j_a}(t) -  C_{j_b, j_a}(t).\vspace{-0.1cm}$$

Moreover, asymptotic results on value functions continue to hold on their approximations.\\

This kind of approximations has been used in \cite{barzykin2021market, barzykin2022dealing} with great success in terms of risk management. We investigate the quality of the approximation in terms of FTPs below.

\subsubsection{FTP in practice}

To compute FTPs as proposed in Section~\ref{MMmodel}, we still have to choose the risk aversion of the theoretical market maker. A natural way to choose $\gamma$ is to calibrate it to composite bid and ask prices, \textit{i.e.} assuming that the quotes of the theoretical market maker correspond to the market composite bid and ask prices when inventory is equal to $0$. \\

The optimal strategy of the theoretical market maker is obtained by solving numerically the HJB equation, using two different methods: (a) an implicit Euler scheme, and (b) the quadratic approximation technique. Depending on the numerical method we use, $\gamma$ calibrated to composite bid and ask prices takes different values.\footnote{The values of $\gamma$ vary across bonds. This comes in part from our choice of a simple market making model to illustrate our concepts.} However, in terms of FTP, the results obtained with the two numerical methods are almost identical, as shown in Table~\ref{ftp} (FTP (a) corresponds to the Euler scheme and FTP (b) to the quadratic approximation).\\

\begin{figure}[h!] 
  \begin{subfigure}[b]{0.5\linewidth}
    \centering
    \includegraphics[width=\linewidth]{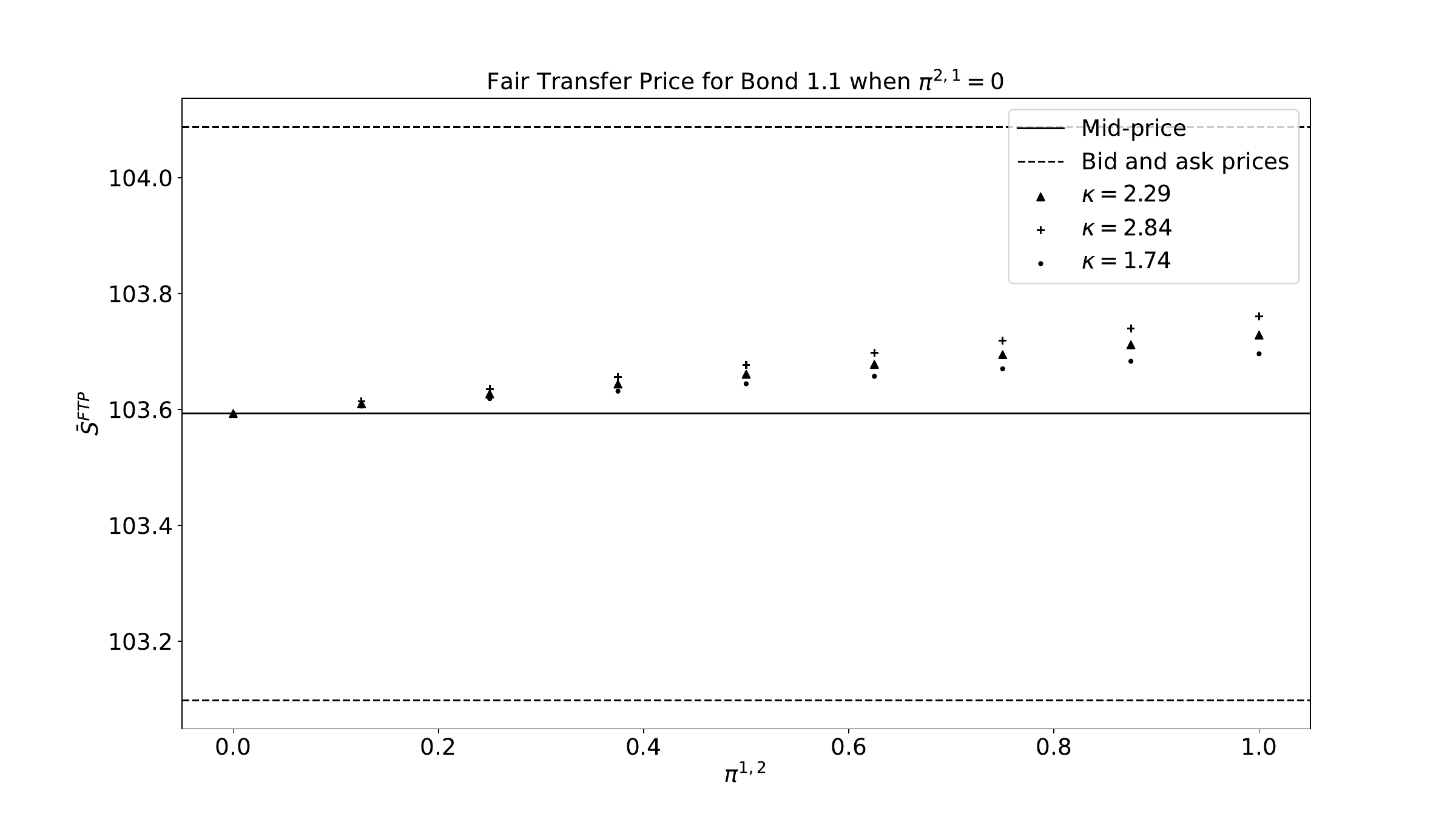} 
    \caption{Bond 1.1} 
    \label{ftp_pi10_00:1} 
    \vspace{4ex}
  \end{subfigure}
  \begin{subfigure}[b]{0.5\linewidth}
    \centering
    \includegraphics[width=\linewidth]{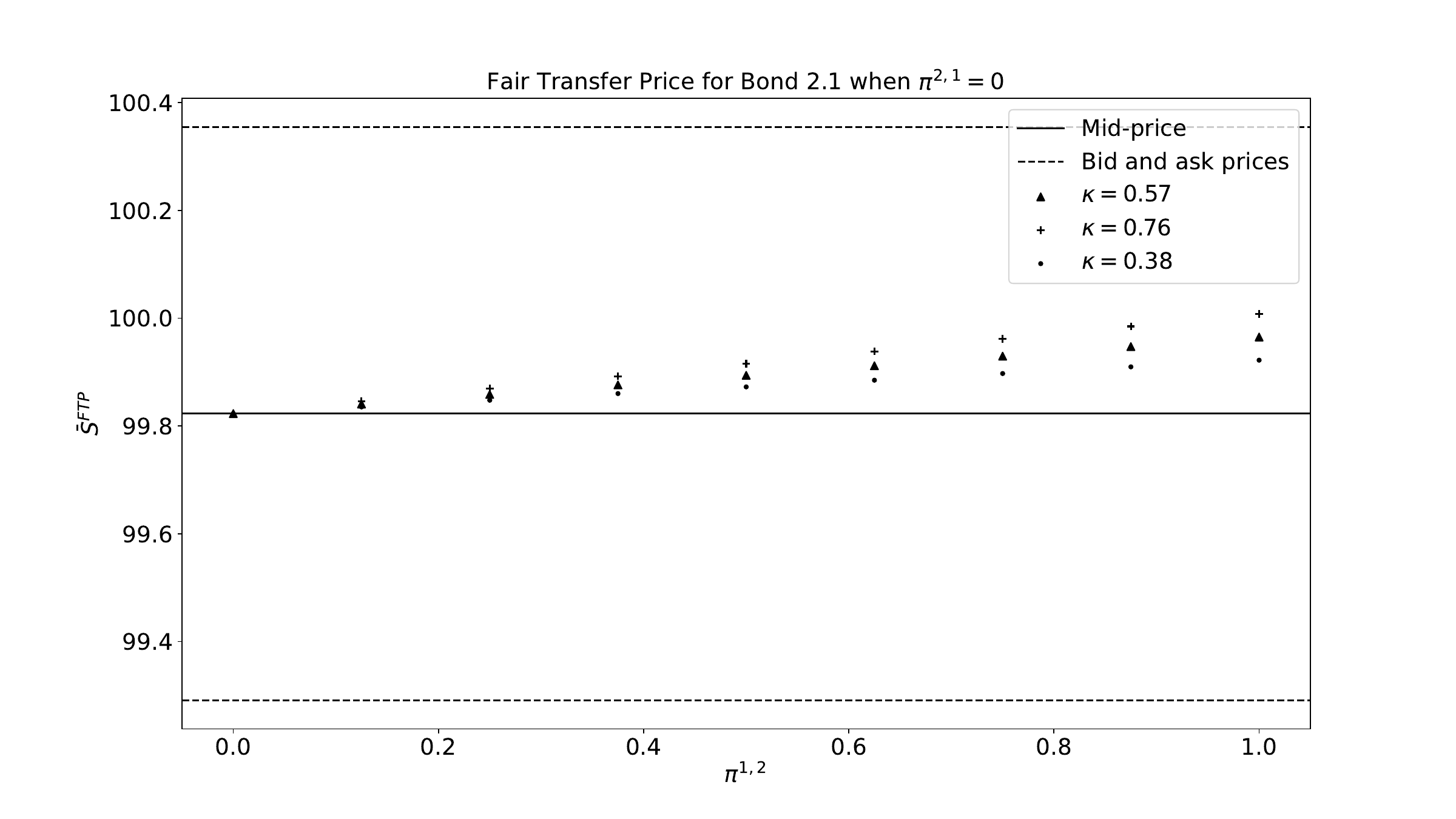} 
    \caption{Bond 2.1} 
    \label{ftp_pi10_00:2} 
    \vspace{4ex}
  \end{subfigure} 
  \begin{subfigure}[b]{0.5\linewidth}
    \centering
    \includegraphics[width=\linewidth]{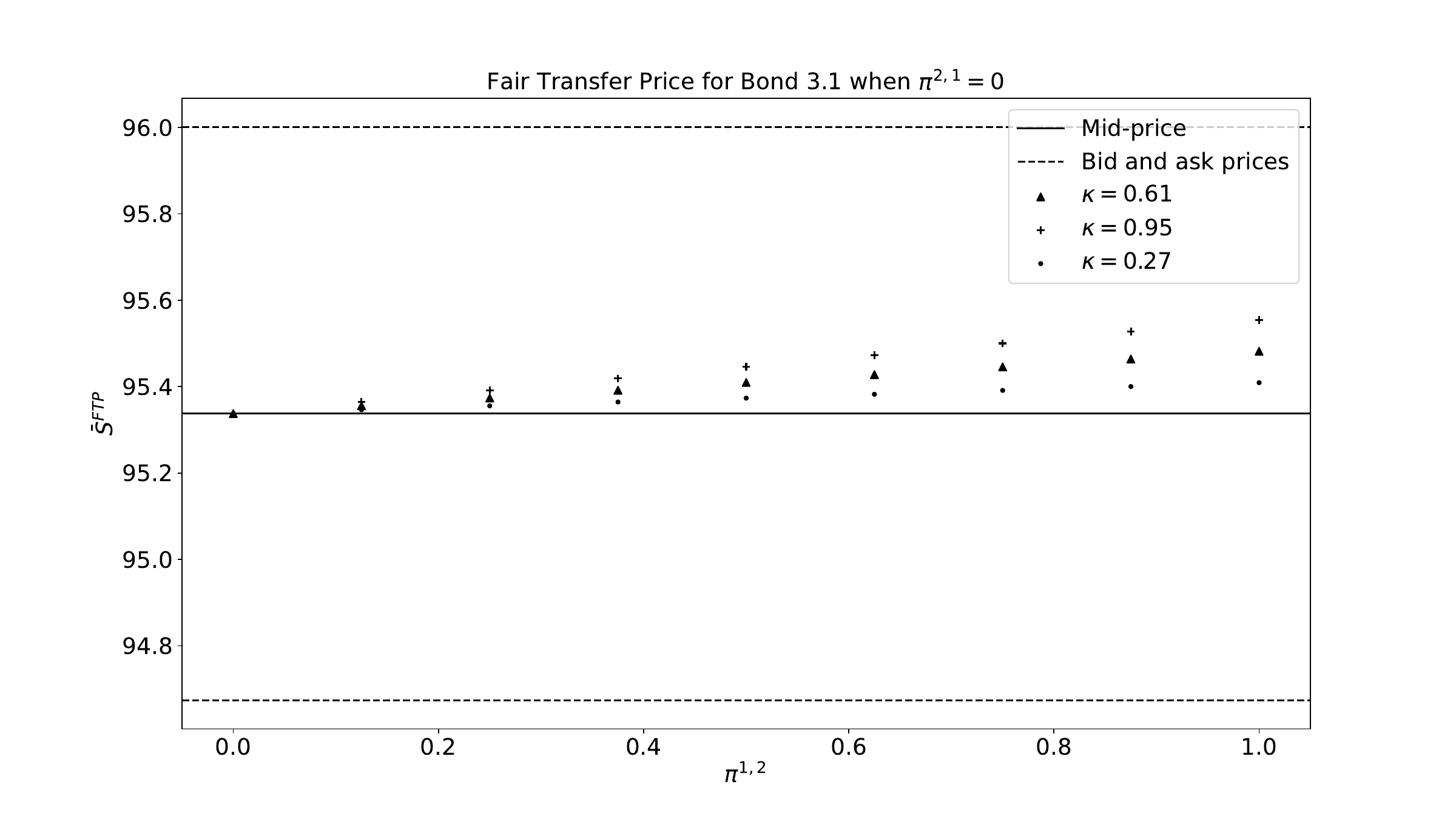} 
    \caption{Bond 3.1} 
    \label{ftp_pi10_00:3} 
  \end{subfigure}
  \begin{subfigure}[b]{0.5\linewidth}
    \centering
    \includegraphics[width=\linewidth]{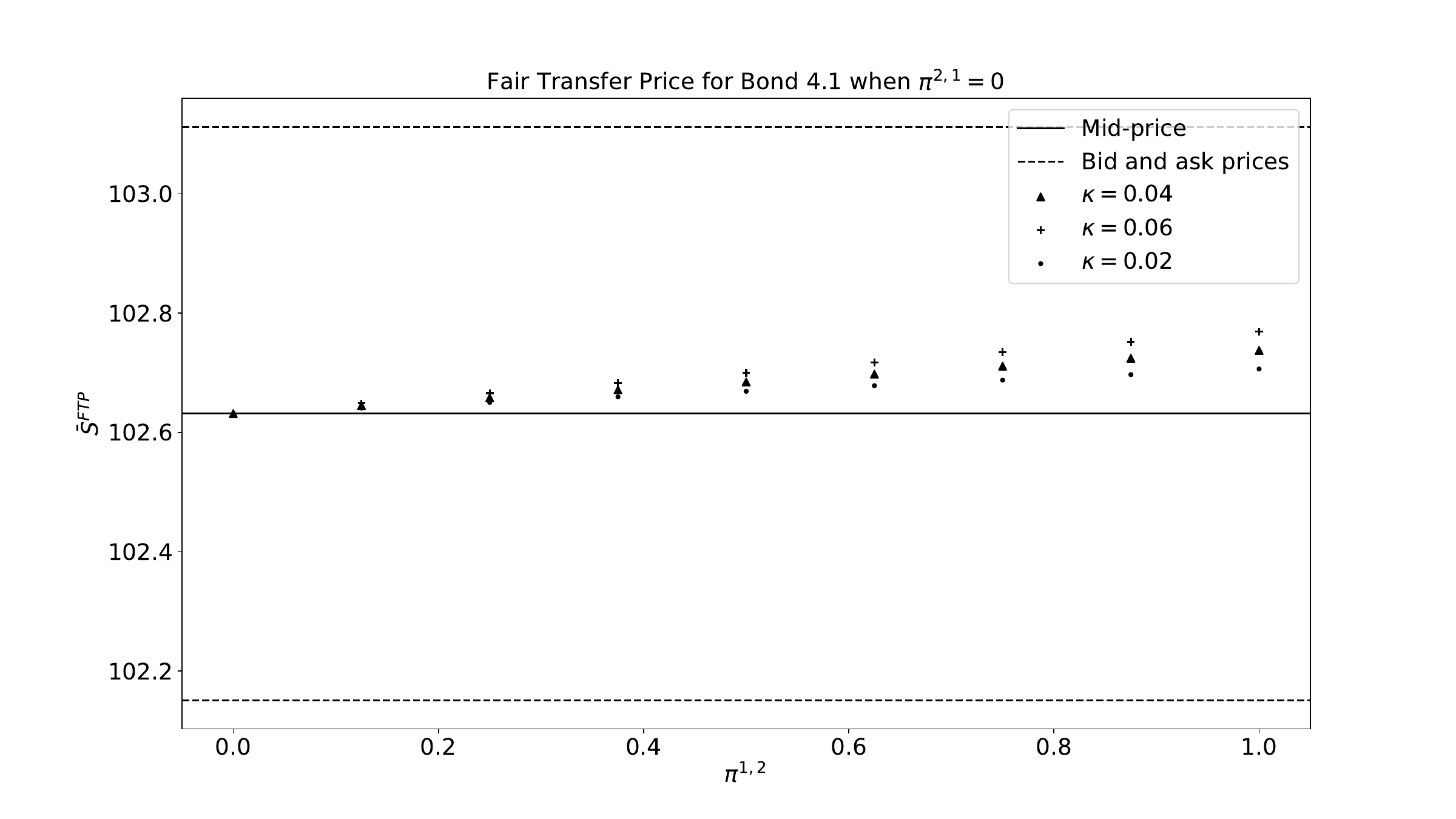} 
    \caption{Bond 4.1} 
    \label{ftp_pi10_00:4} 
  \end{subfigure} 
  \caption{FTP for different bonds as a function of $\pi^{1,2}$ when $\pi^{2,1} = 0$.}
  \label{ftp_pi10_00} 

\end{figure}

\begin{figure}[h!] 
  \begin{subfigure}[b]{0.5\linewidth}
    \centering
    \includegraphics[width=0.9\linewidth]{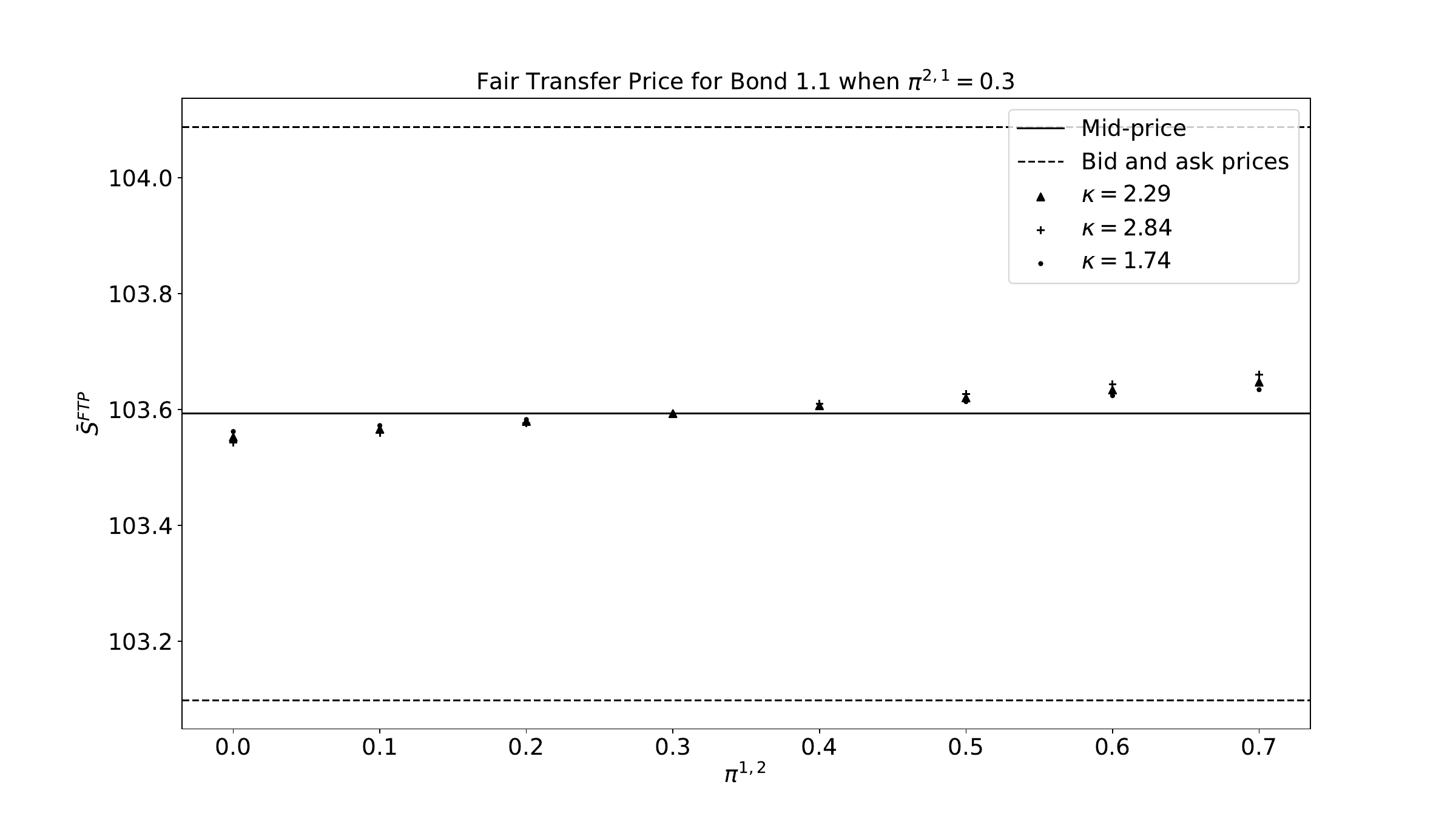} 
    \caption{Bond 1.1} 
    \label{ftp_pi10_03:1} 
    \vspace{4ex}
  \end{subfigure}
  \begin{subfigure}[b]{0.5\linewidth}
    \centering
    \includegraphics[width=0.9\linewidth]{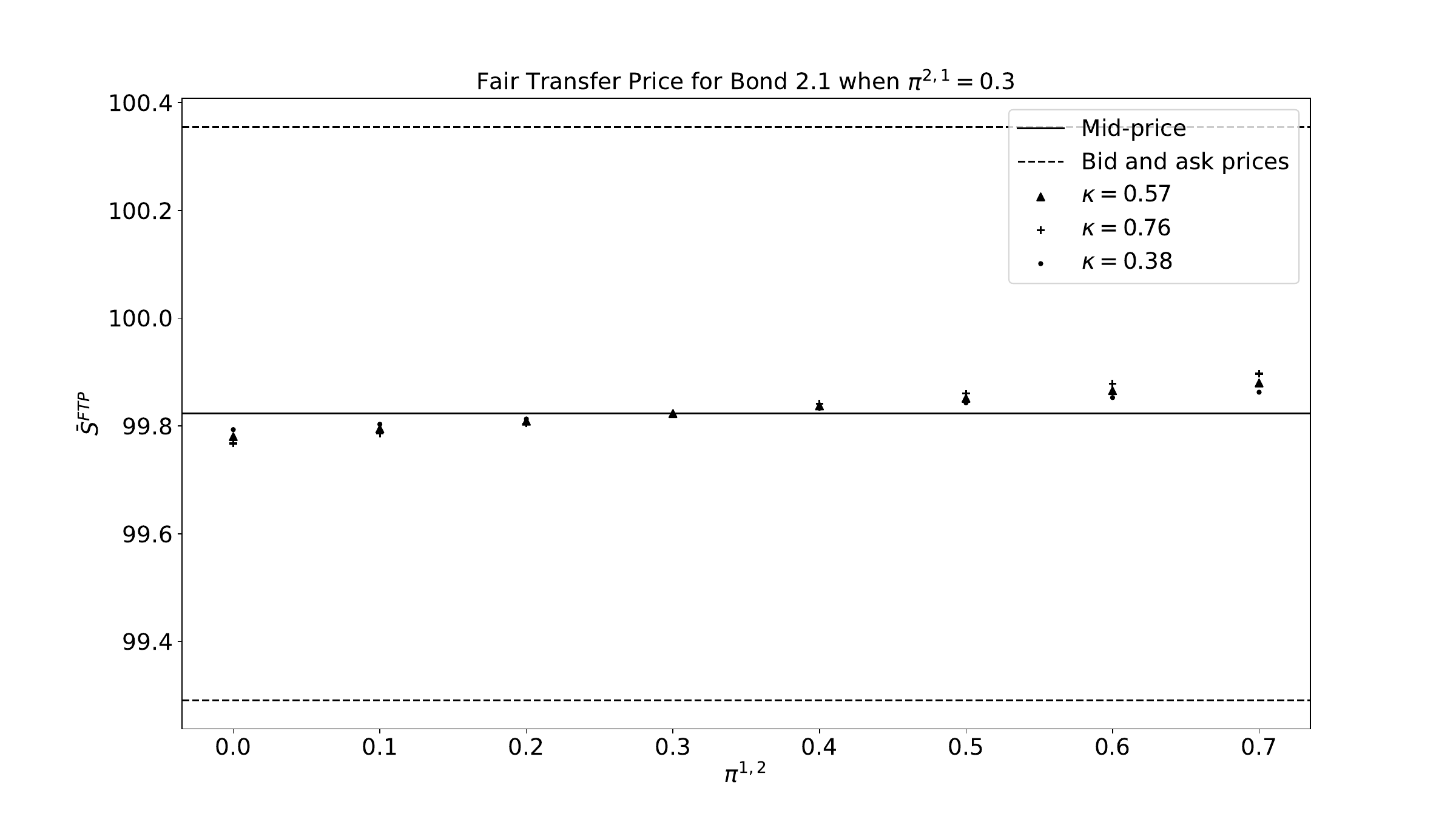} 
    \caption{Bond 2.1} 
    \label{ftp_pi10_03:2} 
    \vspace{4ex}
  \end{subfigure} 
  \begin{subfigure}[b]{0.5\linewidth}
    \centering
    \includegraphics[width=0.9\linewidth]{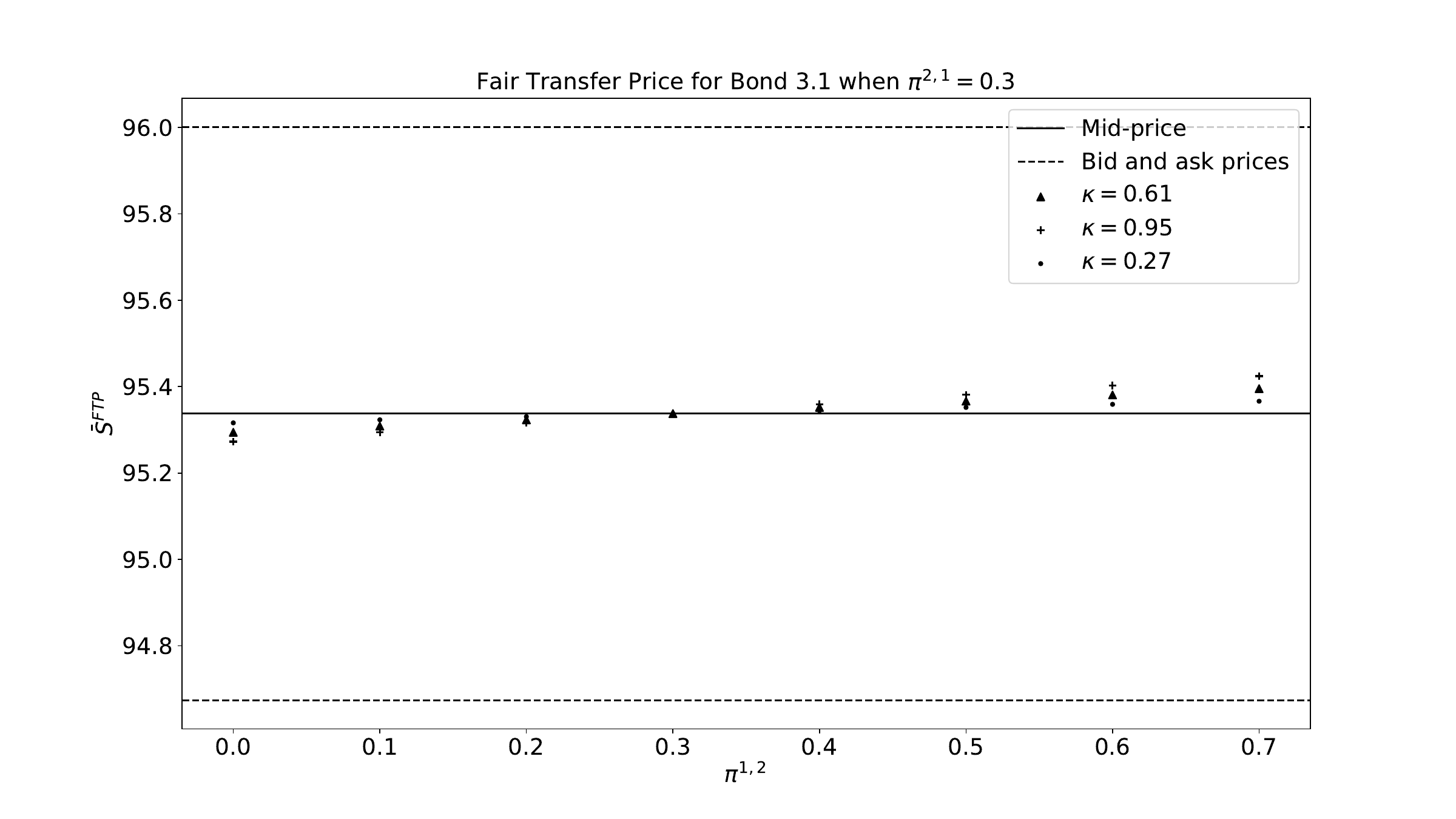} 
    \caption{Bond 3.1} 
    \label{ftp_pi10_03:3} 
  \end{subfigure}
  \begin{subfigure}[b]{0.5\linewidth}
    \centering
    \includegraphics[width=0.9\linewidth]{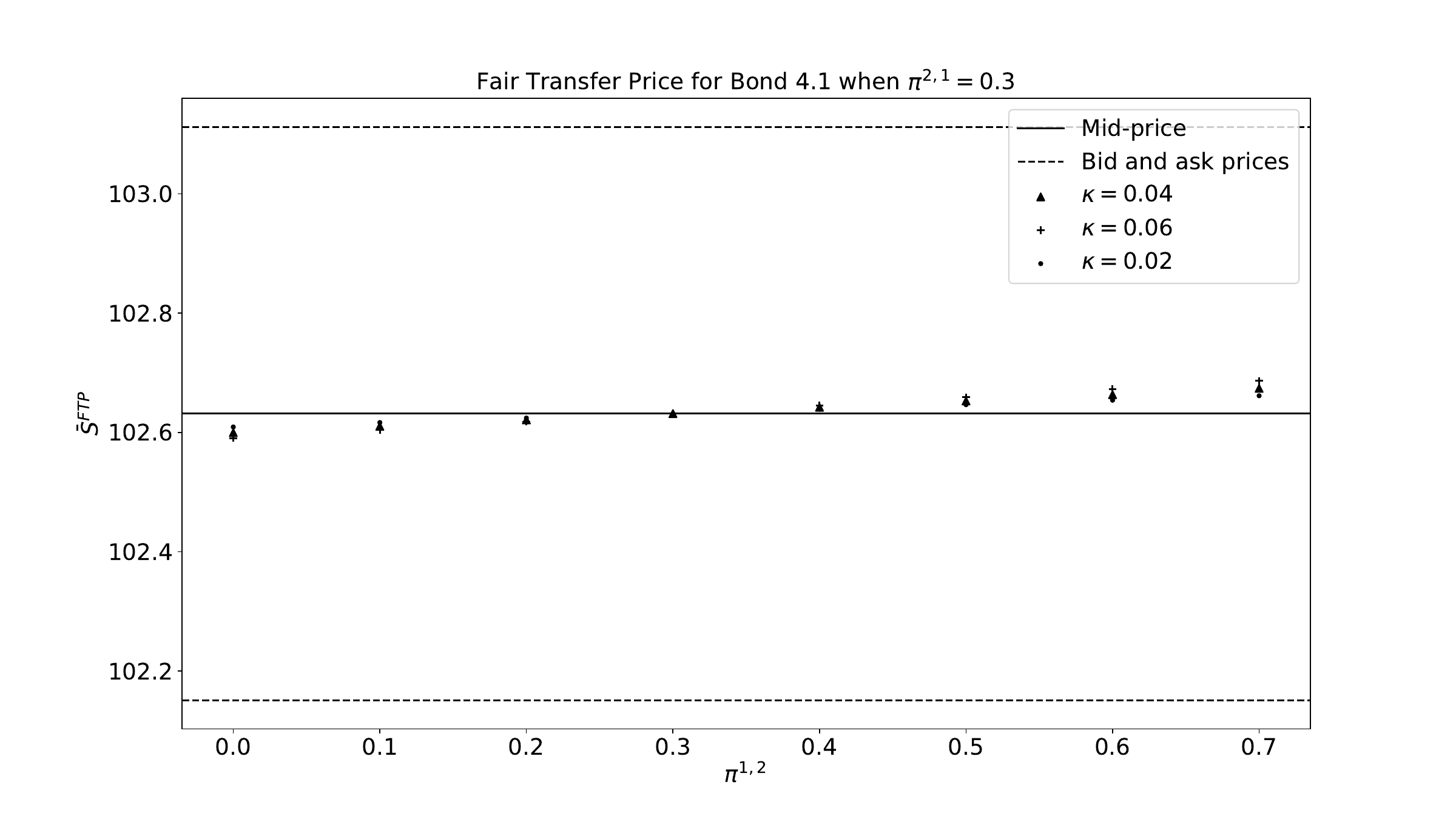} 
    \caption{Bond 4.1} 
    \label{ftp_pi10_03:4} 
  \end{subfigure} 
  \caption{FTP for different bonds as a function of $\pi^{1,2}$ when $\pi^{2,1} = 0.3$.}
  \label{ftp_pi10_30} 

\end{figure}

As with micro-prices, one can never be certain in practice to be in any given state, and the FTP has to be computed as an expectation over the different possible states, depending on the current estimate. Therefore the FTPs exhibited in Table~\ref{ftp} correspond to bounds for the FTPs that would be used in practice. Notice that the adjustments given by FTPs are of lower magnitude than those suggested by micro-prices. As with micro-prices, we study how FTPs evolve depending on the probabilities. Figure \ref{ftp_pi10_00} documents FTPs as a function of $\pi^{1,2}$, when $\pi^{2,1} = 0$ while Figure \ref{ftp_pi10_30} documents FTPs as a function of $\pi^{1,2}$, when $\pi^{2,1} = 0.3$. We see that adjustments are always small. This is linked to the fact that, even when the market is imbalanced, market makers can slightly skew their quotes to deter risk-increasing trades and transform requests into trades when trades would result in a less risky position (less inventory in absolute value in our case). This strongly relies on our implicit assumption that S-curves are the same independently of the liquidity regime. However, we found no empirical evidence of the influence of intensities on fill rates.\\

\setlength{\tabcolsep}{3pt}
\begin{table}[h!]
\begin{center}
\begin{tabular}{c | c c c c c c c c } 
 \hline
Bond &  $\gamma$ (a) &  $\gamma$ (b)  & Bid price & Ask price & $\pi^{\!2,1}\!\!=\!\!1\!\!\!$ : FTP (a)  & FTP (b) & $\pi^{\!1,2}\!\!=\!\!1\!\!\!$ : FTP (a)  & FTP (b)\\ [0.5ex] 
 \hline
1.1 & $4.5\cdot 10^{-9}$ & $5.1\cdot 10^{-9}$ & 103.098 &  104.088 & 103.458 & 103.458 & 103.728 & 103.729 \\ [0.4ex]  
1.2 & $8.9\cdot 10^{-9}$ & $9.1\cdot 10^{-9}$ & 96.514 &  97.600 & 97.092 & 97.092 & 97.122 & 97.122 \\ [0.4ex]  
1.3 & $4.4\cdot 10^{-8}$ & $5.2\cdot 10^{-8}$ & 98.631 &  99.661 & 99.038 & 99.037 & 99.254 & 99.255\\ [0.4ex]  
1.4 & $8.5\cdot 10^{-7}$ & $1.6\cdot 10^{-6}$ & 93.049 &  95.325 & 94.167 & 94.172 & 94.207 & 94.202 \\ [0.4ex]  
2.1 & $6.1\cdot 10^{-8}$ & $6.9\cdot 10^{-8}$  & 99.291 &  100.355 & 99.682 & 99.681 & 99.964 & 99.965 \\ [0.4ex] 
2.2 & $7.0\cdot 10^{-8}$ & $8.3\cdot 10^{-8}$  & 98.603 &  99.936 & 99.106 & 99.104 & 99.433 & 99.435 \\ [0.4ex] 
2.3 & $1.1\cdot 10^{-7}$ & $1.2\cdot 10^{-7}$  & 98.815 &  100.483 & 99.554 & 99.553 & 99.743 & 99.744 \\ [0.4ex] 
2.4 & $1.3\cdot 10^{-7}$ & $1.6\cdot 10^{-7}$ & 97.570 &  100.235 & 98.824 & 98.824 & 98.981 & 98.981 \\ [0.4ex] 
3.1 & $4.9\cdot 10^{-7}$ & $5.6\cdot 10^{-7}$  & 94.674  & 96.001 & 95.195 & 95.193 & 95.480  &  95.482\\ [0.4ex]
3.2 & $6.1\cdot 10^{-7}$ & $7.6\cdot 10^{-7}$  & 91.860 & 92.927 & 92.364 & 92.365 & 92.423  &  94.422\\ [0.4ex]
3.3 & $7.0\cdot 10^{-7}$ & $9.6\cdot 10^{-7}$  & 96.484  & 97.790 & 97.104 & 97.107 & 97.169  &  97.166\\ [0.4ex]
3.4 & $4.3\cdot 10^{-7}$ & $7.7\cdot 10^{-7}$  & 94.220  & 95.458 & 94.815 & 94.824 & 94.860  &  94.851\\ [0.4ex]
4.1 & $1.2\cdot 10^{-7}$ & $1.3\cdot 10^{-7}$  & 102.151 &  103.112 & 102.523 & 102.525 & 102.740 & 102.738 \\ [0.4ex]  
4.2 & $1.3\cdot 10^{-7}$ & $1.7\cdot 10^{-7}$  & 104.327 &  105.242 & 104.691 & 104.701 & 104.878 & 104.868 \\ [0.4ex]  
4.3 & $1.8\cdot 10^{-7}$ & $2.2\cdot 10^{-7}$ & 104.293 &  105.355 & 104.697 & 104.706 & 104.951 & 104.942 \\ [0.4ex]  
4.4 & $1.5\cdot 10^{-8}$ & $1.6\cdot 10^{-8}$  & 107.991 &  108.884 & 108.377 & 108.377 & 108.498 & 108.498 \\ [0.4ex]  
 \hline
\end{tabular}
\end{center}
\caption {FTP for the different bonds in imbalanced states.}
\label{ftp}
\vspace{5mm}
\end{table}

\newpage
\section*{Conclusion}

In this paper, we developed a new approach based on the use of a bidimensional Markov-modulated Poisson process to model liquidity in OTC markets relying of requests for quote. The statistical estimation procedure we proposed is based on an EM algorithm and can be used either at the asset level or at a more macroscopic level. Although asymmetric states are hard to identify with great confidence, we showed on corporate bond data that flow imbalances contain information about the evolution of the price. We used flow asymmetries to generalize the notion of micro-price proposed by Stoikov in the context of markets organized around limit order books. We also coined a new concept inspired by the recent OTC market making literature: Fair Transfer Price. It is related to the quotes proposed by a market maker who takes flow imbalances into account and, therefore, projects liquidity asymmetries onto the price space. We noticed that the price adjustments associated with FTP are often small, smaller than those associated with micro-prices.\\

\section*{Acknowledgment}

This research has been conducted with the support of J.P.~Morgan and under the aegis of the Institut Louis Bachelier. The ideas presented in this paper do not necessarily reflect the views or practices of J.P.~Morgan. The authors would like to thank Morten Andersen (J.P.~Morgan), Gabriele Butti (J.P.~Morgan), and Nabil Nouaman (J.P.~Morgan) for the numerous and insightful discussions they had with them on the subject. The paper was presented at several conferences and seminars, including the 17th Financial Risks International Forum, the London Mathematical Finance Seminar Series, the LPSM seminar ``Mathématiques financières et actuarielles, probabilités numériques'', the Imperial Finance and Stochastics seminar, and the EWGCFM meeting at Khalifa University (Abu Dhabi). The audience at these talks should be warmly thanked.

\section*{Data availability statement}

Due to confidentiality reasons, the data used in this article cannot be made publicly available.\\

\newpage

\appendix

\section{Two important extensions}\label{ext}
\label{annexe}
\subsection{An exchangeability assumption  to impose symmetry in the asymmetries}
\label{MMPPex}
In the estimation procedure proposed in Section 2, we considered two sets \(\{\lambda^{1,b}, \ldots, \lambda^{m_b,b}\}\) and \(\{\lambda^{1,a}, \ldots, \lambda^{m_a,a}\}\): one for the bid and one for the ask. Even if one considers \(m^b = m^a = m\), when estimating intensities on real data, there is no chance that the estimated parameters will coincide between the bid and the ask. However, if the parameters are close and/or if there is no reason to believe that there is a structural asymmetry between the bid and the ask, it makes sense to impose that both sides share a unique set of intensities \(\{\lambda^{1}, \ldots, \lambda^{m}\}\).\\

In this appendix, we further assume some form of symmetry in liquidity asymmetries: there may be periods when liquidity is higher on one side than the other, but the exact opposite could have happened with the same probability. In mathematical terms, this corresponds to a point-in-time exchangeability assumption. In our Markovian setup, this means that the transition matrix \(Q\) of the Markov chain \((\lambda^b_t, \lambda^a_t)_t\) is also that of the Markov chain \((\lambda^a_t, \lambda^b_t)_t\). These assumptions are essential to build a model where prices are driven by imbalances. In particular, they guarantee that the price process does not drift indefinitely in such a model.\\

A natural question is of course that of estimating the intensities (or equivalently the diagonal matrix $\Lambda = \textrm{diag}(\lambda^{1}, \ldots, \lambda^{m})$) and the transition matrix $Q \in M_{m^2}$ using a set of RFQs at the bid and at the ask.\\

The likelihood computed in Section \ref{LLsec} is of course valid in the specific case we consider here with $\Lambda^b = \Lambda^a = \Lambda$, but the EM algorithm has to be adapted to take into account the constraints imposed by the symmetry assumptions. The log-likelihood \eqref{llc2} now writes
\begin{eqnarray*}
&&\mathcal{L}(Q,\Lambda | t_1, \ldots, t_N, \frak s_1, \ldots \frak s_N, \tau_1, \ldots, \tau_{P+1}, s^b_0, \ldots, s^b_P, s^a_0, \ldots, s^a_P)\\
&=& \log\left((\pi_0)_{(s^b_0-1)  m +s^a_0}\right) + \sum_{\substack{1\le j_b \le m\\1\le j_a \le m}}\sum_{\substack{1\le k_b \le m\\1\le k_a \le m\\ (k_b, k_a) \neq (j_b,j_a)}} \tilde{n}^{(j_b,j_a),(k_b,k_a)}\log(Q_{(j_b-1)m+j_a,(k_b-1)m+k_a})\nonumber\\
&& - \sum_{\substack{1\le j_b \le m\\1\le j_a \le m}} \left(\left(\sum_{\substack{1\le k_b \le m\\1\le k_a \le m\\(k_b, k_a) \neq (j_b,j_a)}} Q_{(j_b-1)m+j_a,(k_b-1)m+k_a}\right) + \lambda^{j_b} +\lambda^{j_a}\right)\tilde{T}^{(j_b, j_a)} \nonumber\\
&&+ \sum_{\substack{1\le j_b \le m\\1\le j_a \le m}} \tilde{n}^b_{(j_b,j_a)} \log(\lambda^{j_b}) + \sum_{\substack{1\le j_b \le m\\1\le j_a \le m}} \tilde{n}^a_{(j_b,j_a)} \log(\lambda^{j_a})
\end{eqnarray*}
and the matrix $Q$ verifies $Q_{(j_b-1)m+j_a,(k_b-1)m+k_a} = Q_{(j_a-1)m+j_b,(k_a-1)m+k_b}$ for $1 \le j_b,k_b \le m$ and $1 \le j_a,k_a \le m$.\\ 

Subsequently, the $M$-step (\textit{i.e.} the update) is modified and becomes
$$ \widehat{\Lambda}_{j,j} \leftarrow \frac{\sum_{k=1}^{m}\mathbb{E}_{\widehat{\Lambda}, \widehat{Q}, t_1, \ldots t_N, \frak s_1, \ldots \frak s_N}\left[\tilde{n}_{(j,k)}^{b}\right] + \sum_{k=1}^{m}\mathbb{E}_{\widehat{\Lambda}, \widehat{Q}, t_1, \ldots t_N, \frak s_1, \ldots \frak s_N}\left[\tilde{n}_{(k,j)}^{a}\right]}{\sum_{k=1}^{m}\mathbb{E}_{\widehat{\Lambda}, \widehat{Q}, t_1, \ldots t_N, \frak s_1, \ldots \frak s_N}\left[\tilde{T}^{(j,k)}\right]+\sum_{k=1}^{m}\mathbb{E}_{\widehat{\Lambda}, \widehat{Q}, t_1, \ldots t_N, \frak s_1, \ldots \frak s_N}\left[\tilde{T}^{(k,j)}\right]}\quad \text{ for } 1 \le j \le m,$$
and, for $1 \le j_b,k_b \le m$ and $1 \le j_a,k_a \le m$ with $(j_b, j_a) \neq (k_b,k_a)$,
$$\widehat{Q}_{(j_b-1)m+j_a,(k_b-1)m+k_a} \leftarrow \frac{\mathbb{E}_{\widehat{\Lambda}, \widehat{Q}, t_1, \ldots t_N, \frak s_1, \ldots \frak s_N}\left[\tilde{n}^{(j_b,j_a),(k_b,k_a)}\right] \! + \!  \mathbb{E}_{\widehat{\Lambda}, \widehat{Q}, t_1, \ldots t_N, \frak s_1, \ldots \frak s_N}\left[\tilde{n}^{(j_a,j_b),(k_a,k_b)}\right]}{\mathbb{E}_{\widehat{\Lambda}, \widehat{Q}, t_1, \ldots t_N, \frak s_1, \ldots \frak s_N}\left[\tilde{T}^{(j_b,j_a)}\right] + \mathbb{E}_{\widehat{\Lambda}, \widehat{Q}, t_1, \ldots t_N, \frak s_1, \ldots \frak s_N}\left[\tilde{T}^{(j_a,j_b)}\right]}$$ 
where the expectations are the same as in Section \ref{EMsubsec} with $\widehat{\Lambda^b} =\widehat{\Lambda^a} = \widehat{\Lambda}$.

\subsection{A multi-asset extension}\label{MMPPmulti}

In what follows, we consider a set of \(d\) assets and propose a one-factor liquidity model that echoes, in some sense, the CAPM. More precisely, we consider a Markov chain similar to the one used above, but we assume that the intensity process of asset \(i\) is given by
\[
(\lambda^i_t)_t = (\lambda^{i,b}_t, \lambda^{i,a}_t)_t = (\beta^{i,b} \lambda^b_t, \beta^{i,a} \lambda^a_t)_t.
\]

In other words, \((\lambda_t)_t = (\lambda^b_t, \lambda^a_t)_t\) represents an aggregate, while asset-level sensitivities to this aggregate are represented by coefficients \((\beta^{i,b})_i\) and \((\beta^{i,a})_i\).\footnote{For identifiability reasons, we consider the normalization \(\sum_{i=1}^d \beta^{i,b} = \sum_{i=1}^d \beta^{i,a} = 1\).}\\

To compute the likelihood of a sequence of RFQ times $t_1< \ldots< t_N$ corresponding to RFQs in assets $i_1, \ldots, i_N$ and sides $\frak s_1, \ldots, \frak s_N$ where the sides are encoded as elements of $\{b,a\}$ as above, let us introduce two counting processes $(N^{RFQ,i,b}_t)_t$ and $(N^{RFQ,i,a}_t)_t$ for each asset $i$, and the function 
$$\mathcal{G} : t \mapsto (\mathcal G^{(j_b-1) m_a+j_a, (k_b-1) m_a+k_a}(t))_{1\le j_b,k_b\le m_b,1\le j_a,k_a\le m_a}$$
where $$\mathcal G^{(j_b-1) m_a+j_a, (k_b-1) m_a+k_a}(t) = \mathbb P(\forall i, N^{RFQ,i,b}_t = 0, N^{RFQ,i,a}_t = 0,  \lambda_t = (\lambda^{k_b,b}, \lambda^{k_a,a})  | \lambda_0 = (\lambda^{j_b,b}, \lambda^{j_a,a}))$$

Using the same reasoning as in Section 2, we obtain for $h>0$, $1\le j_b,k_b\le m_b$ and $1\le j_a,k_a\le m_a$:
\begin{eqnarray*}
&&\mathcal G^{(j_b-1) m_a+j_a, (k_b-1) m_a+k_a}(t+h)\\
&=& \mathbb P(\forall i, N^{RFQ,i,b}_{t+h} = 0, N^{RFQ,i,a}_{t+h} = 0,  \lambda_{t+h} = (\lambda^{k_b,b}, \lambda^{k_a,a})  | \lambda_0 = (\lambda^{j_b,b}, \lambda^{j_a,a}))\\
&=& \sum_{l_b=1}^{m_b} \sum_{l_a=1}^{m_a}  \mathbb P(\forall i, N^{RFQ,i,b}_{t+h} = 0, N^{RFQ,i,a}_{t+h} = 0,  \lambda_{t+h} = (\lambda^{k_b,b}, \lambda^{k_a,a}), \lambda_{t} = (\lambda^{l_b,b}, \lambda^{l_a,a})  | \lambda_0 = (\lambda^{j_b,b}, \lambda^{j_a,a}))\\
&=& \sum_{l_b=1}^{m_b} \sum_{l_a=1}^{m_a} \mathcal G^{(j_b-1) m_a+j_a, (l_b-1) m_a+l_a}(t) \mathbb P\left(\forall i, N^{RFQ,i,b}_{t+h} = 0, N^{RFQ,i,a}_{t+h} = 0,  \lambda_{t+h} = (\lambda^{k_b,b}, \lambda^{k_a,a})\right.\\
&& \qquad\qquad\qquad\qquad\qquad\qquad\qquad\qquad\qquad \left.\Big| \forall i, N^{RFQ,i,b}_{t} = 0, N^{RFQ,i,a}_{t} = 0,  \lambda_{t} = (\lambda^{l_b,b}, \lambda^{l_a,a})\right) \\
&=& \mathcal G^{(j_b-1) m_a+j_a, (k_b-1) m_a+k_a}(t) \left(1+Q_{(k_b-1) m_a+k_a, (k_b-1) m_a+k_a}h +o(h)\right)\\
&&\qquad\qquad\qquad\qquad\qquad\qquad\qquad\times\prod_{i=1}^d\left(1-\beta^{i,b}\lambda^{k_b,b} h + o(h)\right)\left(1-\beta^{i,a}\lambda^{k_a,a} h + o(h)\right)\\
&&+ \sum_{1\le l_b \le m_b, 1\le l_a \le m_a, (l_b,l_a) \neq (k_b,k_a)}  \mathcal G^{(j_b-1) m_a+j_a, (l_b-1) m_a+l_a}(t) \left(Q_{(l_b-1) m_a+l_a, (k_b-1) m_a+k_a}h + o(h)\right).
\end{eqnarray*}

This leads to the following differential equation:
\begin{eqnarray*}
&&\frac{d\ }{dt}\mathcal G^{(j_b-1) m_a+j_a, (k_b-1) m_a+k_a}(t)\\
&=& \mathcal G^{(j_b-1) m_a+j_a, (k_b-1) m_a+k_a}(t) \left(Q_{(k_b-1) m_a+k_a, (k_b-1) m_a+k_a} - \sum_{i=1}^d \beta^{i,b}\lambda^{k_b,b} - \sum_{i=1}^d \beta^{i,b}\lambda^{k_a,a}\right)\\
&& + \sum_{1\le l_b \le m_b, 1\le l_a \le m_a, (l_b,l_a) \neq (k_b,k_a)}  \mathcal G^{(j_b-1) m_a+j_a, (l_b-1) m_a+l_a}(t) Q_{(l_b-1) m_a+l_a, (k_b-1) m_a+k_a}
\end{eqnarray*}
which, in matrix form, writes $$\mathcal G'(t) =  \mathcal G(t) \left(Q - \sum_{i=1}^d \beta^{i,b} \Lambda^b \otimes I_{m_a} - \sum_{i=1}^d \beta^{i,a} I_{m_b} \otimes \Lambda^a\right).$$

As $\mathcal G(0) = I_{m_bm_a}$, we conclude that $$\mathcal G(t) = \exp\left(\left(Q - \sum_{i=1}^d \beta^{i,b} \Lambda^b \otimes I_{m_a} - \sum_{i=1}^d \beta^{i,a} I_{m_b} \otimes \Lambda^a\right)t\right) = \exp\left(\left(Q - \Lambda^b \otimes I_{m_a} - I_{m_b} \otimes \Lambda^a\right)t\right)$$ thanks to the normalization choice.\\

If we assume that $\lambda_0$ is distributed according to $\pi_0$, then, using the same reasoning as above, the likelihood writes

\begin{eqnarray*}
&&\mathcal{L}(Q,\Lambda^b,\Lambda^a| t_1, \ldots, t_N, i_1, \ldots i_N, \frak s_1, \ldots \frak s_N)\\
&=& \pi'\left(\prod_{n=1}^N \exp\left(\left(Q - \tilde{\Lambda}^b - \tilde{\Lambda}^a\right)(t_n-t_{n-1})\right) \beta^{i_n,\frak s_n} \tilde{\Lambda}^{\frak s_n}\right)e\\
&=& \left(\prod_{i=1}^d (\beta^{i,b})^{K^{i,b}}\right) \left(\prod_{i=1}^d (\beta^{i,a})^{K^{i,a}}\right) \pi'\left(\prod_{n=1}^N \exp\left(\left(Q - \tilde{\Lambda}^b - \tilde{\Lambda}^a\right)(t_n-t_{n-1})\right)\tilde{\Lambda}^{\frak s_n}\right)e
\end{eqnarray*}
where $K^{i,b}  = \text{Card}(\{n, i_n = i, \frak s_n = b\})$ and $K^{i,a}  = \text{Card}(\{n, i_n = i, \frak s_n = a\})$.\\

From this expression we deduce that (i) we can merge RFQs at the bid across assets and RFQs at the ask across assets to estimate the parameters of $Q$, $\Lambda^b$ and $\Lambda^a$ using the EM algorithm of Section \ref{EMsec} or that of Appendix \ref{MMPPex}, and (ii) we can separately estimate the $\beta$ coefficients. Regarding the former, our EM algorithms can be used on merged data. The latter (the estimation of the sensitivities) is trivial: maximizing  $\prod_{i=1}^d (\beta^{i,b})^{K^{i,b}}$ subject to $\sum_{i=1}^d \beta^{i,b} =1$ indeed boils down to setting $\beta^{i,b}$ proportional to $K^{i,b}$, \textit{i.e.} $\beta^{i,b} = \frac{K^{i,b}}{\sum_{j=1}^d K^{j,b}}$ -- and similarly we obtain $\beta^{i,a} = \frac{K^{i,a}}{\sum_{j=1}^d K^{j,a}}$.

\end{document}